# Cross-Scale

## Multi-scale coupling in space plasmas

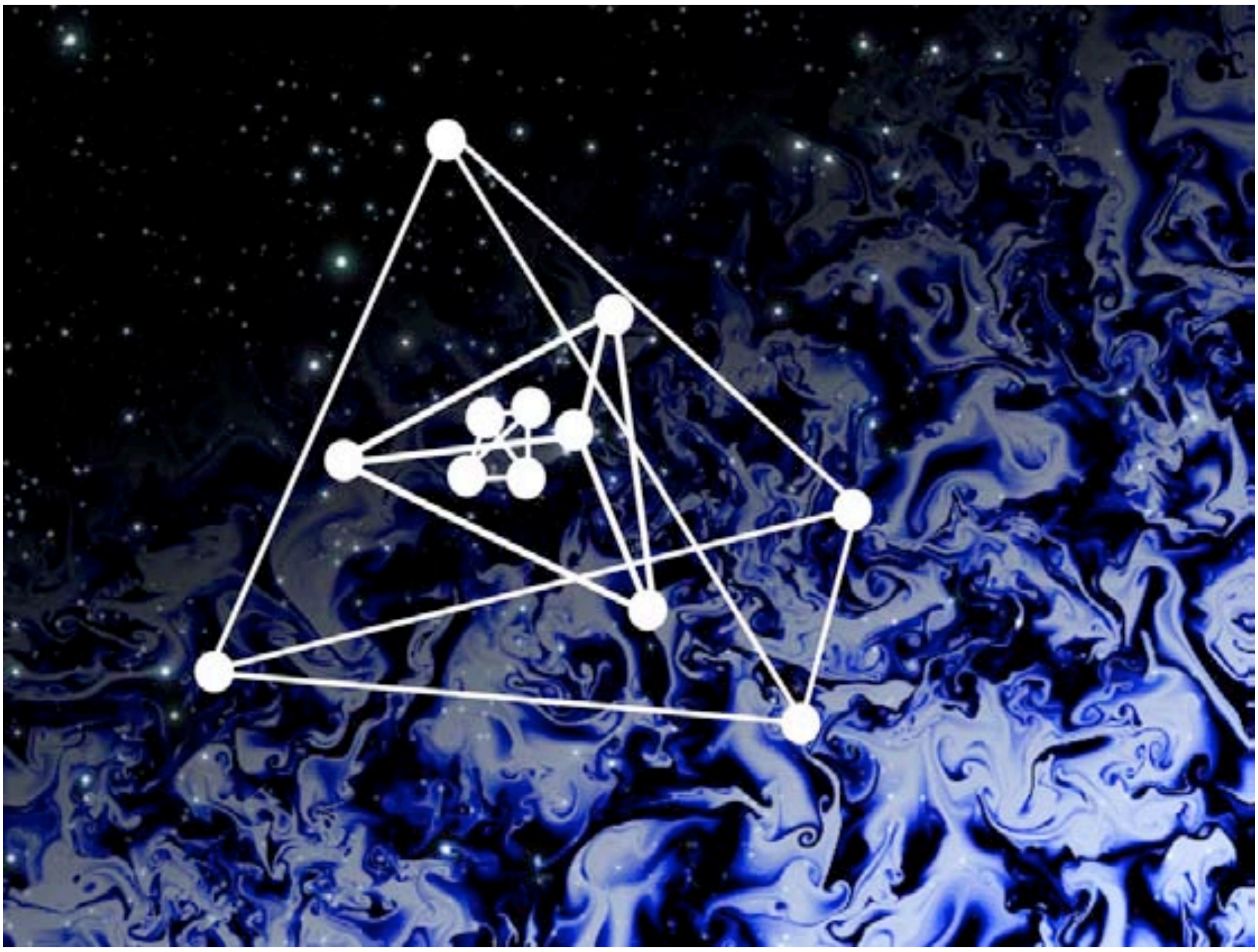

## Assessment Study Report

# Mission Description

| | | | |
|---|---|---|---|
| **Key scientific objectives** | Detailed in situ multi-spacecraft exploration of universal plasma phenomena occurring in near-Earth space, in particular:<br>o *Shock processes*<br>o *Reconnection*<br>o *Turbulence*<br>These processes involve structured, 3-dimensional, time-varying interactions across multiple length scales (electron, ion and magnetohydrodynamic fluid). The standalone 7 spacecraft ESA mission examined in this study will enable coupling between pairs of scales to be probed simultaneously. International collaboration would expand this to cover all scales at the same time. | | |
| **Strawman reference payloads** | Combination of the following state-of-the-art particle and fields instruments with strong heritage: DC and AC magnetometers, 2D electric field instruments (wire booms), electron density sounder, electron and ion electrostatic analyzers, ion composition analyzers, high energy particle detectors. | | |
| **Transfer to operational orbit** | • Launch by Soyuz-Fregat 2-1B from Kourou, October 2017<br>• Direct insertion into 200 km × 5.3 $R_E$ (33.800 km), i=4°, 3570 kg<br>or 219 km × 3.8 $R_E$ (23.867 km), i=4°, 3703 kg<br>• Transfer to operational orbit by dedicated dispenser-like transfer vehicle<br>• Use of lunar resonance optional (~355 m/s saving) | | |
| **Initial operational orbit**<br>(no active orbit control) | 10 $R_E$ × 25 $R_E$<br>Orbital period: 103.9 h (4 days 7h)<br>Inclination: 14°, argument of perigee = 205° (October tailbox crossing) | | |
| **Radiation environment** | Without lunar resonance: 59 krad (1.5mm), 6.2 krad (4mm)<br>with lunar resonance: 66 krad (1.5mm), 11 krad (4mm) | | |
| **Operational lifetime** | • 1 year commissioning and early science operations<br>• 2 years science operational phase<br>- electron scale spacecraft moving to fluid scale after 1 year<br>• 2 years extended science operation | | |
| **Spacecraft Modules** | **Transfer stage** | **Electron-scale transferred to Fluid scale after 1 year** | **Ion-scale** |
| | **Two options presented, second option values given in []** | | |
| Number of spacecraft | 1 | 4 (3 ⇒ fluid) | 3 (+1 e/I shared) |
| Spacecraft separation | N/A | 2 – 100 km<br>( fluid 3000 – 15000km) | 50 – 2000 km |
| Stabilization | Slow spin<br>[3 axis] | 15 rpm spin | 15 rpm spin |
| spacecraft Δv requirements | 1633 m/s [1420 m/s] | 70-199 [114] m/s | 14.2 [42] m/s |
| Design lifetime | several weeks<br>[6 month on case of lunar resonance] | 5 year | 5 year |
| Platform dry mass (excl. P/L)<br>Incl. margin | 261.1 kg<br>[558 kg] | 210.5 kg<br>[161.6 kg] | 210.5 kg<br>[159.2 kg] |
| Model P/L mass / power | - | 15-33kg / 15-60W | 28-33kg /44-60W |
| Total mass (incl. propellant and 20% system margin) | 1687 kg<br>[2001 kg] | 255 kg average<br>[161.1-179 kg] | 255 kg<br>[170.1-172.8 kg] |
| Maximum power | 0W<br>[1402 W] | 220-236 W<br>[221 W] | 226-236 W<br>[221 W] |
| Telemetry band | S-band | X-band [S-band] | X-band [S-band] |
| Continuous compressed science data bit rate per spacecraft | few kbps | 170-1300 kbps | 90 -170 kbps |
| Average downlink capacity | 800 kb/s | | |
| **Key mission drivers** | • Multiple spacecraft assembly, integration and verification (AIV)<br>• Multiple spacecraft mission operations<br>• Payload accommodation/interface requirements<br>• Data downlink volume | | |
| **Key technological challenges** | • Inter spacecraft synchronization / localization<br>• Star mapper upgrade to 15 RPM<br>• Flash-memory (size & space qualification) | | |

# Foreword

Driven by the support and interest of the international space plasma community to examine simultaneous physical plasma scales and their interactions, the Cross-Scale Mission concept was submitted and accepted as an ESA Cosmic Vision M-class candidate mission. The Cross-Scale mission proposal comprised 10 ESA spacecraft to join 2 spacecraft from its sister mission, SCOPE, provided by JAXA, to form 3 nested tetrahedra to address the science of cross-scale coupling in plasma.

In late 2007 a Science Study Team (SST) was set up for Cross-Scale, eventually consisting of representatives of the European, Japanese, American and Canadian Plasma Physics communities, whose initial activities involved the formation of the initial payload definition document based on the science requirements with the aid of the community. At the same time ESA's internal Concurrent Design Facility (CDF) carried out its own study in preparation of the Industrial studies, examining mission feasibility and identifying critical areas of mission design and any technology developments.

In September 2008 the two industrial studies began, officially starting Cross-Scale's assessment phase, while in parallel instrument consortia teams began studying the payload components of the mission. Based on information from the CDF and SST, it was decided that the 10 ESA spacecraft concept from the original proposal was a significant cost driver and beyond the single Soyuz launch capacity, so the industrial contractors were informed via the ESA study team and SST to focus the studies on examining the feasibility of a 7 spacecraft ESA mission. This approach was to identify the key requirements of an independent ESA mission without discounting the feasibility of cooperation with international partners. Indeed, the full SCOPE concept, comprising 5 spacecraft, was successfully progressing at the same time and in January 2009 passed its mission design review to move into Phase A study with a collaboration between JAXA and CSA. The benefits of such a SCOPE Cross-Scale collaboration are clear, but have not been the focus of this assessment study.

This report presents an overview of the assessment study phase of the 7 ESA spacecraft Cross-Scale mission. Where appropriate, discussion of the benefit of international collaboration with the SCOPE mission, as well as other interested parties, is included.

# Authorship and Acknowledgements

Cross-Scale Science Study Team:

Lead Scientist: Steve Schwartz, Imperial College, London

Stuart D. Bale, Space Sciences Laboratory, University of California, Berkeley

Masaki Fujimoto, ISAS/JAXA, Japan

Petr Hellinger, Institute of Atmospheric Physics and Astronomical Institute, Prague, Czech Republic

Mona Kessel, NASA, USA

Guan Le, NASA, USA

William Liu, CAS, Canada

Philippe Louarn, CESR, Toulouse, France

Ian Mann, University of Alberta, Canada

Rumi Nakamura, IWF, Graz, Austria,

Chris Owen, Mullard Space Science Laboratory, UCL, Surrey, UK

Jean-Louis Pinçon, LPC2E-CNRS, Orléans, France

Luca Sorriso-Valvo, LICRYL-CNR, Rende, Italy

Andris Vaivads, Swedish Institute of Space Physics, Uppsala, Sweden.

Robert F. Wimmer-Schweingruber, Institut für experimentelle und angewandte Physik, Universitaet Kiel, Germany

ESA Study Manager: Peter Falkner, ESA/ESTEC

ESA Deputy Study manager and Payload Manager: Arno Wielders, ESA/ESTEC, The Netherlands

ESA Study Scientist team: C. Philippe Escoubet (2007-2008), Matt Taylor and Arnaud Masson, ESA/ESTEC, The Netherlands

The following people are acknowledged for their valuable contributions:

Peter Cargill, Patrick Chaizy, Luke Drury, Melvyn Goldstein, Mike Hapgood, Harald Kucharek, Benoit Lavraud, Hermann Opgenoorth, Goetz Paschmann, Yoshifumi Saito, Michelle Thomsen, Ellen Zweibel

ESA planning and coordination office:

Marcello Coradini, Philippe Escoubet (2009 onwards), Fabio Favata and Timo Prusti

# Table Of Contents

# 1 Executive Summary

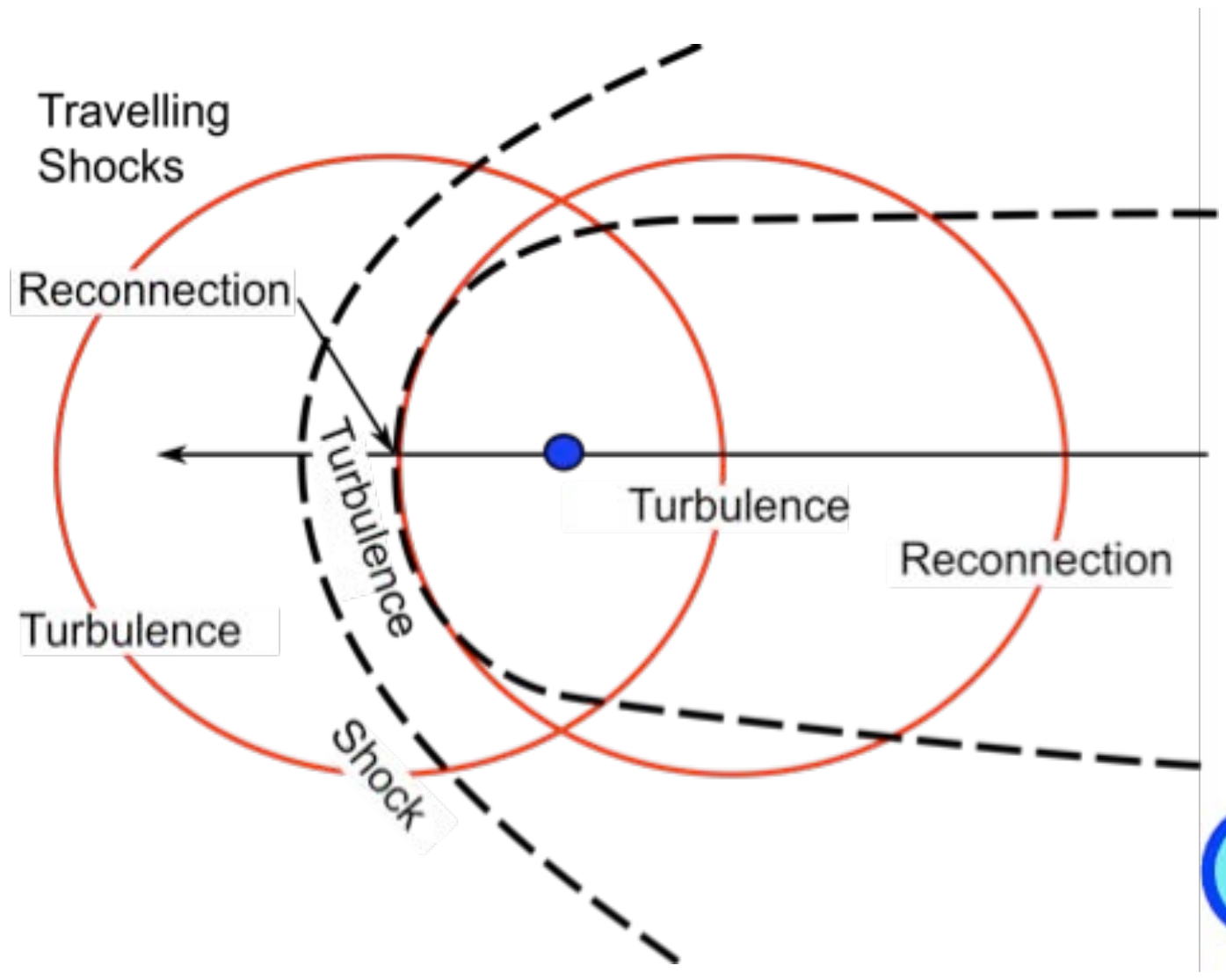

Regions around the Earth where shock waves, reconnection and turbulence are at work. The red lines indicate the Cross-Scale near equatorial orbit, which facilitates sampling of all these regions. The dashed lines are the magnetopause and bow shock.

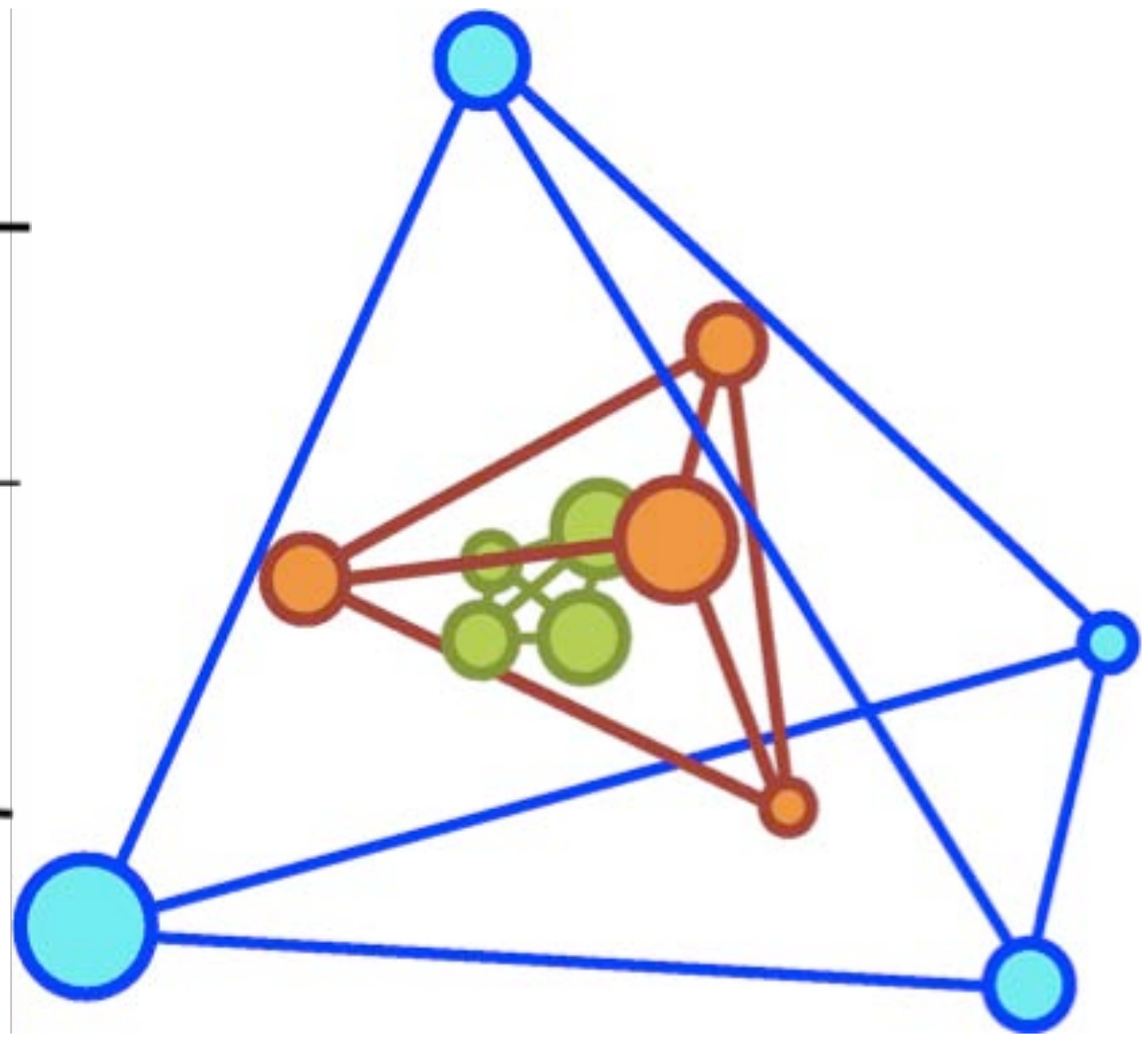

The Cross-Scale concept of a multi-scale configuration of spacecraft, in this case showing the optimum 12 spacecraft at each vertex of the three nested tetrahedra.

## 1.1 Science Objectives

Cross-Scale is an M-class mission dedicated to the study of three fundamental physical processes behind some of the most energetic phenomena in the Universe, from radio galaxy jets, supernovae remnants, and cosmic rays to solar flares and magnetic storms.

**Shock waves** guide strong flows around obstacles or at interfaces between two flow regimes (e.g. supernovae remnants, stars, magnetized planets). They are important locations for the transfer of directed bulk flow energy into heat, with an attendant acceleration of energetic particles (e.g. coronal mass ejections).

**Magnetic reconnection** releases stored magnetic energy to the plasma in the form of both heat and accelerated particles (e.g. solar flares). It couples neighbouring plasmas and allows an exchange of material between previously isolated regions. Moreover, the consequent change in magnetic topologies provides a coupling between plasma regions which often drives the global scale dynamics of the system (e.g. magnetic storms and disruption of magnetic confinement devices).

**Turbulence** transports energy from large scales at which it is input to small scales where it is dissipated with important implications for the physics of, for example, the solar wind, magnetotail, and interstellar medium. In the process, it interacts strongly, and often selectively, with plasma particle populations as a source and/or sink of energy.

To date, through in situ data taken within the heliosphere, remote sensing of the solar surface and more distant astrophysical objects, and theoretical modeling, we know that shocks, reconnection and turbulence operate and play pivotal roles throughout the Universe. The explosive nature of solar flares, together with early theoretical difficulties in matching observed timescales, forced a recognition that understanding the temporal variability would be necessary to even estimate the efficiency of reconnection in processing a given amount of magnetic energy. With dual spacecraft studies (ISEE, AMPTE) and more recently the first multispacecraft mission (Cluster) we now appreciate that, in addition, the spatial topology of the reconnected fields and the acceleration region at collisionless shocks is highly structured, braided, and hence spatially complex at every scale. Progressing from a qualitative realization that shocks, reconnection and turbulence are fundamental plasma phenomena to a quantitative analysis capable of matching existing observations and of extrapolation to other scenarios cannot be accomplished without taking the critical step to unravel how the key temporal and spatial scales conspire with one another. Central to the conspiracy in a collisionless plasma is the coupling amongst electron, ion, and fluid-level behaviour through the action of the self-consistent electromagnetic fields.

Near-Earth space is a unique laboratory for quantifying the physics of these three processes, as unlike more distant regimes, we are able to sample it directly by plasma and fields experiments on satellites. Cross-Scale will be the first space mission capable of quantifying the highly variable and structured coupling between these three key physical scales (with enough spacecraft in a nested formation carrying state-of-the-art plasma instrumentation).

Most of the visible universe, including our solar system, is in the highly ionized plasma state and therefore Cross-Scale's focus has a very broad application. Improvement of our understanding of these fundamental processes of plasmas (shocks, reconnection, and turbulence) will provide us with the tools to understand more general aspects of space and plasma physics. In addition to the intrinsic physics of the plasma universe, Cross-Scale's focus on fundamental processes have direct bearing on human endeavours, including hazards to human exploration due to high energy radiation and energy generation in laboratory devices.

## 1.2 Science Questions

Cross-Scale will target compelling and fundamental questions, including:

How do shocks accelerate and heat particles?

- What mechanisms accelerate particles at shocks ?
- How is the energy incident on a shock partitioned?
- How do shock variability and reformation influence shock acceleration?

How does reconnection convert magnetic energy?

- What initiates magnetic reconnection?
- How does the magnetic topology evolve?
- How does reconnection accelerate particles and heat plasma?

How does turbulence control transport in plasmas?

- How does the turbulence cascade transfer energy across physical scales?
- How does the magnetic field break the symmetry of plasma turbulence?
- How does turbulence generate coherent structures?

These address directly the Cosmic Vision question "How does the Solar System work?" by studying basic processes occurring "From the Sun to the edge of the Solar System". By quantifying the fundamental processes involved, the advances made by the mission will extend beyond the Solar System to plasmas elsewhere in the Universe.

## 1.3 Mission Strategy and Profile

Cross-Scale will meet its scientific objectives by employing the 3D concepts and techniques, pioneered by Cluster, of a tetrahedral configuration to separate spatial and temporal variations over the scale of the tetrahedron. The Cross-Scale spacecraft will be flown to form two nested tetrahedra with logarithmically separated scales. By sharing a corner, this is accomplished with 7 spacecraft. This will enable Cross-Scale to measure spatial and temporal variations simultaneously on two separate scales for the first time. Focused instrumentation on each scale then enables the dynamic coupling across those key scales to be determined. Two reconfigurations of the constellation over the mission lifetime will re-deploy the spacecraft to different sized tetrahedra, enabling in pairwise fashion the exploration of the coupling amongst all three key scales (electron, ion, and fluid). Over the two-year mission in an eccentric orbit the spacecraft will encounter various collisionless shocks, explore regions of both spontaneous and strongly-driven reconnection, and investigate both nascent and highly evolved plasma turbulence.

The simultaneous coverage of all three scales requires a minimum of 10 spacecraft, and would maximize the science return in terms of both efficiency (no need to wait for reconfigurations) and comprehensive datasets. This goal can be achieved by combining Cross-Scale with its sister mission SCOPE [1]. SCOPE is a JAXA mission comprised of five spacecraft built by JAXA and the Canadian Space Agency.

## 1.4 Spacecraft and Payload

The Cross-Scale spacecraft are identical in bus design (for economy) with flexible resources to accommodate instrument suites tailored to the relevant scale. The SCOPE spacecraft complete the instrumentation and locations needed to meet the optimum science requirements across all three scales simultaneously. Dedicated hardware will determine the inter-satellite ranges at the smaller scales.

The ESA Concurrent Design Facility and two industry consortia have studied the technical aspects of the mission. While some aspects of the spacecraft are challenging (inter-spacecraft communications, large onboard memory, centralized processing, autonomous operation and data retrieval), there are no new technologies that must be established. Construction and calibration of the instruments is a large task (a factor ~ 2 larger than previous missions) but builds, together with the operation of a fleet of spacecraft, on recent European experience and success.

## 1.5 Operations

Limited instrument modes and minimal station-keeping are planned to restrict the ground segment resources required by the mission. A data segment is planned to ensure that the data are processed and calibrated, with high-quality science data made available in a timely fashion for the widest possible community.

## 1.6 Technology, Programmatics, and Status

There are no technologies that need to be developed or proven for a launch in 2017, or earlier. The mission is thus low-risk for high-science return. The considerable international interest in mission participation, in particular from Japan and Canada via the SCOPE mission, but also from USA and Russia, highlights the universal recognition of the main science objectives, with the potential to provide complementary measurements and redundancy to maximise and guarantee the science return.

Early in 2009, SCOPE passed its mission design review and has now moved into a joint Phase A study, where JAXA will provide the launcher, the thruster module and the mother-daughter spacecraft pair, while the Canadian Space Agency (CSA) will provide the three spacecraft of the outer-formation. This joint study will end with the joint System Requirements Review in 2010.

## 1.7 Conclusion

Cross-Scale is an innovative, next-generation mission with high-quality science objectives that attack fundamental physical phenomena in collisionless plasmas. It taps directly into European leadership in multi-point in situ space plasmas, and is built around flight-proven technologies.

# 2 Scientific Objectives

## 2.1 Universal Plasma Processes

A small number of phenomena dominate the behaviour and effects of plasmas throughout the Universe: collisionless shocks, magnetic reconnection and plasma turbulence. Shock waves resulting from supernova explosions and other energetic flows accelerate cosmic rays to high energies. Magnetic reconnection plays a pivotal role in the release of stored energy in phenomena as diverse as solar flares and γ-ray-rich "magnetars." Turbulence in astrophysical disks allows accretion to proceed by transporting angular momentum; it also channels energy from the largest scales through a cascade to the smallest where it dissipates.

### 2.1.1 Shocks

Shock waves are formed whenever a supersonic flow encounters an obstacle of some kind, including other material. The basic process involves a transition from supersonic to subsonic flow, so that information can propagate sufficiently upstream of the obstacle to deflect the oncoming flow. To do so, a shock must decelerate, deflect, compress, and heat the incident material. Classically, the shock transition occurs abruptly, on scales related to the dissipative (i.e., collisional) processes. In astrophysical contexts, however, the incident material is usually a highly-ionised plasma in which collisions are negligibly rare. Under these circumstances, shock waves may still exist, but the absence of collisional coupling can result in highly non-equilibrium physics that can be traced to the multiple scales associated with multiple species and fields.

Collisionless plasma shocks are some of the most spectacular, visually-striking and energetic phenomena in the Universe. Generated by supernovae, stellar winds, or the rapid motion of objects such as neutron stars, they have a number of important effects. Supernova shock waves may trigger the collapse of galactic nebulae and hence the formation of planetary systems. They are responsible for heating and deflecting the surrounding plasma, and blow large-scale magnetic bubbles out of galactic disks. Collisionless shocks also accelerate particles to extraordinarily high energies. A primary source for galactic cosmic rays is widely believed to be shocks driven by supernova remnants, where Mach numbers can reach 400 or more. Such shocks are strongly modified by the energetic particles they accelerate. However, a key, and unknown, ingredient in this process is the injection of suprathermal particles into the acceleration mechanism. That injection occurs over a thin "sub-shock" region at which the final bulk deceleration, corresponding to Mach numbers in the range 5-40, takes place.

The interaction of the fast-moving solar wind with the Earth's magnetosphere results in a bow shock. Its curvature means that regions of quasi-parallel (magnetic field parallel to the shock normal) and quasi-perpendicular shock front co-exist, and that the Mach number varies along the shock surface. Magnetohydrodynamic Mach numbers can reach 20, comparable to those in many astrophysical scenarios and directly relevant to the sub-shock regions that fuel the injection of suprathermal particles into the larger Fermi acceleration mechanism.

The terrestrial bow shock is therefore an accessible example of an astrophysical shock. The age, large scale, and strength of interplanetary shocks provide access to a wider range of parameters over which to explore the relative efficiencies of ion and electron acceleration. The bow shock accelerates copious amounts of ions accompanied by a relatively small number of largely coherently accelerated electrons. Astrophysical shocks are known electron accelerators (as observed directly by the X-ray and synchrotron emission) and presumed ion accelerators (e.g., cosmic ray protons). Moreover, the acceleration at supernova remnants is most intense at a pair of diametrically opposite locations suggesting that the external magnetic field orientation at the shock remains a key parameter under interstellar conditions. Thus high-resolution observations at interplanetary shocks, which sometimes show appreciable electron acceleration, offer breakthroughs in understanding what factors control the relative efficiencies of electron and ion acceleration.

The ability to measure particle distributions and electromagnetic fields around and within the shock front makes it possible to study the details of the collisionless shock transition and accompanying plasma phenomena. Historically, *in situ* measurements at the bow shock and interplanetary shocks have challenged existing theory and have led to significant advances in computational modeling.

### 2.1.2 Reconnection

Magnetic reconnection is a fundamental plasma physics process which breaks down the barriers between neighbouring plasmas, releasing energy from their magnetic fields, transferring material and momentum between those plasmas, and accelerating a part of the plasma population to high energies.

In highly-electrically-conductive space plasmas, the constituent charged particles are unable to convect transverse to the local magnetic field direction; the magnetic flux is "frozen in" to the plasma material. Thus two such plasmas in regions of magnetic flux with different sources generally cannot mix. They are separated by a sheet of electrical current that exerts the forces acting to keep the plasmas apart. The plasmas remain isolated from one another, even if pressed together by external forces or momentum. However, exceptions to this principle arise in situations in which magnetic reconnection occurs. A key consequence of magnetic reconnection is the linkage of magnetic fields from the two neighbouring plasmas, allowing material to flow between the previously isolated regions (e.g., from a stellar atmosphere to planetary magnetosphere or between different regions of the solar corona – see Figure 1). The reconnection process locally disrupts the current sheet, changes the topology of the magnetic fields, and releases its energy, e.g., into the bulk flow and heat of the plasma.

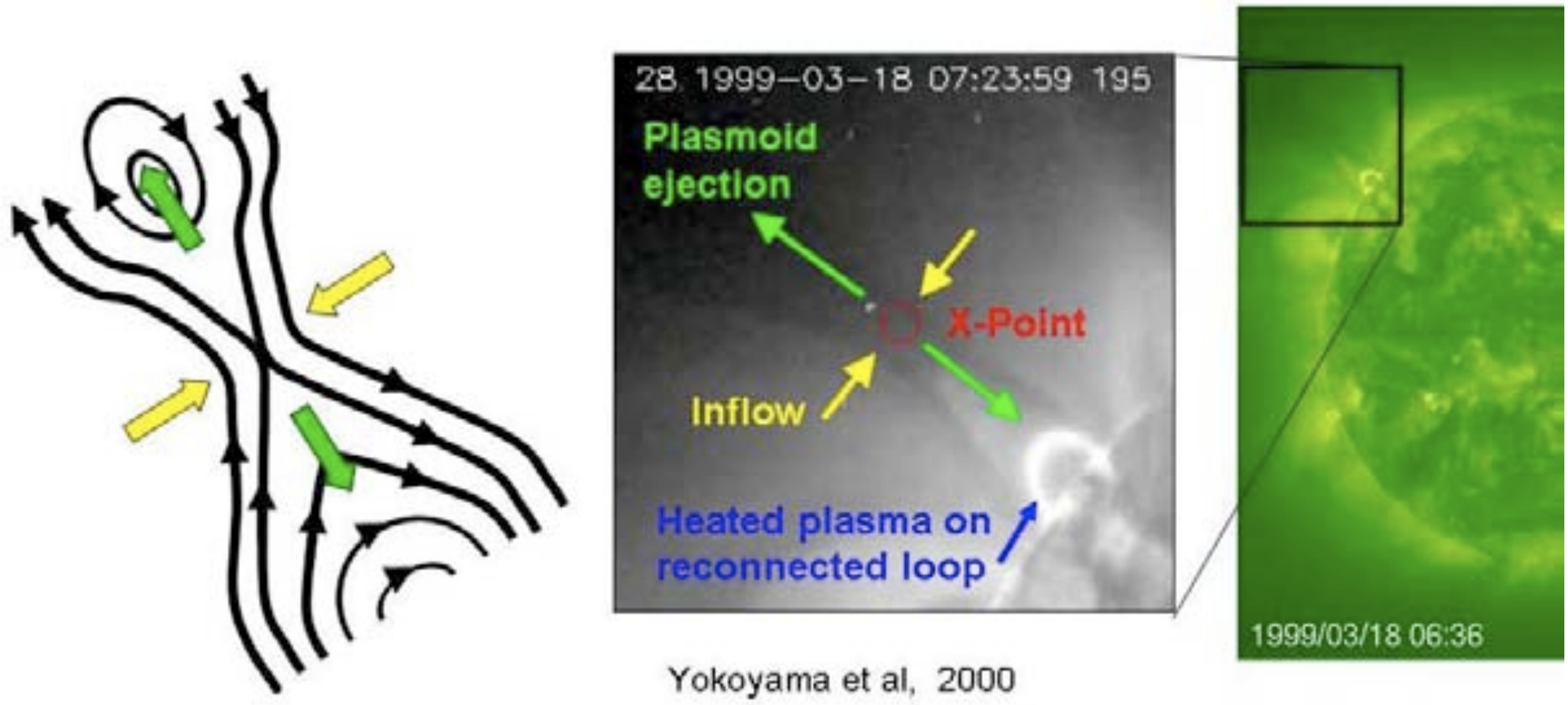

*Figure 1. The image on the right shows a reconnection event occurring in the Sun's corona. A study by Yokoyama et al. [2] identified coronal plasma in- and outflow (yellow and green arrows) as expected for a reconnection event, as well as heated material appearing on closed magnetic loops formed by reconnection.*

Many in situ measurements made by spacecraft at current sheets in the vicinity of Earth (principally the magnetopause and the magnetotail current sheet) have provided compelling evidence that reconnection does indeed occur in collisionless space plasmas. However, these have yet to provide clear indications of how it happens, or indeed an adequate picture of the conditions needed to initiate and maintain the process. For example, observations suggest that a necessary, but not sufficient, condition for sustained magnetic reconnection is a significant shear angle between the magnetic field vectors in the two adjacent plasma regions. The change in magnetic topology probably takes place through a complex 3D structure that involves ion, intermediary ("Hall"), and electron scales that determine, in ways not understood, which field lines and how much magnetic flux are involved.

The Universe has numerous situations where reconnection is expected to play significant roles in their dynamical evolution, including stars (exotic and otherwise) and planetary systems at all stages of their life cycles [3],[4]. Reconnection also governs the interactions of those systems with their surrounding media and often controls the dynamics of the systems themselves. In the interstellar medium, such topological reconfigurations enable the large-scale galactic magnetic field structure to emerge from the turbulent eddies that provide the initial dynamo action that generates it.

### 2.1.3 Turbulence

Turbulence occurs in many astrophysical plasmas such as the interstellar medium, accretion disks, stellar winds, supernova remnants, and collapsing nebulae. It is dynamically important in the transport of energy, mass and momentum in many of these environments. The accurate prediction of turbulent properties, and their effects on the surrounding plasma, is therefore key to the quantitative analysis of many astrophysical and solar system scenarios, as well as laboratory plasmas such as tokamaks.

The richness of plasma wave modes raises questions about the basic scaling of the transfer or cascade of energy from one scale to the next, and even whether the cascade proceeds in this manner at all scales. Magnetic turbulence has long been believed to strongly influence cosmic ray propagation and energisation, although progress has been traditionally hampered by numerous assumptions that have been made concerning the 3D nature of the turbulence. The larger-scale magnetic field raises questions about both the anisotropy and the spatial variability of the turbulence.

Measurements of turbulence in solar system plasmas have provided several key insights into its behaviour which are difficult or impossible to obtain any other way. We know that the turbulent energy appears to follow a power-spectrum that is indistinguishable (for unknown reasons) from the simpler fluid description up to kinetic scales, at which other spectral shapes and an apparent dissipation scale have recently been reported. We also know that the turbulence is not isotropic and have been able, within limited scales, to identify the parent plasma modes that hold part of the turbulent energy. Such advances both validate and challenge theoretical and numerical models.

Despite these successes, many key questions remain unanswered concerning the nature of the turbulent cascade, particularly near the ion and electron kinetic scales: How is it driven to an anisotropic state by the magnetic field and its local environment? How does it spontaneously generate structures in a previously uniform plasma? How is the energy ultimately dissipated; which plasma constituents receive and share this energy? Does the dissipation occur at thin, 3D vertical structures within the turbulent plasma? These issues must all be quantified if the large-scale effects of space plasma turbulence are to be accurately incorporated into the physical models of their environs.

### 2.1.4 Multi-scale Coupling

Critically, most astrophysical plasmas are collisionless, which means that their constituents can be far from equilibrium with each other. Shocks, reconnection, and turbulence are controlled by dynamics which are therefore coupled on 3 fundamental scales simultaneously: the electron kinetic, ion kinetic, and fluid scales. It is the nonlinear interaction of 3D, time-varying structures on these 3 scales which thus yields the consequences of these processes. The resulting nonlinear dynamics provides diverse and exotic mechanisms for momentum and energy flow and redistribution.

This complex, three dimensional nature of plasma structures has long been recognised. Previous, existing and upcoming missions have been designed to measure this 3D structure using multiple spacecraft. A minimum of four spacecraft are necessary to determine 3D structure: ESA's Cluster and NASA's MMS missions both use four spacecraft for this task. A fundamental restriction of multi-spacecraft measurements, however, is that they are sensitive only to scales of the order of the spacecraft separation. With four spacecraft, multiple scales can be probed by varying this separation, but only one scale can be measured at any time. Plasmas are not just three-dimensional: they also exhibit time-varying structure on many temporal scales, simultaneously. As a result the cadence of the measurements needs to be tuned to the temporal scale of the phenomena. Different scales are affected by different physical processes. It is the interplay of these which results in the complexity of shocks, reconnection, turbulence and other phenomena, and consequently in their large-scale effects.

To understand the interplay of forces and dynamics within such regions and hence predict their effects, it is essential to measure the time dependent behaviour in 3D at the relevant scales. For example, examining the relative heating mechanisms for ions and electrons at shocks and within reconnection regions requires simultaneous measurements at both ion and electron scales to measure the coupling between these two species. Four spacecraft at each scale, with scale-appropriate instrumentation and temporal resolution, can resolve these processes provided that the fluid-scale context can be inferred by passage through the corresponding region, by cataloguing separately the fluid-ion scale coupling with two pairs of tetrahedra, or ideally by the simultaneous measurement of the 3D fluid scale. This logic suggests an optimum complement of 12 spacecraft (3 independent tetrahedra on each of the three scales) that can be reduced, with some loss of contextual symmetry, by sharing a common corner to 10 spacecraft, and a minimum complement of 7 spacecraft to undertake at different stages the pairwise scale-coupling investigations.

The near Earth environment provides a perfect laboratory in which to investigate the cross-scale coupling of shocks, reconnection and turbulence. The contrast with astrophysical observations, all of which rely on remote sensing, cannot be overstated. In-situ measurements tell us how the charged particles behave (particles instruments are able to reconstruct the velocity space distribution function) and how the electromagnetic fields develop (magnetic and electric field measurements complete the complex wave-particle interplay of plasma physics). They provide crucial information for the fundamental understanding of the Plasma Universe that cannot be obtained by any other scheme.

While simulations of collisionless plasmas have revealed a great deal about their dynamics and complexity, it is not possible to simulate the key phenomena of interest in sufficiently large simulation boxes to resolve the three logarithmically-spaced physical scales in 3D. Indeed, this goal will not be achieved within the time-scale of ESA's Cosmic Vision 2015-2025 programme, assuming that computing capability continues to increase at its historical rate.

Laboratory plasmas are also incapable of probing over the necessary scales. The only way to validate theoretical concepts and limited numerical experiments concerning these phenomena, and so to study them in sufficient detail, is to measure them directly in space. Therefore near-Earth space, an accessible natural laboratory that contains examples of all the phenomena of interest, is the obvious target for such a mission, which we call Cross-Scale.

## 2.2 How do shocks accelerate and heat particles?

Cross-scale coupling is integral to shock processes in collisionless plasmas. Small-scale electron dynamics result in a highly-structured, fluctuating electric field within the shock ramp. At scales around an order of magnitude larger, ions gyrate through the ramp, with trajectories determined by the fluctuating electric field.

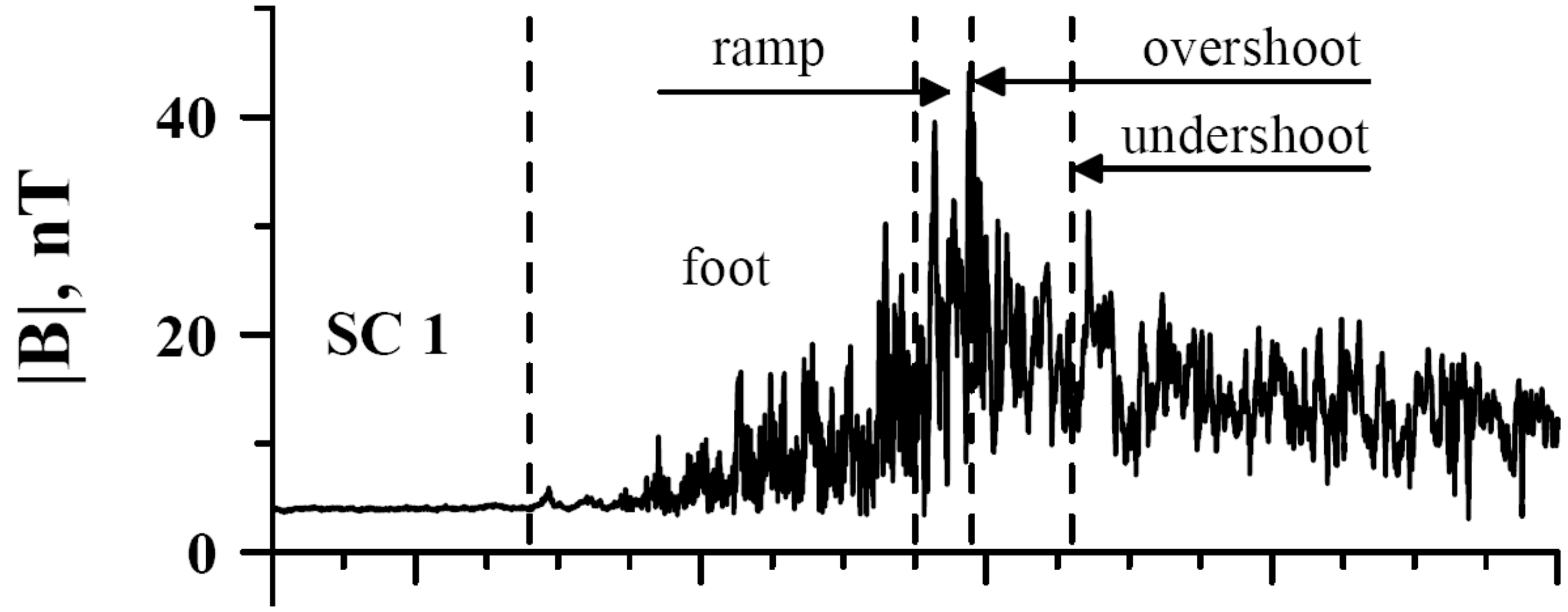

*Figure 2. The magnetic field structure at a typical collisionless shock. When observed by multiple spacecraft, this structure is shown to be highly variable on both the ion (foot) and electron (ramp) scales (From [5]).*

Reflected and gyrating ions then generate the even larger-scale reformation and rippling of the shock front, which in turn affect the small-scale dynamics of the electrons - as well as being pivotal for the acceleration of inflowing particles to high energies. Even a simple pass through a collisionless shock reveals the inherent structure that should be contrasted with the idealized discontinuity usually associated with simple shock waves.

Collisionless shocks both heat the main plasma constituents and accelerate sub-populations of ions and electrons to high energies. This acceleration is possible because the physics involves more than one scale. Acceleration is a part of the larger question about how shocks partition the bulk flow energy incident upon them, which again is distributed over fluid, ion, and electron scales. Finally, variations in the driving flows, and instabilities within the transition scales, lead to transient phenomena that can significantly alter or enhance the acceleration efficiency.

### 2.2.1 What mechanisms accelerate particles at shocks?

#### *Diffusive acceleration*

The velocity difference between the upstream and downstream plasma is a natural agent for particle energisation via the Fermi process in which particle scattering centres converge with respect to each other, resulting in a net gain after one complete shock crossing cycle. The total energy gain then depends on number of crossings and the relative velocities of the scattering centers, which are determined by the global shock structure together with the efficiency of the particle scattering.

This process involves some initial injection of particles and effective scattering. At collisionless shocks, some particles are reflected from the shock. Their propagation back into the upstream plasma is unstable to the generation of waves, which in turn scatter them. This is a highly nonlinear process, which initiates further acceleration. However, many aspects of this process involve spatially variable structures, within the shock and in both the upstream and downstream plasma.

Fundamental to modern acceleration theory is the scattering length of the energised particles. Repeated reflection and scattering (the so-called “first order Fermi” process) can accelerate particles to very high energies. Recent studies [6] have for the first time evaluated the scattering mean free path for shock-accelerated particles under unusually steady conditions. Simultaneous measurements of the resonant plasma waves and particles over ion and fluid scales are required to quantify the wave modes and particle interactions, and hence scattering properties, over a range of plasma conditions. Only in this way can we extrapolate shock acceleration efficiencies to other astrophysical environments.

Quasi-parallel shocks, at which the magnetic field is nearly aligned with the normal to the local shock surface, are associated with Short Large Amplitude Magnetic Structures (SLAMS) which grow in the generally turbulent foreshock/shock region accompanied by energetic particles. Their polarisation suggests they grow due to a hot ion instability, essentially feeding off the particle pressure gradients, whereas other foreshock turbulence is beam-driven. Thus the relationship between the SLAMS and the general turbulence field is not clear. The basic presumed structure is sketched in the case study described in Figure 3.

#### *Coherent acceleration*

The surface of a shock is believed to be rippled by local ion and current instabilities, but the properties of these ripples, e.g., amplitude and wavelength, are unknown. The ripples provide time-varying fields which can trap some particles, enabling them to “surf” the shock front and systematically pick up energy from the large-scale motional electric field. Such surfing is potentially important for both ion and electron acceleration, and may “inject” suprathermal particles into the first order Fermi process, the efficiency of which is dramatically improved if fed a pre-accelerated population. In order to quantify the effects of shock ripples, measurements are required of ripples of the shock surface at fluid scales; of ion distribution variations around and within the ripples, with variations on the scale of an ion gyroradius; and of electron heating and acceleration at the smallest scales.

Again, the origin and evolution of the cycle of turbulence and particle distributions needs to be explored by measurements at the disparate scales. SLAMS exhibit internal structure down to electron scales, and their overall size, shape and porosity are correspondingly difficult to determine without multi-scale observations.

Yet these characteristics are critical to our picture of quasi-parallel shocks. Large-scale measurements of the orientation of SLAMS, energetic particle gradients and foreshock waves must be combined with ion-scale structures within SLAMS, with electron-scale fine structure, and with 3D electric field measurements.

### 2.2.2 How is the energy incident on a shock partitioned?

In the frame of the shock, the upstream, incoming plasma carries kinetic, thermal and electromagnetic energy into the shock front. This energy is then partitioned into a number of forms. At modest Mach numbers, most is distributed between the downstream kinetic and thermal energies of the ions and electrons. Knowledge of the fraction of energy distributed into each of these plasma constituents is essential for identifying the key physical processes and for deducing the shock parameters, heating and acceleration efficiencies from the partial remote sensing observations of astrophysical shocks. For example, the energy taken up by electron heating at astrophysical shocks, together with the actual shape of electron energy spectra, is responsible for the observed X-ray emission. The ion heating must be inferred from those observations. It is possible to measure this partitioning directly at the terrestrial bow shock. We know that this partitioning varies considerably with shock parameters and is also highly variable in space and time. Therefore, the application of these measurements to astrophysical shocks is not straightforward without a detailed knowledge of the physical processes which govern the energy redistribution.

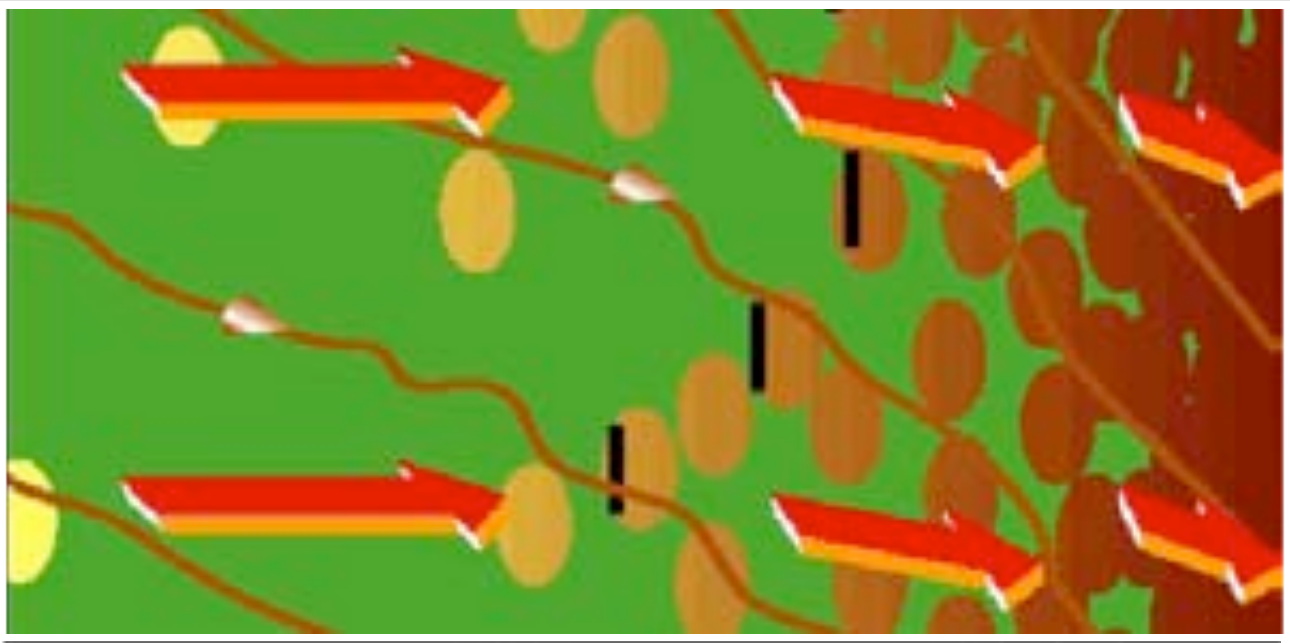

**The quasi-parallel shock transition, (upper panel) as envisaged in early work (after [7]) based on limited multi-spacecraft data. Studies showed that the coherent Short Large Amplitude Magnetic Structures (SLAMS) sweep up suprathermal particles (bottom left, from [8]) during the acceleration process, in contrast to the traditional picture of diffusive first-order Fermi acceleration with monotonic pressure profiles.**

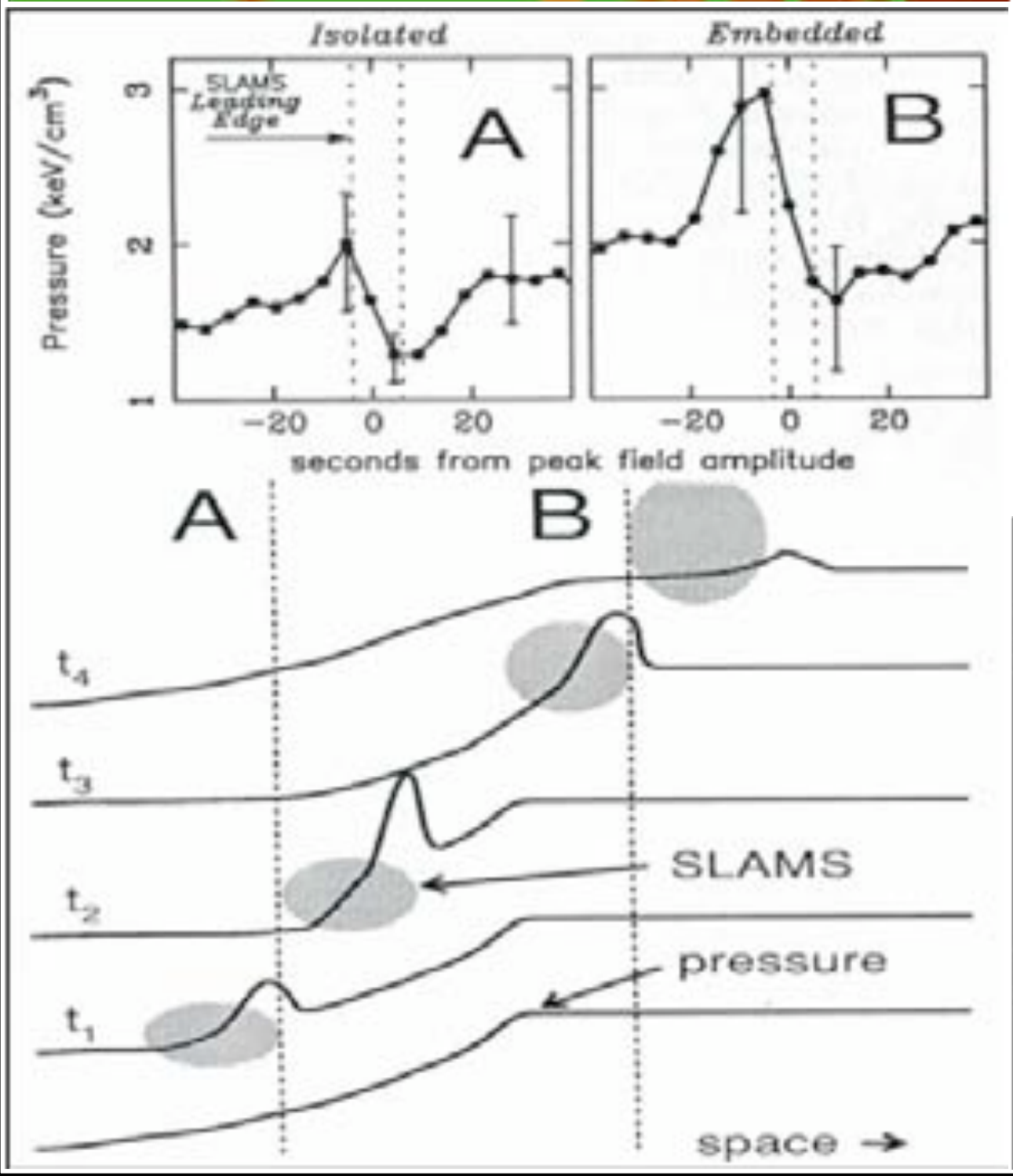

**More recent results (below) at 1/10$^{th}$ (top) and 1/100th (bottom) of the dimension inferred from [7] reveal considerable sub-structure, calling into question some of the key assumptions about the scales associated with shocks and Fermi acceleration under quasi-parallel conditions. The variability and small-scale structure has important, and to date elusive, implications for the efficiency of the quasi-parallel shock in terms of energy partition and ion acceleration (after [9] and [10]).**

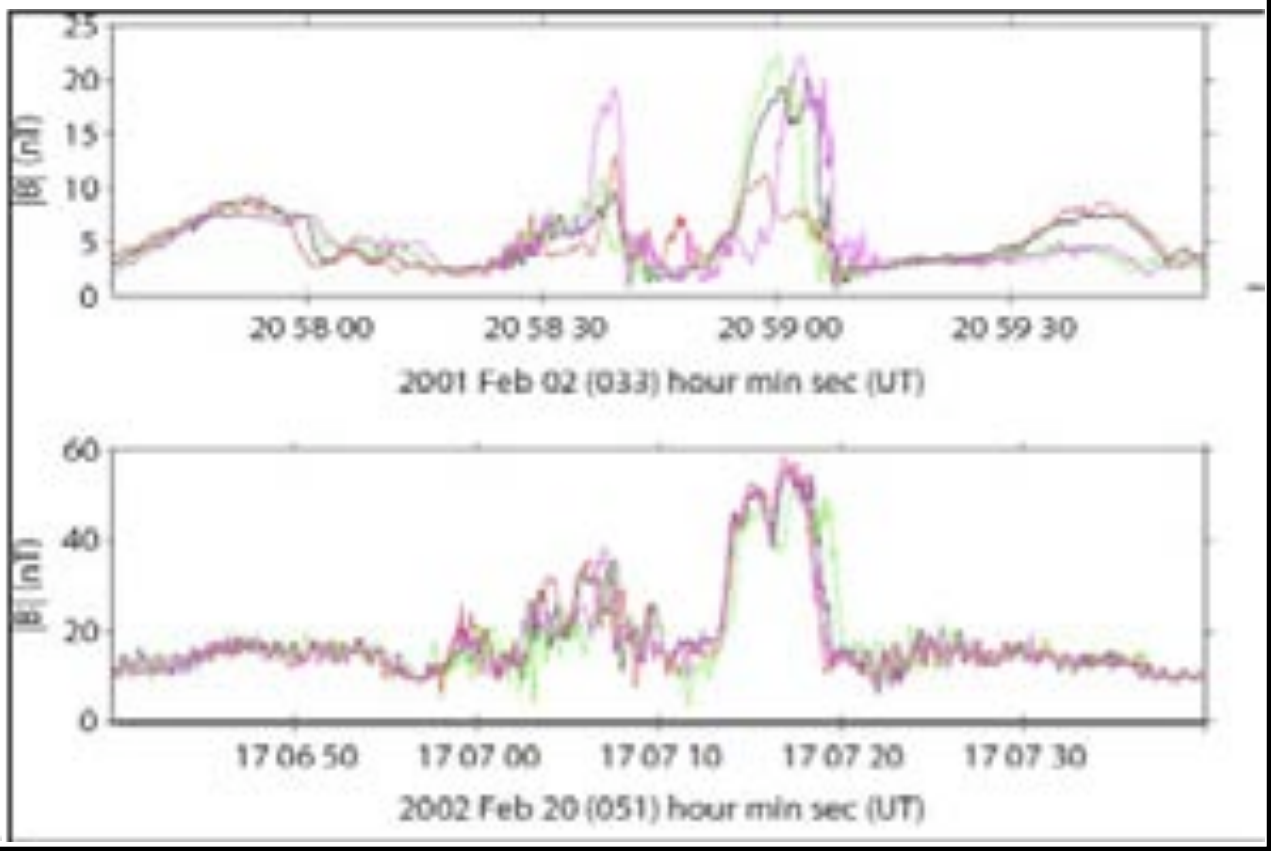

*Figure 3. Case Study 1: SLAMS and Particle Acceleration: A universal shock process in operation.*

## *Ion reflection*

The rise in electric potential through the shock layer is responsible for decelerating the incoming ions, some of which are reflecteid at shocks above a critical Mach number. This reflection initiates the spread in velocities that will ultimately account for the rise in non-directed motion (the “heating” required of the shock). However, the potential is the integral of the electric field across the shock - and this electric field is known to be highly-structured and variable on all measured scales down to that of the electron scales [11]. Figure 4 demonstrates large-amplitude, small-scale electric field spikes within the main shock transition. However, the role of such spikes and their variability requires electron-scale measurements of the 3D electric field together with larger-scale multi-point measurements to accurately determine the shock orientation, motion, and feedback processes.

## *Electron heating*

Just as ion trajectories within the shock are affected by small-scale electric field structures, electron trajectories and small scale electric field structures are affected in turn by the large-scale shock profile, which is constantly varying as a result of reformation and ion dynamics. Very little is currently known about the effect of large-scale shock variability on the formation, size and lifetime of small-scale electric field structures, such as those illustrated in Figure 4, but without this knowledge, the 3D electric field structure, and final electron distribution functions, cannot be determined. The shock potential, which is responsible for the ion reflection/heating process described above, is predominantly a thermo-electric field that is intrinsically related to the electron pressure gradient, or more correctly the electron kinematics in a collisionless plasma.

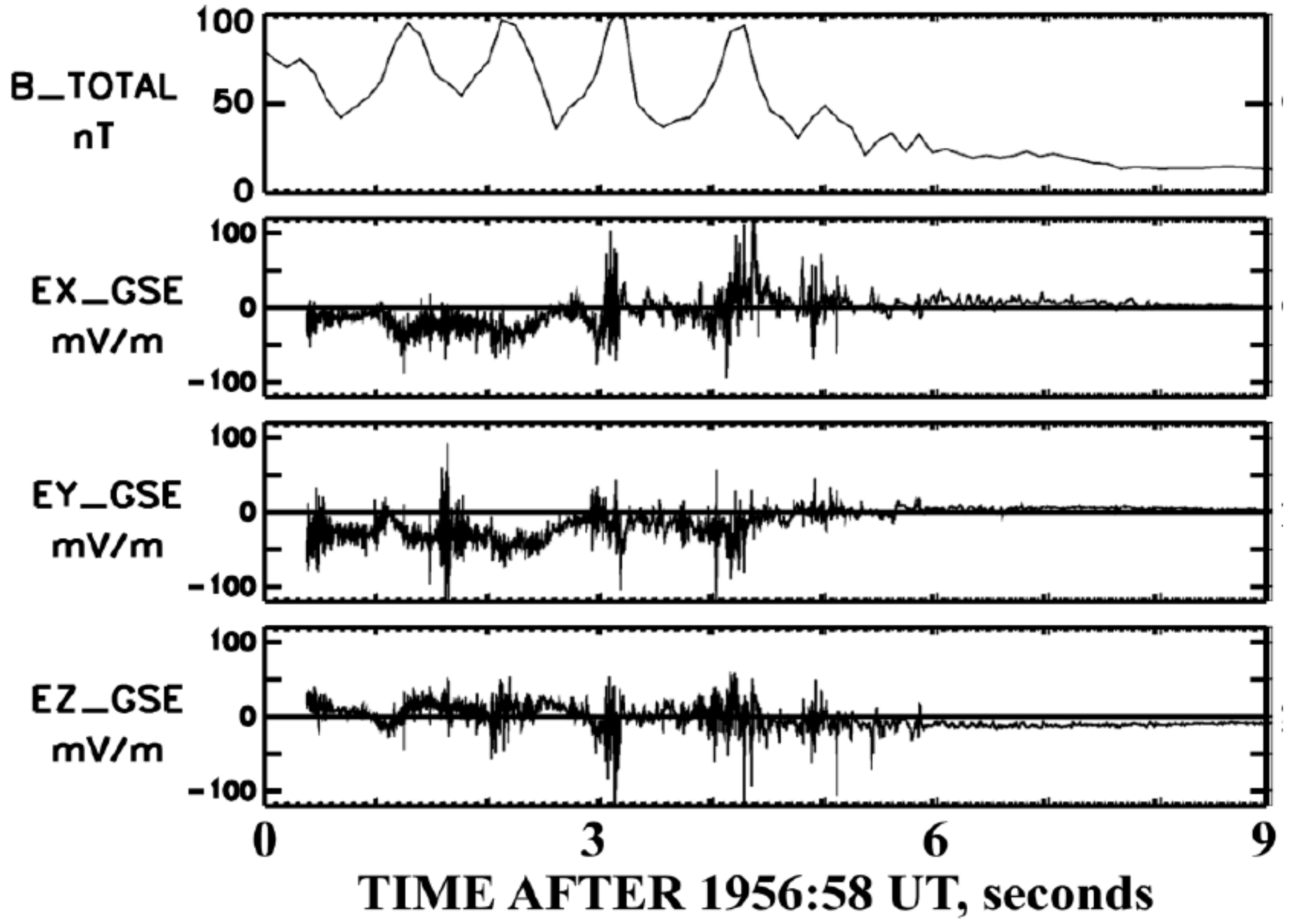

*Figure 4. Recent measurements of the 3D electric field at a high-Mach number collisionless shock which reveal the presence of electric field spikes (lower panels) on scales much smaller than the ion scale variation in the magnetic field shown in the top panel. The size and variability of such spikes is not known. Full 3D measurements are required to transform such electric fields into the cross-shock electric potential that controls both the ion and electron dynamics (from [12]).*

Thus the electron dynamics and ion heating are strongly interlinked processes. Of key importance is the nature of the electron motion: Magnetised or unmagnetised? Dominated by steady fields or high frequency fluctuations or turbulence? What is the feedback from the ion scales? Electrons with different energies and/or gyrophases may behave differently, requiring good coverage of the electron distributions over several decades of energy and resolved in pitch angle.

### *Heavier ion species heating*

Observations of heavier ions show that they are efficiently heated and accelerated at shocks. This is somewhat puzzling, since the main shock fields are self-consistently tuned to process the dominant momentum and energy carriers, which in the case of the solar wind are protons. The resolution of this puzzle almost undoubtedly rests in some interplay between the electric fields (at small scales), the (non-steady) proton reflection, and larger-scale variations. Some measurement of the heating and non-thermal components of several ion species (with different charge to mass ratios) will identify the physical mechanisms and their parametric dependences. Heavier ions are also believed to influence many aspects of shock variability, including both the upstream waves and the internal shock structure, making characterisation of the ionic content of the plasma a key ingredient in both the heating and acceleration aspects of shock physics.

## 2.2.3 How do shock variability and reformation influence shock acceleration?

### *Internal variability*

Supercritical shocks are fundamentally variable in time and space. They exhibit reformation, a quasi-periodic variation in the shock profile on scales comparable to the proton gyroradius. This results in a non-planar, and varying, shock profile, with important consequences for how particles are deflected, heated and accelerated.

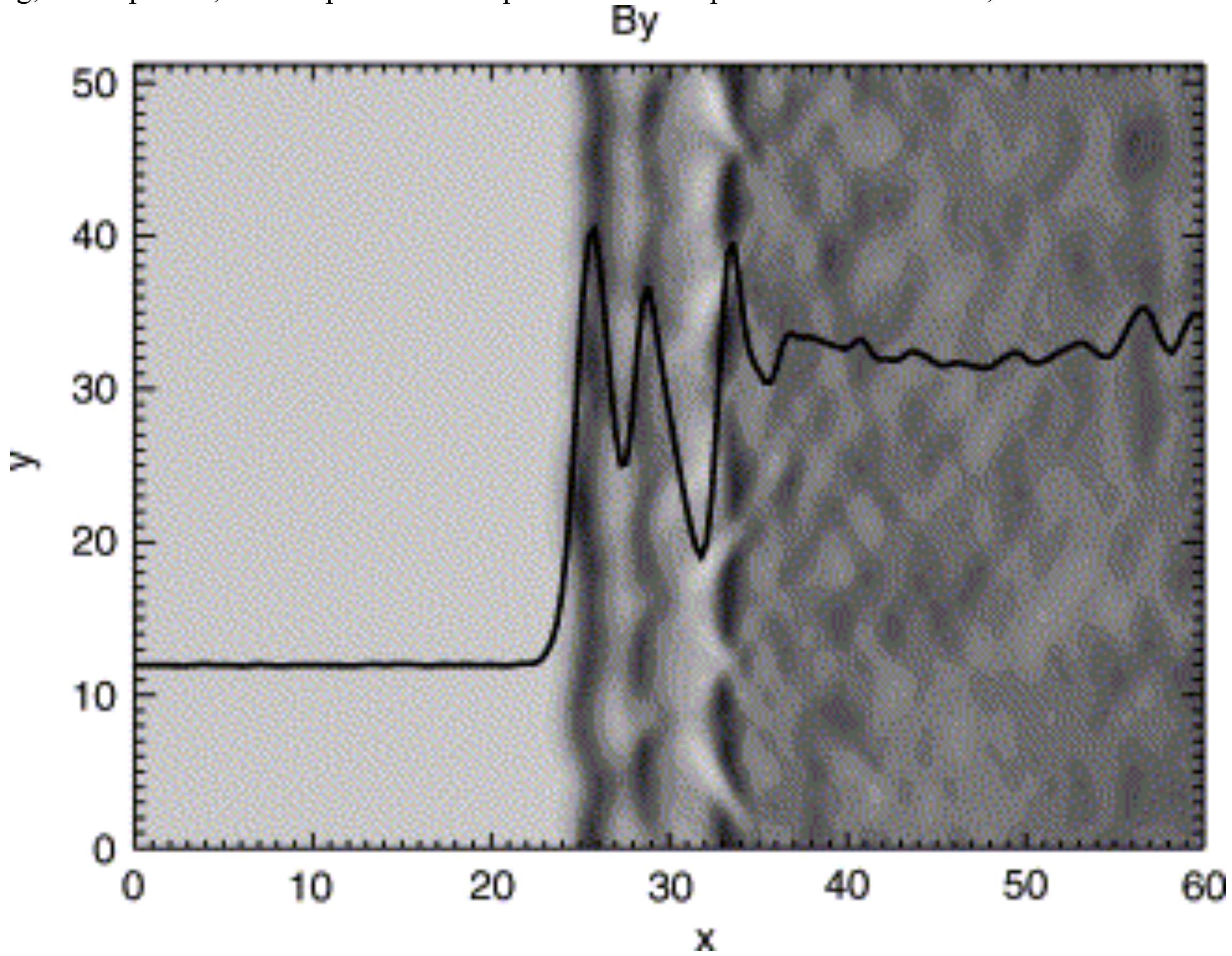

*Figure 5. Two-dimensional magnetic field maps of a simulated shock overlain with the average shock profile. Note the "rippled" appearance of the shock surface and region immediately downstream (to the right) (from [15]). These ripples vary in time, influence the ion dynamics/reflection, and serve as sites of possible electron trapping and acceleration.*

Computer simulations ([13],[14],[15]) such as that shown in Figure 5 illustrate the difficulty faced by two, or even four, spacecraft trying to disentangle the overall structure and dynamics of the shock surface ([13],[16]). The trajectory of an individual ion through the shock ramp is controlled by the instantaneous electric field that it encounters – but since this field is structured and variable, different ions follow markedly different trajectories. The number of ions reflected by the shock, which is directly related to the partition of energy problem, is known to vary (see Figure 6) but determining the feedback between that variability and the shock energy partition requires simultaneous multi-scale measurements.

This variability leads to many complex trajectories within the shock, some of which result in ions being ejected into the upstream plasma. There they seed the shock acceleration process and modify the upstream plasma conditions. Thus the impact of shocks on the ambient medium occurs over an extended region. A knowledge of the intricate feedback between very fine scale electric field structures and ion dynamics, and the resulting variability in the shock profile and structure, will lead to a definitive solution to the "injection problem" that is at the heart of shock acceleration as invoked for cosmic-ray production. Is this a simple consequence of non-ideal ion reflection or is it controlled by other parameters, such as the electron heating, plasma beta, and magnetic field geometry? The shock-accelerated particles influence the upstream flow, decelerating and deflecting it while increasing the energy density not associated with the directed incident bulk flow. Thus the accelerated particles modify the Mach number, plasma beta, and even magnetic field orientation found at the main shock transition. In more extreme circumstances at cosmic-ray modified shocks, the Mach number can be decreased by one to two orders of magnitude.

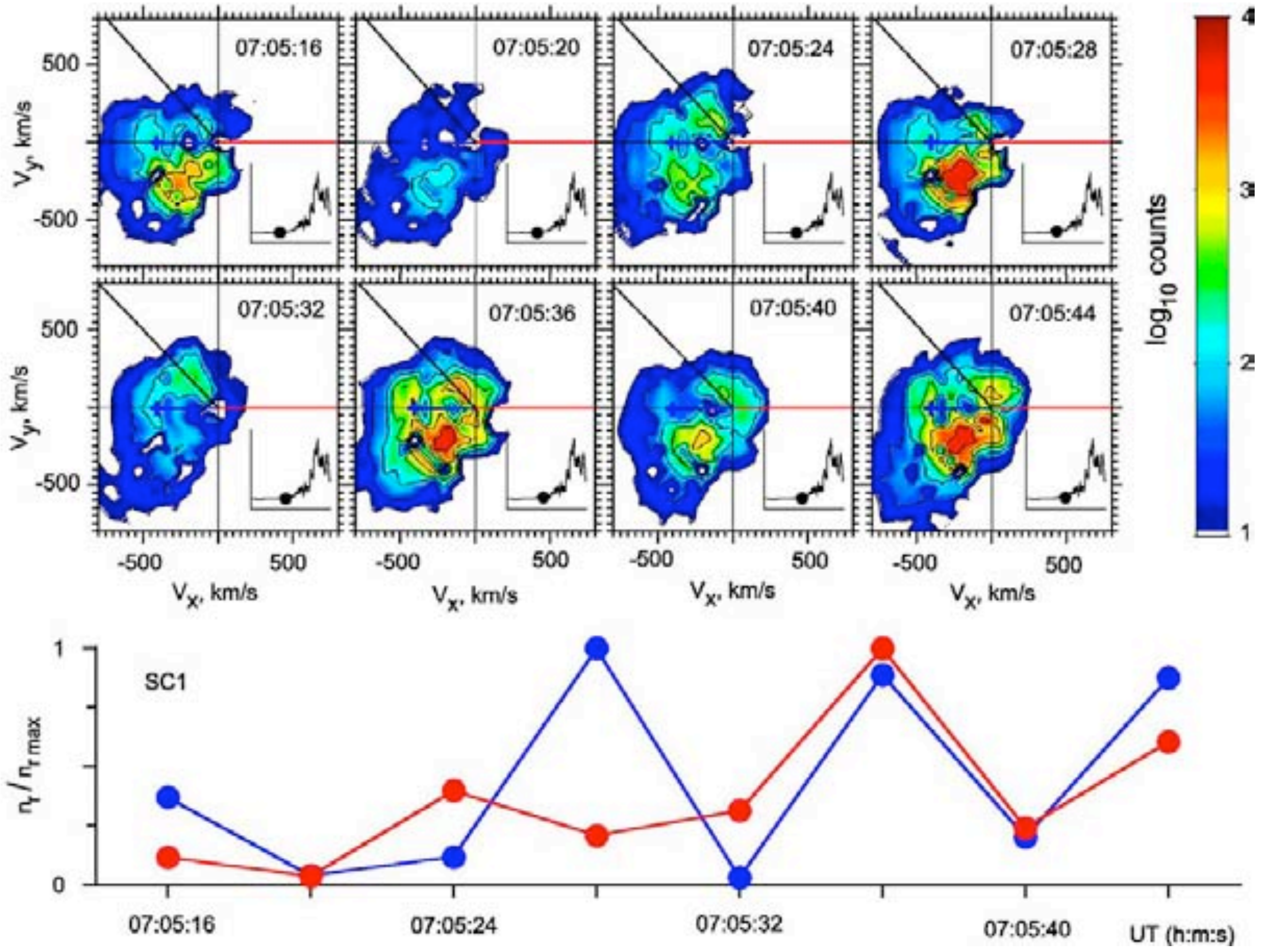

*Figure 6. Illustration of complex ion distributions near the upstream edge of a high Mach-number shock (top panels show 2D cuts of the ion velocity space distributions) and the variability of the total number density of reflected ions observed there (bottom) separated into two gyrophase populations. Quantitative assessment of the role of small scale fields, variability, and large-scale control and consequences await full multi-scale measurements (from [17]).*

Shock variability also has a strong influence on the acceleration of electrons. In the "fast Fermi process," incident electrons experience energy-gain during reflection that is a strong function of the angle between the magnetic field and local (on an electron scale) shock normal. Trapping within ripples and smaller variations enables electrons to reside close to the shock where they gain energy from the shock electric fields. Both these processes rely heavily on internal and variable shock structure.

### *Externally-induced variability*

Several circumstances conspire to complicate the physics at real shocks. Some, described above, are intrinsic variability. Others are due to external variations, which can have profound effects on the shock dynamics. For example, Hot Flow Anomalies [7] are formed when a plasma discontinuity with suitable orientation and parameters impinges on a shock front. Despite the overall pressure balance through such a discontinuity, and its thin structure, the particle dynamics at the interaction region can give rise to dramatic explosive events that create a hot cavity upstream of and/or attached to the shock. At the bow shock, these are observed to have some of the hottest, most fully thermalized particle distributions seen in situ anywhere in the solar system. It is not at all clear how particles are accelerated within the cavity of a hot flow anomaly, nor the role played by electron dynamics. As sources of energetic particles and possible triggers of further events, these structures should be the subject of further study. Measurements of the large, fluid-scale cavity shape and evolution must be made simultaneously with ion distributions within and around the cavity, as well as fine-scale electric field and electron variations, in order to understand the dynamics and effects of these phenomena.

Shock-shock interactions occur often in real systems, such as travelling interplanetary shocks colliding with planetary bow shocks, or forward shocks catching up with slower or reverse travelling shocks in strongly variable astrophysical systems. Theoretically, such interactions provide conditions for extreme heating and particle acceleration, and will be studied in detail for the first time by Cross-Scale.

## 2.3 How does reconnection convert magnetic energy?

Magnetic reconnection is expected to occur when magnetic fields are sheared across relatively thin current layers [18]. In such current sheets the kinetic effects of the particle populations become important, and the onset of reconnection is expected to occur on the distance and time scales of the relevant electron and ion gyromotions. Microscale processes control the change of topology of the magnetic field, eventually affecting the large-scale plasma mixing and converting magnetic energy to plasma energy. On the other hand, large-scale (MHD) processes control the location and formation of thin current sheets, and thus directly affect how reconnection initiates and evolves. It is therefore essential to follow both the large-scale and kinetic scale processes of the plasma to understand the onset, the evolution, and the result of magnetic reconnection.

Current sheets where reconnection takes place can be either large-scale boundaries such as solar wind discontinuities, or form dynamically on small scales, such as in plasma turbulence (as discussed in the case study presented in Figure 8) and within shear layer vortices [19]. Reconnection plays important roles in a variety of astrophysical contexts, as well as in the disruption of confined laboratory plasmas. The re-organisation of the solar atmosphere and release of energy associated with a solar flare requires a fast reconnection process, and motivated early reconnection theories. In turbulent, highly structured coronal plasma, multiple sites of small-scale reconnection are thought to be the solution to the "coronal heating problem." In the interstellar medium, reconnection is required to untangle the turbulent magnetic geometry to enable the galactic dynamo to maintain the global galactic field. Magnetic stresses and subsequent reconnection in the atmospheres of magnetars are responsible for some of the most extreme gamma-ray outbursts found in sky. Magnetic reconnection also occurs in ground laboratory plasma devices. These include both dedicated devices to study reconnection and in, for example, tokamaks as an unwanted effect. In particular, the occurrence of magnetic reconnection puts severe limits on the plasma regimes that are attainable in tokamaks. Better understanding of the magnetic reconnection process is therefore very important for control of fusion plasmas.

## *Quasi-steady 2D reconnection*

The most popular model of steady-state reconnection has anti-parallel magnetic fields either side of a current sheet. Plasma inflow and outflow vectors define a plane that is perpendicular to the current sheet and that also contains the anti-parallel magnetic fields (see Figure 7); the reconnection electric field is normal to this plane.

The change in magnetic topology occurs inside the so-called electron diffusion region, which is very thin, with an extent normal to the current sheet of order an electron gyroradius or less. This region is surrounded by an ion diffusion region no broader than an ion gyro-radius along the current sheet normal.

In their respective diffusion regions the ion/electrons are no longer magnetised, i.e. they no longer gyrate about the magnetic field and are not "frozen-in" to the magnetic flux. In the ion diffusion region, the reconnection electric field ensures that "frozen in" electrons continue to flow inwards towards the electron diffusion region but demagnetised ions are not affected in the same way. The electrons accelerate as the magnetic field weakens near the current sheet.

In the vicinity of a reconnection site, the relative motion of ions and electrons generates a "Hall current" and an associated "Hall electric field" acting to oppose the charge separation. As a consequence of this current system the local magnetic field is deflected out of the inflow/outflow plane, generating a characteristic quadrupolar "Hall magnetic field" signature. In this picture, the electron diffusion region is somewhat of a "black box" and processes within it that demagnetise the electrons allowing reconnection to occur are not well understood.

Single spacecraft observations have been interpreted as evidence of Hall magnetic fields, electric fields and outflow jets; these interpretations have been strengthened by 4-spacecraft multi-point Cluster observations showing that the fields and flows are organised in patterns consistent with expectations of this model. A more direct test for unmagnetised ions is needed to verify that the Hall fields are located within the ion diffusion region. This can only be accomplished by a comparison of the ion flow velocities and electric field drift velocities, which will differ there, set in their larger-scale context.

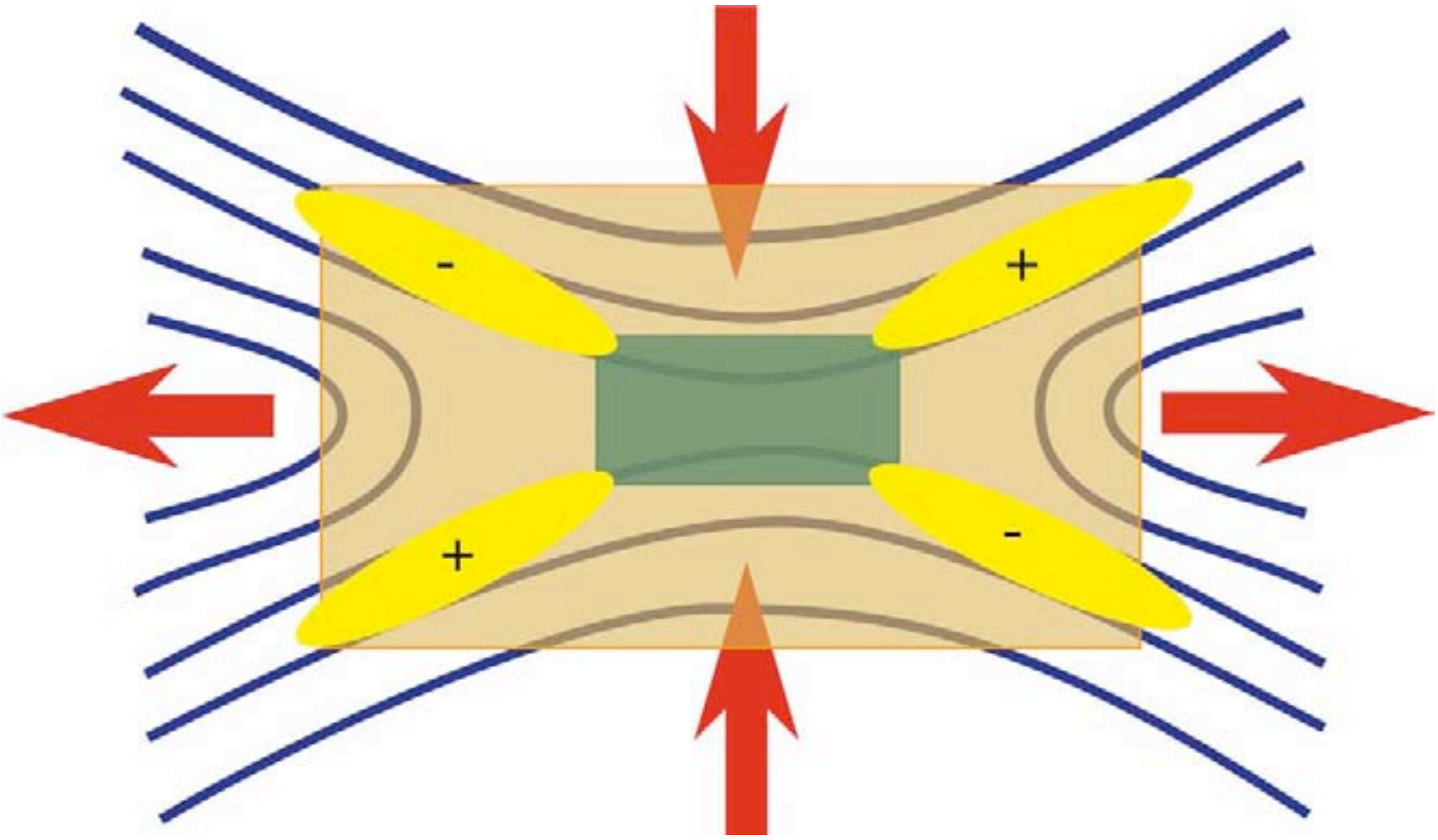

*Figure 7. Reconnection geometry in 2D showing the classic inflow and outflow regions, the change in magnetic topology, together with the ion (beige) and electron (green) diffusion regions and quadrupolar Hall current systems (yellow).*

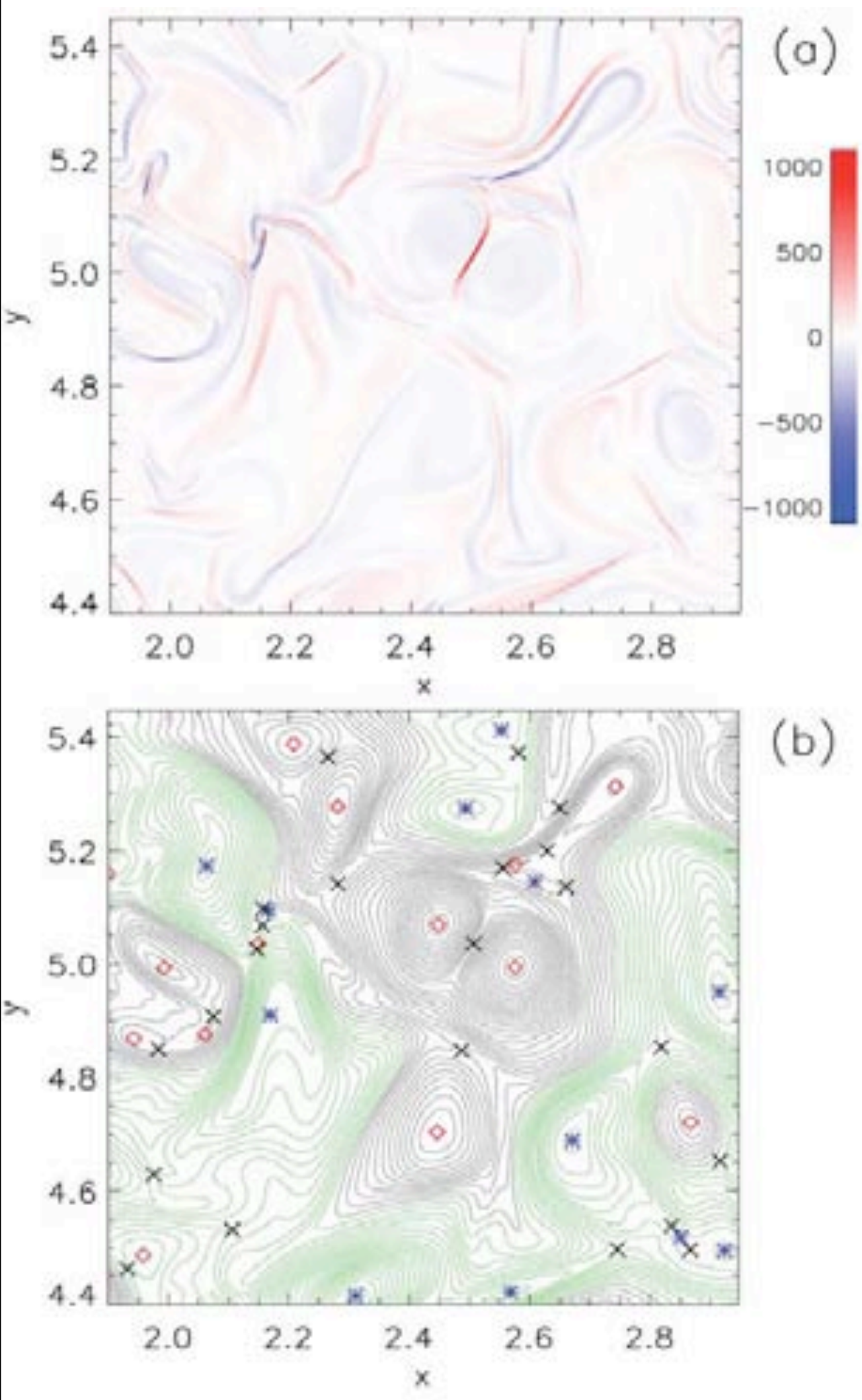

**Magnetic reconnection is a key process in interstellar gas [20]. It is necessary for operation of galactic dynamos and may be an important heating process. Since the Lundquist number in the interstellar medium is large ($10^{15}$ – $10^{21}$, depending on local conditions and scale), reconnection can be important only if it is fast, i.e. collisionless and independent of resistivity.**

**In order for interstellar reconnection to be collisionless, the medium must be structured on very small scales. Recently a number of studies have investigated the effect of turbulence - both MHD and collisionless - on reconnection ([21]-[25]). The panels show typical current sheet and field structures arising in a simulation [25] of a turbulent plasma (black crosses indicate X-points, blue stars and red diamonds indicate regions of maximum and minimum magnetic potential respectively). The scaling of the reconnection rate with system size and Lundquist number, the differences between 2D and 3D systems, the dependence on the turbulent spectrum, the effect of driven vs spontaneous turbulence, the role of collisionless effects, and the long time behaviour are still unclear.**

**However, recent Cluster observations ([26],[27]) have revealed, for the first time, in situ evidence of reconnection in a turbulent plasma ([28],[29]). Observations suggest that reconnection contributes significantly to particle energisation processes at the scales comparable to the small scales of current sheets or even smaller.**

*Figure 8. Case Study 2: Reconnection in a turbulent plasma.*

Such a comparison requires faster plasma measurements than have hitherto been available, 3D electric field measurements, and simultaneous observations at both the electron and ion scales of the inherent 3D structure. This test is an essential and unique Cross-Scale measurement that is central to all aspects of reconnection. Cross-Scale will investigate key specific outstanding questions on magnetic reconnection as highlighted in the following subsections.

The near-Earth plasma environment is unique in that we can make detailed measurements for different types of current sheets, providing insight into their formation and their microphysical processes. We can also examine in detail both the external conditions controlling the occurrence of reconnection and the large-scale consequences it has on the system.

### 2.3.1 What initiates magnetic reconnection?

In order for magnetic reconnection to occur, a region must exist in which the magnetic field can diffuse relative to the plasma and in particular across the current sheet separating two plasma regimes. Spacecraft observations at the terrestrial magnetopause suggest that this does not happen unless the shear angle between the magnetic fields either side of the boundary exceeds ~70°, although recent work on reconnection in the solar wind suggests that it may occur for even smaller shears [30]. However, even at large shear angles, most current sheets remain stable much of the time, without any reconnection. Moreover, the conditions in the large-scale plasma environments either side of reconnecting current sheets have been observed to be quite diverse, showing that reconnection may occur under many different conditions. We do not yet understand why reconnection begins, but an effective approach would be to study why it may begin at a particular point in a current sheet and not at a neighbouring location. A basic requirement for making this comparison is numerous simultaneous measurement points both in and around a current sheet as it begins to reconnect.

Furthermore, according to present theory, a precondition for reconnection is that the current sheet must undergo a thinning process. Quantitative measurements of current sheet thinning have been made by the Cluster mission [31]. Cluster 4-point data can be used to determine the average current through the tetrahedron at a given moment. However, there is a limitation to the usefulness of this: If the tetrahedron is small compared to a current sheet, a local measurement is made but the total current is unknown; if the tetrahedron is large compared to the sheet, the total current is measured, but its distribution within the sheet is unknown.

In order to fully characterize the development of thinning pre-reconnection current sheets we must therefore use a set of nested tetrahedra of spacecraft, across *at least* two different scale sizes, each optimized to simultaneously capture the evolution of the small-scale structure during the thinning process, while monitoring the larger context with the other-scale spacecraft. The largest separations should give information about the magnetic flux gradient across the current sheet, and also characterise the orientation of the current sheet. They also establish important parameters, such as the plasma beta of the inflow regions. The smaller separations should enable measurement of the intensification and thinning of current within the overall sheet structure, and also provide information on the internal variations in plasma characteristics to determine how such currents are supported. It is particularly important to simultaneously monitor the temporal and spatial changes at ion and electron scales, since the electron heating/acceleration processes (e.g., interactions with waves, etc) at an electron-scale current sheet embedded within a broader ion scale current sheet seem to be the key for starting reconnection [32]. Note that in principle, nested tetrahedra also provide multiple sets of 4-point data from which current densities can be calculated, providing valuable information about current and plasma substructures along and across the current sheet.

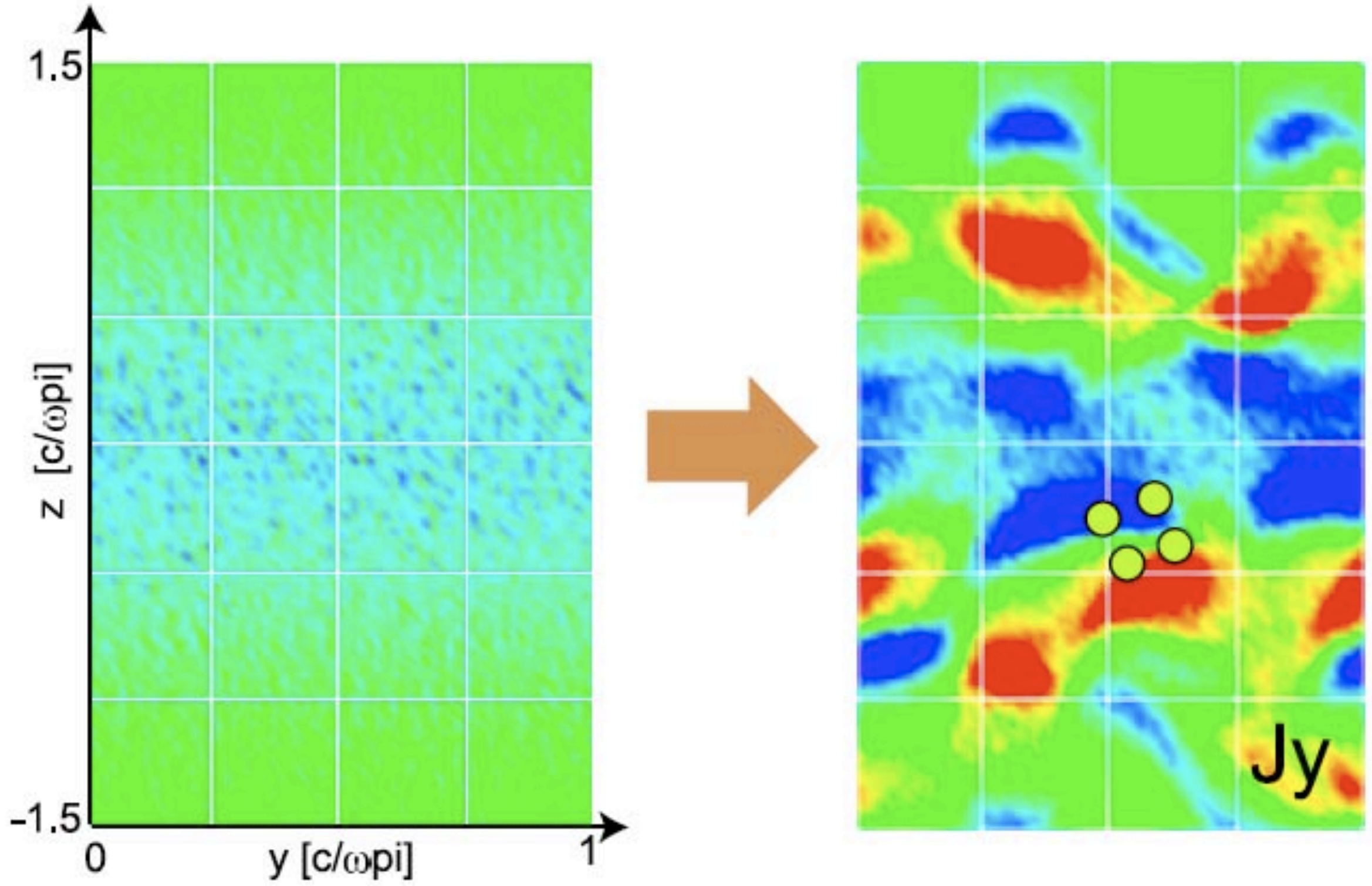

*Figure 9. A simulation of a current sheet during the onset of a reconnection event [34]. The breakup of the current sheet is illustrated. A set of four electron scale Cross-Scale spacecraft is superimposed, to show how it would capture the transition from a stable sheet into rapidly changing fragmented currents which is thought to occur at reconnection onset. Additional spacecraft at larger separations will quantify the orientation and conditions in which that onset is triggered and the spatial scale of the current sheet.*

This set of multi-scale measurements will reveal the critical thickness at which a current sheet begins to break up. Measurements of ion and electron plasma properties put the critical thickness data into context in terms of the ion and electron scale sizes of the system, enabling ready comparison with analytical theory and simulations [33], [34]. Multi-point measurements of plasma waves and particle distributions will facilitate the testing of ideas about current sheet instabilities which may cause the breakup and initiate reconnection (Figure 9).

Internal current sheet instabilities may cause reconnection onset, but another possibility involves the effect of an external driver. Both scenarios may be operating in different situations. Sudden changes of the driving (global scale) electric field in the solar wind are considered to cause a large fraction of reconnection onsets in the magneotail, as well as magnetopause reconnection onset at Earth. Disk accretion is considered to drive the magnetic field reconnection between the magnetospheric field and the disk field in a magnetized young star. A set of simulations using various numerical schemes have investigated driven reconnection scenarios and show consistent outcomes, increasing confidence in the results [35].

Very little information is available about the electron diffusion region. In order to test ideas about how the diffusion occurs it will be necessary to measure the contribution to the reconnection electric field made by very small-scale phenomena including pressure anisotropies associated with non-gyrotropic electron velocity distributions and/or bulk electron flow anti-parallel to the main current. In addition, we also need to identify any waves which may scatter and demagnetize electrons in and around their diffusion region. For example, whistler mode wave activity in the diffusion region may enable more efficient reconnection.

### 2.3.2 How does the magnetic topology evolve?

Once magnetic reconnection has begun, there are many questions about how it is maintained and how it evolves. Reconnection in the solar wind can endure for hours [36]; at the magnetopause it can be either similarly persistent or very bursty on a timescale of minutes; in the magnetotail reconnection seems more often to last only minutes or tens of minutes. However, these time-scales are all long compared to typical electron and ion timescales, showing that reconnection can operate in a quasi-steady state, relative to the timescale of the kinetic processes that are thought to sustain it.

#### *Departures from 2D*

Reconnection topologies are most often more complicated than the picture discussed above; the large scale magnetic fields separated by the current sheet can be sheared, such that there is a component along the direction of the reconnection electric field. Many numerical simulations have studied how this “guide field” scenario might differ from the simpler 2D case, but this cannot be tested without multi-scale observations. Other departures from the symmetrical 2D picture that need to be examined include the differences in flow velocity, plasma density and temperature across the current sheet.

The extents of the ion and electron diffusion regions are unknown, both along the external magnetic field direction and perpendicular to the 2D geometry (i.e., along the neutral- or X-line). They are vital to theoretical ideas about the reconnection rate and particle acceleration. These extents cannot be determined without monitoring the diffusion regions at electron and ion scales simultaneously. Such considerations are even more central to reconnection under 3D geometries. We also need to be able to compare simultaneous observations from regions where reconnection is occurring, with nearby regions where it is not, in order to discover the factors which control the length of the neutral line.

Cross-Scale measurements across a variety of scales are needed to reveal whether there are differences in the way reconnection works under asymmetric plasma conditions including high-beta plasma regimes and skewed rather than anti-parallel fields, in comparison to the highly symmetric configurations. We also need to determine the influence of conditions at the fluid scale on the smaller scale processes. Simulations suggest that reconnection is sensitive to the outermost boundary conditions of the system. Similarly, comparisons of magnetic reconnection with and without significant oxygen ion populations are needed to test models of how the current sheet structure and reconnection rate depends on the plasma composition. It might be expected that an additional scale associated with oxygen gyroradii will play a role in controlling the reconnection process in this case. These measurements require a comparison of the fluid scale properties of the system with at least one of the smaller scales.

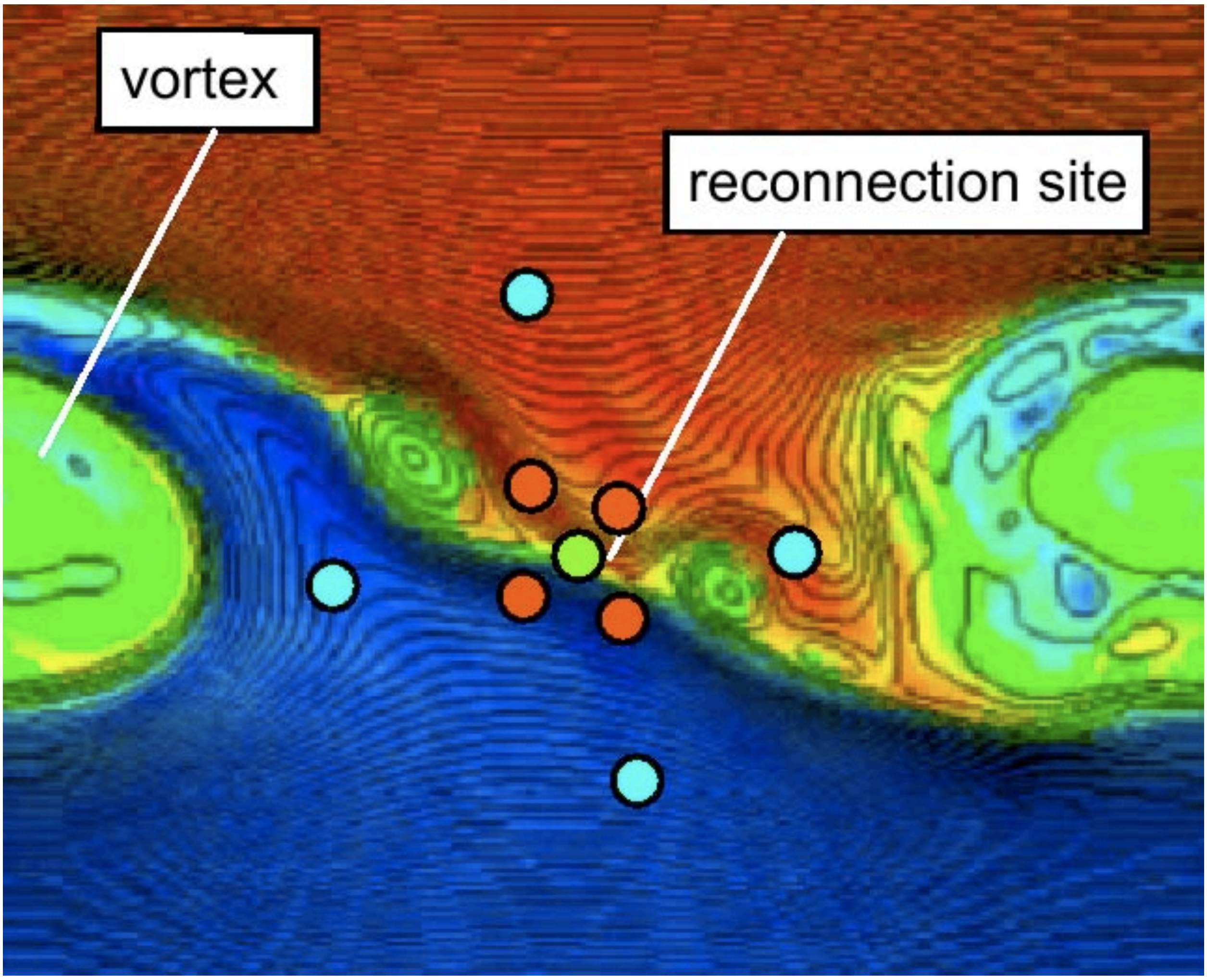

*Figure 10. A simulation of a current sheet during a reconnection event. Note that the situation is more complex than the X-line scenario sketched in Figure 7. Such an ion/fluid scale simulation cannot simultaneously address the electron scale (for that, see Figure 9). A possible Cross-Scale configuration is superimposed showing how the large scale spacecraft measure the in/outflow jets and magnetic fields far from the current sheet; the intermediate/small scales are targeted on the ion/electron diffusion regions respectively.*

## *Turbulent reconnection*

The change in the magnetic topology due to reconnection is even more difficult to analyse within turbulent plasma regions. Turbulence can develop at plasma boundaries (such as vortices at the magnetopause [19]); changes in topology due to reconnection can lead directly to plasma transport across the boundaries and formation of the boundary layer [37] (see Figure 10). Thus spacecraft are required on a large scale to monitor the boundary layer development and on at least one of the smaller ion/electron scales to understand the reconnection mechanisms.

In other situations, such as the magnetosheath, the plasma is turbulent across a large volume; magnetic islands and other coherent magnetic structures can continuously form and disappear due to reconnection. The case study in Figure 8 illustrates one such example and also shows how the Cross-Scale spacecraft, deployed across multiple scales, are needed to analyse the reconnection in that case. Spacecraft at large separation would be able to follow the development of the turbulence while smaller-separation spacecraft would be able to follow the physics of reconnection process itself. Both types of turbulent reconnection are expected in a wide range of astrophysical plasmas (such as the case study in Figure 8).

### 2.3.3 How does reconnection accelerate particles and heat plasma?

Reconnection most probably occurs in a wide variety of contexts beyond the Earth's magnetosphere; most can only be studied by remote sensing of the emissions generated indirectly by energetic particles. Even in the magnetosphere, energised particles measured in situ are often the only sign of a distant reconnection site. In order to interpret such data as reconnection signatures or otherwise, and perhaps to use them to infer the properties of the reconnection site itself, we must understand how particles are accelerated and heated.

Most of the energy released during the reconnection process goes into the energisation of ions and electrons. Observations show evidence of reconnection outflow jets and particle heating, and of the formation of field-aligned beams and energetic tails in the particle distributions. However, the origins of all these kinetic features are not understood in detail.

Inductive electric fields associated with rapid changes in magnetic field may cause strong charged-particle acceleration, such as in a reconnecting current sheet with a non-steady reconnection rate. To verify this with observations we need multiple spacecraft in the vicinity of the reconnection site to provide a combined data set that resolves space-time ambiguities and confirms that the magnetic configuration is rapidly varying. Local electric fields in 3-dimensions and particle data for a broad range of energies at high time resolution are needed to infer where in the system the acceleration occurs.

The reconnection electric field is expected to have a component along the local magnetic field direction in the diffusion region, which will readily accelerate charged particles [38]. A long neutral line can in principle accelerate particles to energies limited only by its length (e.g., 10's of keV in the magnetotail). A test of whether these acceleration mechanisms really operate efficiently in nature requires the Cross-Scale constellation to make measurements of electric and magnetic fields, as well as particle populations, at several points along a significant portion of the length of a neutral line and around it. Figure 11 shows the resulting electron beams and spectra seen in the vicinity of a reconnection region. Proper testing of this model requires simultaneous observations of this kind throughout the diffusion region and in the surrounding plasma.

Strong wave activity is often reported in outflow jets, and energetic (100's keV) electrons are also often seen there, especially near the separatrices. One scenario for generating these electrons proposes that they are heated to a few 10's of keV near the reconnection site, and are further energized to 100's of keV by a combination of diffusion across the reconnection electric field together with strong pitch angle scattering that violates adiabatic invariants [39]. Betatron and Fermi acceleration are also expected on the contracting magnetic field lines in the reconnection outflow jets. Measurements on ion and fluid scales are needed to capture the evolution of particle energy with increasing distance away from the reconnection site, and to measure the magnetic field and other parameters to test whether the degree of energisation is consistent with that expected from these mechanisms. A partial example of such a test is seen in Figure 11.

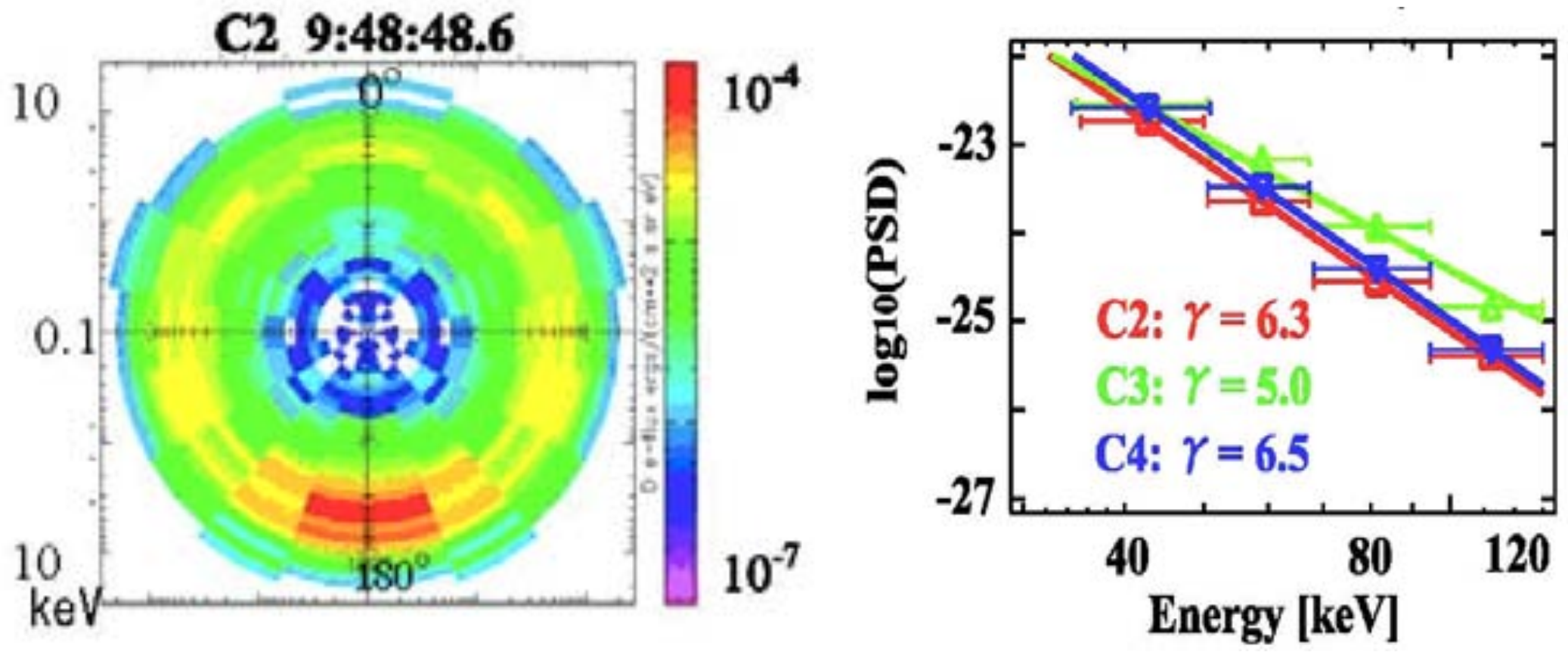

*Figure 11. Electron beam accelerated as part of the reconnection process (left) and the corresponding energy spectra of energised electrons (right) at various locations within the reconnection region close to the diffusion region (C2 red and C3 green) and in the outflow (C4 blue). The acceleration is thought to be due respectively to rapidly varying electron-scale electric field structures [40] and non-adiabatic processes [41].*

Waves with a broad range of wavelengths and frequencies are generated in and around reconnection sites. Wave-particle resonant or diffusive interactions can both tap and enhance particle energy, with different wave types likely to affect different particle populations. Full characterisation of wave vectors and modes requires multipoint measurements on the corresponding scale. When analysing the gamut of relevant plasma waves, e.g., lower hybrid waves, whistler-mode waves, and electron cyclotron waves, we require separations covering scales from ions scales down to the electron scales. Confirmation of relevant wave-particle interactions also requires well-resolved ion and electron velocity distributions at an appropriate cadence.

Significant fluxes of electromagnetic energy liberated at reconnection sites have also been observed to propagate away in the form of waves with an Alfvénic character. Their associated wave-particle interactions are known to cause local particle energisation. The generation mechanism of these waves is not clear. To make significant progress in understanding these waves and their relation to reconnection processes requires spacecraft that are suitably separated at different scales dictated by the nature of the waves: sub-ion scales to study the wave properties across the magnetic field, a few ion scales to study the environment of reconnection site and wave properties along the magnetic field and finally, tens of ion lengths away along the magnetic field to see the development in the wave propagation.

## 2.4 How does turbulence control transport in plasmas?

The ubiquitous nature and important consequences of turbulence in space plasmas make it an important target for the Cross-Scale mission. Turbulence is responsible for the transport of many physical quantities: energy, both between scales and across space; momentum; and energetic particles through the resulting complex, tangled magnetic fields. It covers a vast range of scales, from inter-galactic to below the electron gyroradius.

While significant advances in simulation, theory and observations have been made, the highly complex, non-linear and multi-scale nature of plasma turbulence means that many important questions remain: How is energy transferred between scales, particularly at scales at which kinetic particle dynamics are important? How do the unique properties of plasmas alter the traditional hydrodynamic view of turbulence? What is the origin of the discrete structures that are observed in turbulent plasmas? All of these issues impact directly on the effects of turbulence on the surrounding plasma.

Cross-Scale offers a unique capability to address these fundamental questions about the nature and effects of turbulence in space plasmas. By measuring the 3D turbulent structure at different scales simultaneously (fluid/ion or ion/electron), it will be possible for the first time to measure the energy transfer process as it actually occurs, from the fluid to the different kinetic scales. It will also sample plasma turbulence in a number of dramatically different regimes: the solar wind, where turbulence is well-developed and energy inputs are steady; the highly-disturbed magnetosheath, where fluctuations are strongly driven by shears, shocks and compressions; and the magnetotail, which is dominated by the magnetic field and therefore highly anisotropic.

The following sections describe key space plasma turbulence questions, and how Cross-Scale will address them.

### 2.4.1 How does the turbulent cascade transfer energy across physical scales?

A defining characteristic of turbulence is the nonlinear transfer or cascade of energy between scales, a process that leads turbulence analysis to seek universal scaling laws. However, scaling laws can only exist where the energy transfer process is the same over a range of scales, such as within the plasma fluid (magnetohydrodynamic) regime. Some of the most important effects of turbulence, including plasma heating, particle scattering and the acceleration of stellar winds, are intimately related to fluctuations on the gyroscales of ions or electrons. On these scales, the turbulence is much more complex and is not scale invariant. In addition, many of the assumptions that are made to study fluid range turbulence, e.g., that the plasma flows past spacecraft much faster than the internal wave speeds, or that the plasma is incompressible, cannot be used: many more wave modes are present and they interact in a highly nonlinear manner.

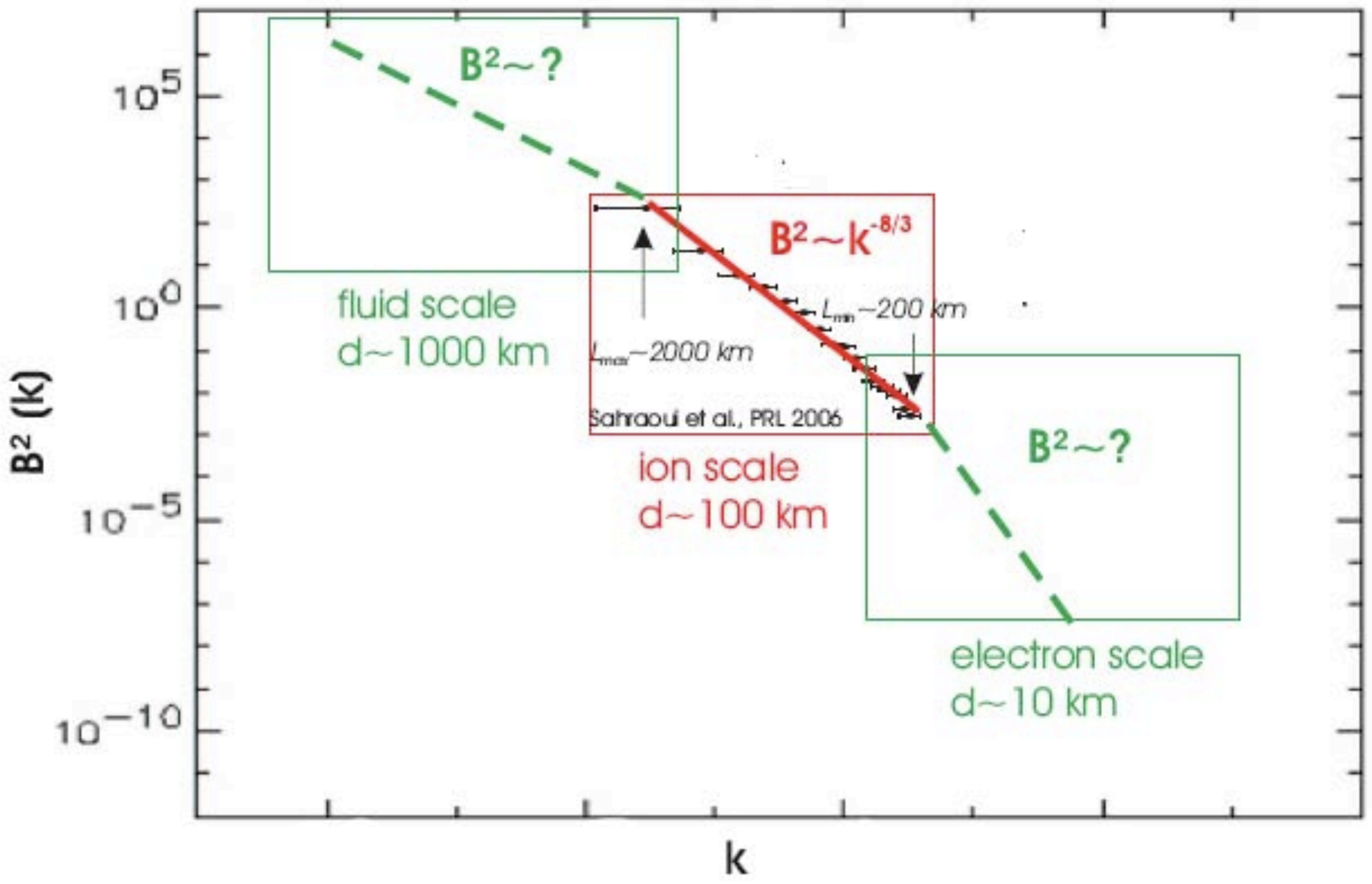

*Figure 12. A wave vector spectrum of turbulent magnetic energy in the highly disturbed magnetosheath derived using multi-point measurements to filter the time series data into their joint space-time spectra. The fitted power law is shown in red for the range of scales, covering around a decade, which were accessible to the four Cluster spacecraft at this time. Recent measurements in the solar wind confirm that the electron scale does indeed show further steepening [43]. Coupling to the fluid or electron scale spectra (green) will be achievable using Cross-Scale's multi-scale strategies.*

The four Cluster spacecraft have made it possible, using novel analysis techniques, to measure the wave vectors of turbulent energy in space plasmas for the first time [42]. However, this analysis is restricted to an order of magnitude in scales around the average spacecraft separation.

The small-scale termination of the turbulence cascade is controlled by dissipation processes. While these are rather straightforward in collisional fluids where viscosity is dominant, in collisionless plasmas there are several possible mechanisms, on both ion and electron scales. There are many wave modes, both electromagnetic and electrostatic, that can participate in the transfer of turbulent energy into particle heating. The multi-scale nature of this situation means that we have no satisfactory theoretical or observational understanding of it – and indeed it may be different in different plasma environments. However, without such a picture, we cannot hope to predict and quantify the effects of turbulent heating on space plasmas.

It is essential that the key physical scales of the cascade – fluid, ion and electron – are measured simultaneously with sufficient high-quality instrumentation (electric and magnetic fields, ions and electrons) if we are to determine the properties of the active turbulent cascade. Cross-Scale will target these scales, with 7 spacecraft enabling 2 scales to be covered simultaneously and the optimum 12 spacecraft configuration measuring this entire range for the first time (see Figure 12). In addition, this will make it possible for the first time to track the energy flow from fluid to ion and electron scales and determine the direct effect on the particles themselves, such as heating, resonances and wave generation. Only in this way can we study the process by which turbulent energy is finally partitioned between fields and particles.

Theory suggests that in addition to the familiar energy cascade to small scales, a "reverse" cascade exists in quantities such as the magnetic helicity, making it possible to generate larger scale magnetic structures. However, such a cascade cannot be measured without multi-scale data. With Cluster, it was possible to glimpse how these complex mechanisms could be studied. Cross-Scale will facilitate a powerful extension of these multi-spacecraft techniques, offering the opportunity to actually quantify the multi-scale interactions and therefore the key to turbulence. Accordingly, Cross-Scale will measure the magnetic helicity spectrum and its development, with the aim of identifying unambiguously this reverse cascade for the first time.

## 2.4.2 How does the magnetic field break the symmetry of plasma turbulence?

Neutral fluid turbulence is usually isotropic. Only near a boundary layer or other compression or shear is the isotropic symmetry broken so that turbulent properties become different in different directions. In contrast, the presence of a local magnetic field direction means that plasma turbulence is always anisotropic, even far from boundaries or shears. Indeed, theoretical work suggests that the turbulent cascade progresses differently parallel and perpendicular to the local field, a result that has been confirmed with spacecraft data in the solar wind [44]. Recently, multi-spacecraft data have made it possible to study anisotropy in the highly disturbed and turbulent magnetosheath plasma upstream of the magnetopause boundary layer (Figure 13). These reveal simultaneous anisotropies relative to two directions – the local magnetic field and the boundary layer – a scenario unique to plasma turbulence.

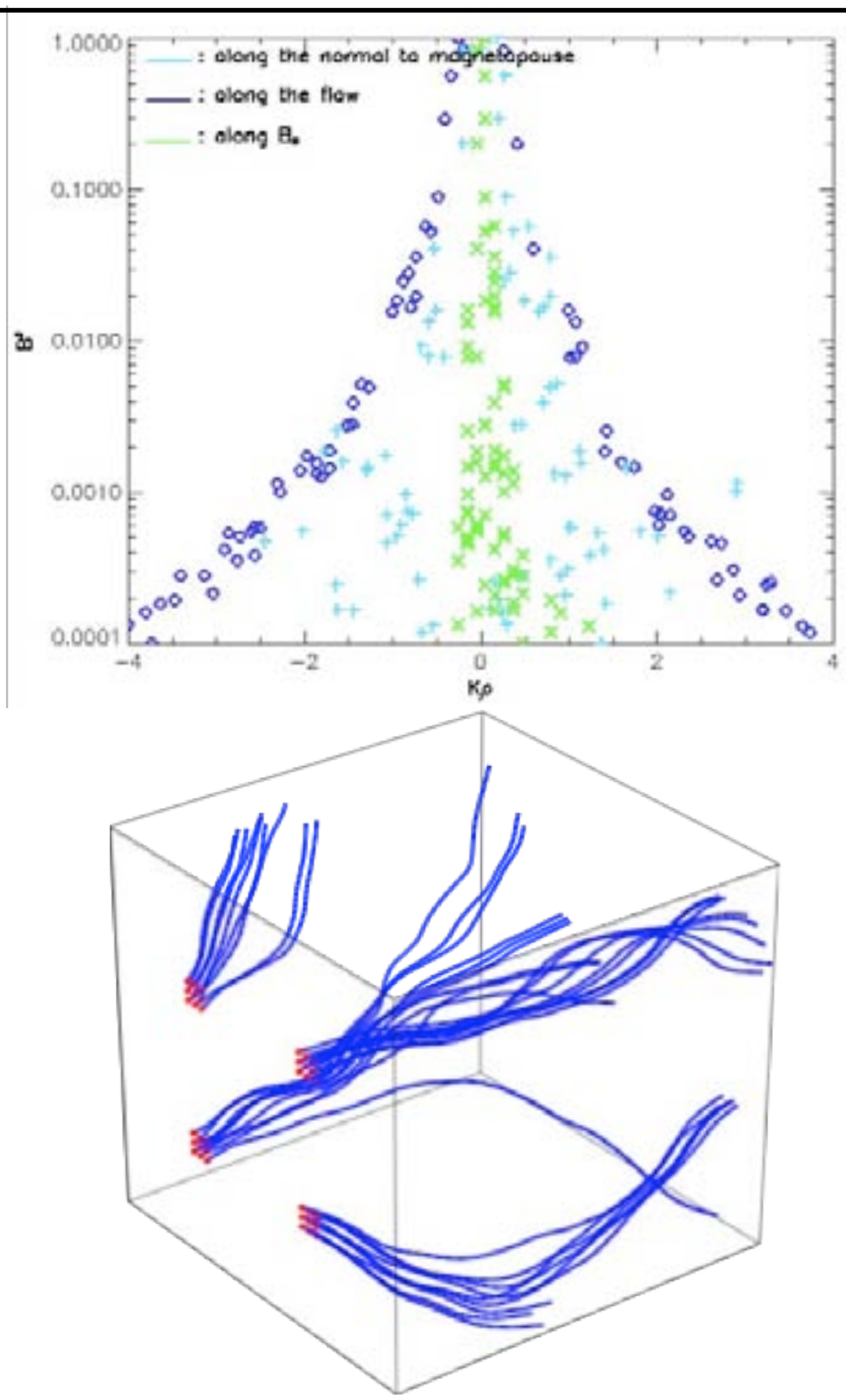

**Measurements of turbulence in space plasmas have revealed the presence of significant anisotropy. Recently, multi-spacecraft data demonstrated that turbulence in the highly disturbed plasma downstream of the Earth's bowshock scales differently in different directions [42] (upper panel). The power in wavevectors parallel to the field scales in a dramatically different way to that along the flow or perpendicular to the local plane of compression. This anisotropy appears to be due both to compression of the plasma next to the magnetopause boundary layer, and to the fundamental differences in energy transfer parallel and perpendicular to the magnetic field. The anisotropy is of more than academic interest. Such anisotropy can have dramatic effects on the large scale connectivity of the magnetic field, as shown in the simulation box (lower panel), where field lines "wander" perpendicular to the field. The transport of energetic particles along this braided structure results in an enhanced diffusion process, which has important implications for the acceleration and propagation of, for example, cosmic rays, throughout the Universe. Cross-Scale will make it possible to study for the first time how this anisotropy develops and is transferred between scales.**

*Figure 13. Case Study 3: Anisotropy and particle transport.*

The anisotropy of plasma turbulence has important consequences for the transport of mass and energy through boundary layers. For instance, the velocity shear layer between the streaming solar wind and the stagnant magnetospheric plasma can become Kelvin-Helmholtz unstable and lead to highly anisotropic turbulent vortical flows, as revealed by recent 4-spacecraft Cluster observations. Structures with a scale comparable to the ion gyroradius can be created through cascading, or through secondary instabilities excited in the vortex. These small-scale dissipative structures can mix the plasma contained in the vortex, and lead to diffusion through the shear layer.

The anisotropy of solar wind turbulence has important implications for the propagation of energetic particles throughout the solar system, as well as the large-scale connectivity of the magnetic field: the turbulence can "braid" or tangle the magnetic field, resulting in field-line wandering far from the average direction (see Figure 13). These effects have been successfully used to reconcile scattering mean free paths of solar energetic particles with measured turbulence levels [44]. However, it is not clear how this anisotropy develops from the large scale where energy is injected, and how it passes between the fluid, ion and electron scales.

Cross-Scale will let us track the development of anisotropy between scales for the first time, providing information on how energy is transferred throughout space and time in the turbulent cascade. By relating this anisotropy to that observed in sheared or compressed hydrodynamic turbulence, we will gain insight into anisotropy as part of the universal process of turbulence. Recent advances based on the available data are limited to only around a decade in scale and at one location at a time, making it impossible to determine the evolution of anisotropy in scales or in space. Only Cross-Scale can make the multi-scale, multi-point measurements that are required, on both fluid and kinetic scales, to measure, understand and predict the effects of the anisotropy that is unique to plasma turbulence.

### 2.4.3 How does turbulence generate coherent structures?

In both neutral fluids and plasmas, turbulence generates discrete structures. Spacecraft data reveal the presence of large magnetic/density structures known as mirror modes [42],[44],[45], Kelvin-Helmholtz waves [19] and current sheets [46], which appear to be involved in, or the result of, the energy transfer process. Fully-developed turbulence is generally not spatially uniform, and can exhibit strong spatio-temporal "burstiness," commonly known as intermittency (see Figure 14).

Intermittency has consequences on the transport of energy over scales, through the modification of the scaling in the inertial range, and on its dissipation at the small scales [47]. It is frequently proposed to explain the heating of the solar corona. It is also expected to strongly affect scattering and diffusion of plasma particles [48]. Statistical methods brought from hydrodynamics (e.g., probability density functions and structure functions) have been applied to single spacecraft data but they cannot provide information on the 3D nature of intermittent structures in plasma turbulence.

With its 7 spacecraft baseline Cross-Scale will make measurements at two scales simultaneously, in particular both ion and electron, facilitating the analysis of the statistical properties of spatial and temporal inhomogeneity. That analysis will reveal how intermittency is generated and transmitted down the cascade. More than this, however, with Cross-Scale's measuring points we will move beyond statistical analyses and for the first time actually "image" these structures, answering vital questions such as: How large are they? What is their 3D shape? How do they link together?

The recent identification of reconnection driven by turbulence (see Figure 8) demonstrates the importance of current sheets in turbulent flows. Their 3D structure and development is not well known, but without this knowledge we cannot predict the statistical effects of the resulting multiple reconnection sites. Additional observations of bursty reconnection in the Earth's magnetotail suggests that even in this very low $\beta$ plasma (where the strong magnetic field tends to resist the particle pressure), broad-band fluctuations can drive reconnection, emphasising the multi-scale nature of this process. These results show that turbulence and reconnection are intimately linked and a multiscale approach is essential in order to fully quantify their effects.

Cross-Scale will not only measure electron and ion scales simultaneously: the spacecraft formation can also be considered as multiple intermediate-scale tetrahedra at several locations. By performing a similar analysis at multiple locations simultaneously, for the first time it will be possible to measure unambiguously the growth and development of these structures as they travel past the formation.

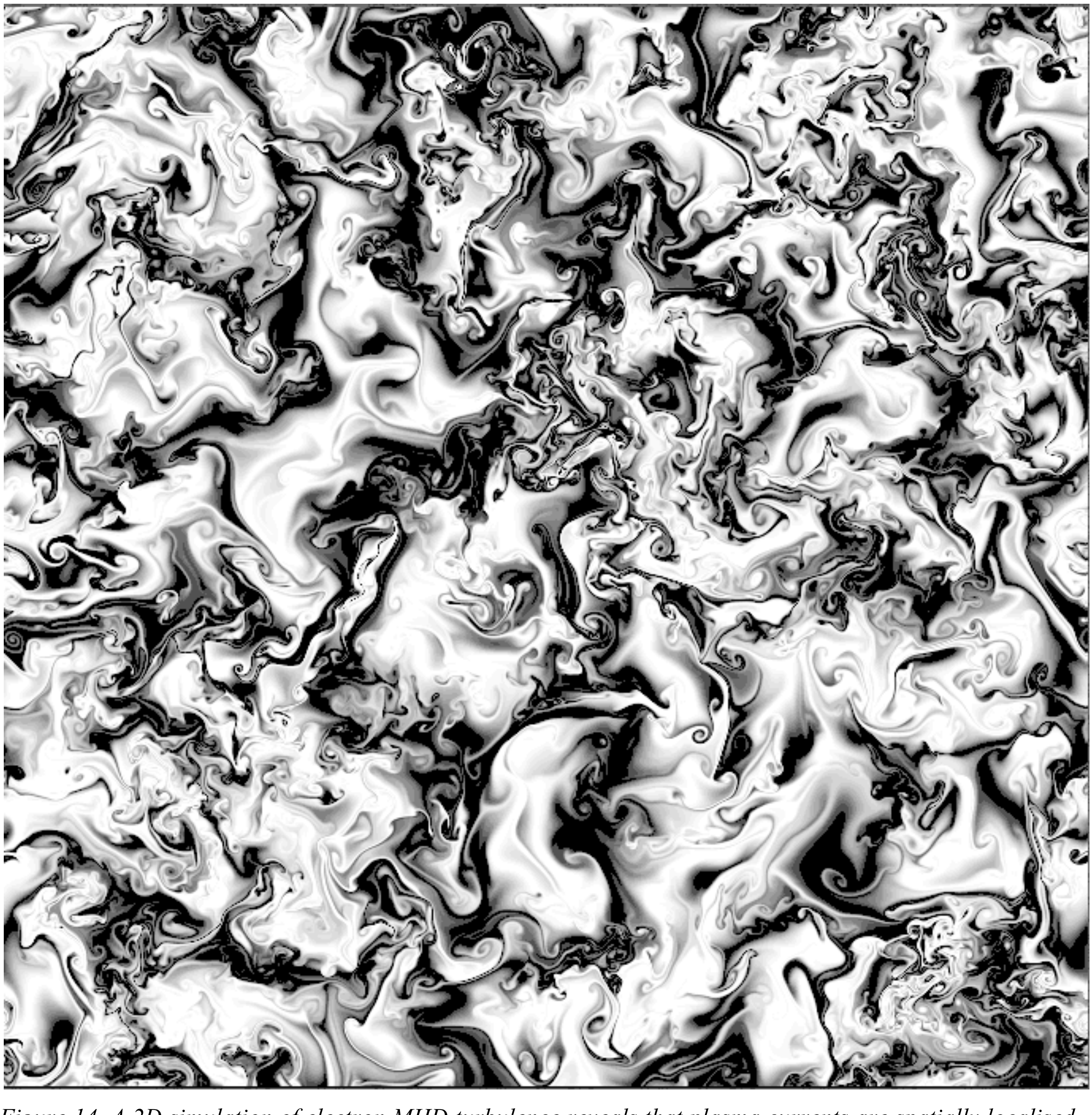

*Figure 14. A 2D simulation of electron MHD turbulence reveals that plasma currents are spatially localised into thin sheets and whorls. Many-point measurements, such as those from Cross-Scale, are required to measure such structures in 3D in space plasmas over a range of scales not accessible to simulation [49].*

# 3 Science Requirements

The Science Requirements for the Cross-Scale mission are varied and complex. The SciRD [50] provides an exhaustive description of these requirements. The matrix in Annex 1 of the SciRD then collects this information into the required measurements at different scales, resolutions, cadences, and other aspects, and identifies the Cross-Scale instruments that will provide those measurements. Finally the summary table in Annex 2 of the SciRD relates each instrument on each spacecraft directly to the science requirements.

Here we provide an overview of the details given in the SciRD. In particular, for each science objective we identify some of the key measurements that are required to meet that objective. Before we go into the detailed measurement requirements, let us consider the environments Cross-Scale will encounter and how those relate to the science objectives. These impose requirements on the number of spacecraft, their deployment, and the capabilities of the instrumentation they need to carry.

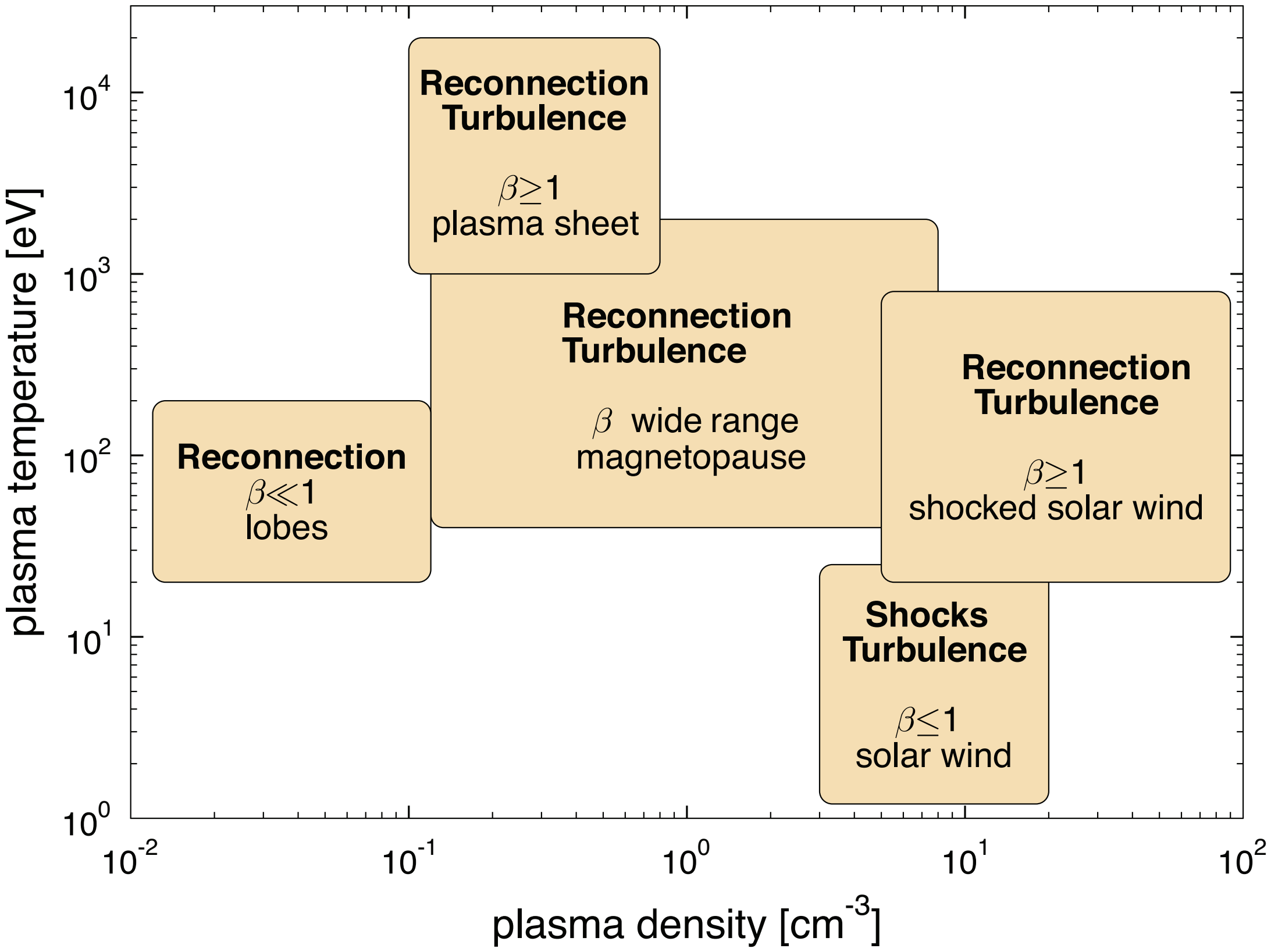

*Figure 15. Cross-Scale science objectives in different plasmas, where β is the ratio of thermal to magnetic pressures.*

Figure 15 shows the processes that Cross-Scale will study in different regions of the near Earth environment. Magnetic reconnection can be found in most regions from the low-density lobes and plasma sheet regions to the high density shocked solar wind. Turbulence phenomena are also found in various regions. Finally shocks are found mainly in the solar wind, which has a more restricted range of temperature and density.

Figure 16 recasts this information into a non-dimensional form using the ratio of thermal to magnetic pressure, β, and the ratio $f_p/f_{ce}$ (where $f_p$ is the plasma frequency and $f_{ce}$ is the electron gyrofrequency). This enables a more relevant comparison of different plasma regimes outside the solar system. The figure demonstrates that plasma environments, from the Earth's magnetosphere and the solar corona to supernova remnants and laboratory plasma experiments, differ greatly in the values of physical quantities but overlap in relevant non-dimensional parameters. Thus, we expect that results obtained in one-plasma environment to be applicable to another. However, we note that when making realistic comparisons between different environments other parameters such as collisionality need to be considered.

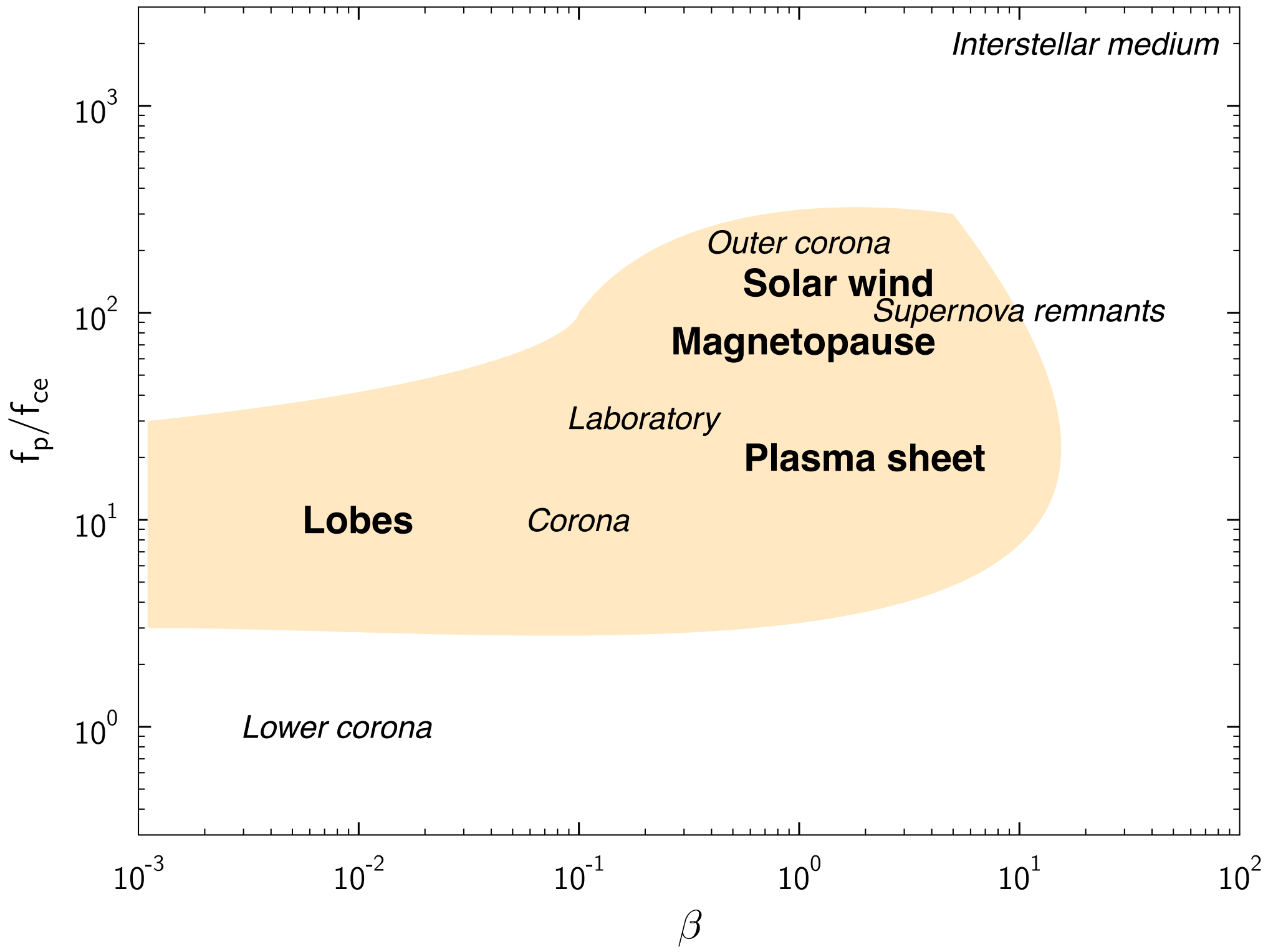

*Figure 16. Plasma regimes in non-dimensional parameter space, $f_p/f_{ce}$ (proportional to the ratio of the speed of light to the local Alfven speed) vs $\beta$ (ratio of thermal to magnetic pressure). The shaded region indicates the plasma regimes that will be sampled by Cross-Scale.*

Figure 17 gives us the various scale lengths and time scales present in the near-Earth environment. The smallest lengths and shortest timescales correspond to electron dynamics, followed by ion scales to eventually the fluid scales that set the context and driving conditions for the phenomena of interest. The coupling between the various scales should be addressed by Cross-Scale by measuring at least two scales simultaneously.

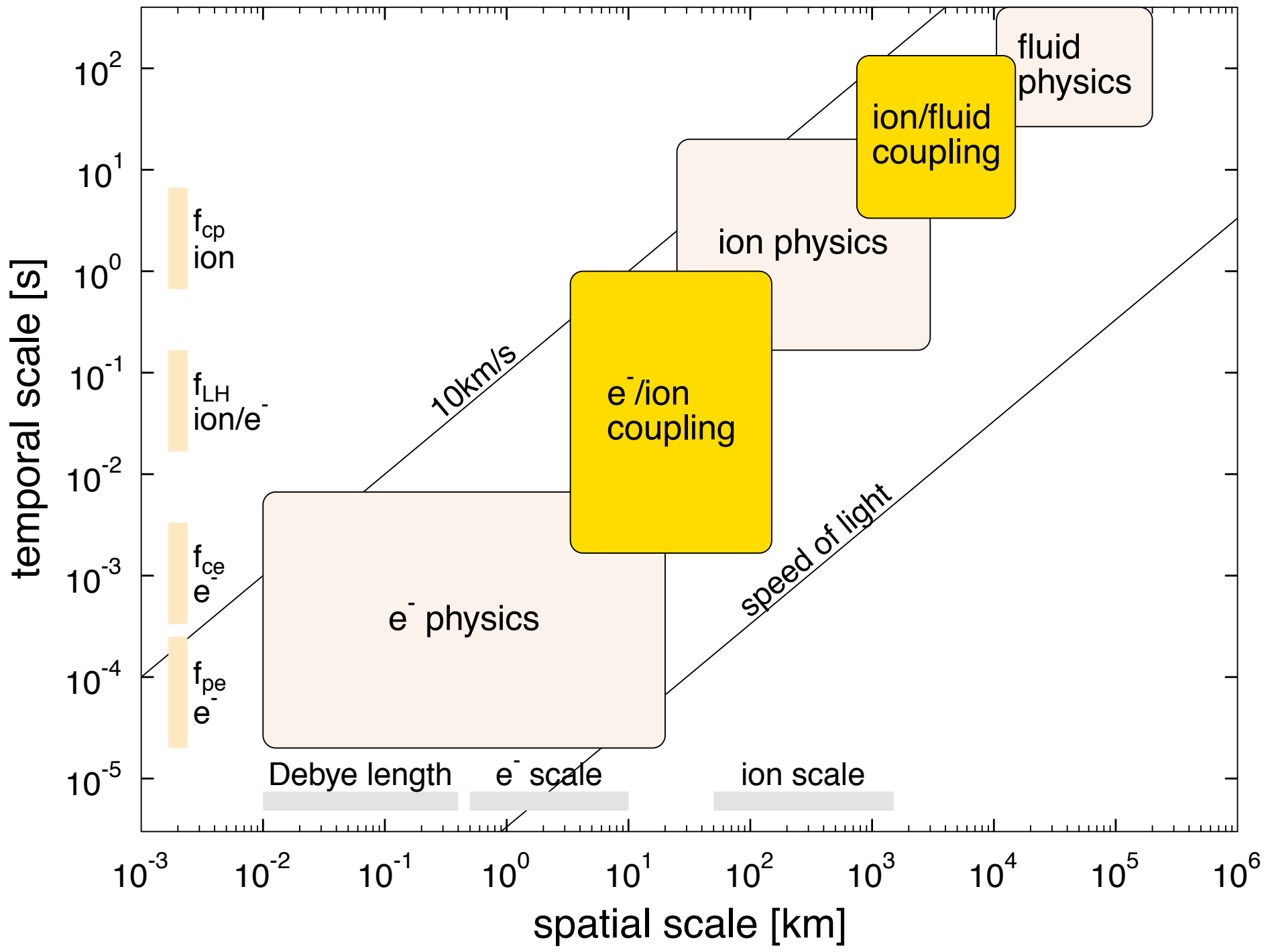

*Figure 17 Characteristic scales to be studied by Cross-Scale.*

Cross-Scale needs to quantify the answers to the main science objectives:

- How do shocks accelerate and heat particles?
- What mechanisms accelerate particles at shocks ?
- How is the energy incident on a shock partitioned?
- How do shock variability and reformation influence shock acceleration?
- How does reconnection convert magnetic energy?
- What initiates magnetic reconnection?
- How does the magnetic topology evolve?
- How does reconnection accelerate particles and heat plasma?
- How does turbulence control transport in plasmas?
- How does the turbulence cascade transfer energy across physical scales?
- How does the magnetic field break the symmetry of plasma turbulence?
- How does turbulence generate coherent structures?

The following sections summarise the measurements that are required to answer these questions. The SciRD provides a more detailed discussion of those requirements together with a comprehensive matrix that captures the physical quantities, their resolution, and the potential instruments that could make those measurements.

## 3.1 Collisionless Shocks Science Requirements

### 3.1.1 Context and parameters

To answer the questions of shock heating and acceleration in a way that can be applied universally requires a comprehensive characterisation of the underlying plasma parameters on either side of the shock and within both the electron and ion scale. These include basic plasma moments (density, velocity, temperature, and temperature anisotropies) at cadences comparable to the relevant scales (~2 seconds at the ion scale on 4 spacecraft, ~0.1s at the electron scale). DC electric and magnetic fields at comparable, or slightly higher, cadences establish the global orientation and shock geometry, together with the basic sub-structure.

Additional fluid-scale contextual data would be desirable. In the baseline Cross-Scale configuration of 7 spacecraft the final coupling to the fluid scale would be accomplished in the latter half of the mission when the formation was reconfigured to study fluid-ion scale coupling. In an international partnership, e.g., with SCOPE, all three scales would be available from the outset of the mission.

### 3.1.2 Energised/heated particle populations

Core to both the heating/energy partition problem and particle acceleration are the particle distribution functions, which in a collisionless plasma are typically far from Maxwellian. Thus 3D phase space distributions of thermal (<30 keV) electrons and ions need to be measured within the various scales simultaneously in order to unravel the coupling mechanisms that govern the physics. Again the cadences are set by the scale dynamics. These measurements capture most of the heating and the most numerous accelerated populations. Measurements of energetic electrons and ions, together with ionic species, need to be made by one, and preferably more, spacecraft to follow the acceleration processes to higher energies and to study the role of composition as both an input and output of the acceleration mechanisms.

### 3.1.3 Accelerating fields and variability

Variability and fluctuating, turbulent fields are thought to be the primary acceleration agents beyond the lowest order DC thermalisation. Some aspects, such as those related to shock reformation, are contained within the overall parameters and heating requirements described above. Beyond that, key measurements include a definitive measurement of the full 3D electric field at one (or more) location (to determine, for example, the field-aligned electric field) and 2D electric/3D magnetic spectra up to several kHz simultaneously at multiple scales to determine the high-frequency wave modes and intermittency of turbulence that scatters and potentially selectively accelerates particles.

## 3.2 Magnetic Reconnection Science Requirements

### 3.2.1 Context and Parameters

Magnetic reconnection is expected to occur when magnetic fields are sheared across relatively thin current layers. In such current sheets, the kinetic effects of the particle populations become important within electron and ion diffusion regions which form on the distance and time scales of the relevant electron and ion gyromotions respectively. However, microscale processes also control the change of topology of the magnetic field, eventually affecting the large-scale plasma mixing and converting magnetic energy to plasma energy. Conversely, large-scale (MHD) processes can control the location and formation of thin current sheets, and thus directly affect how reconnection initiates and evolves.

These issues fix a requirement to follow both large-scale and kinetic-scale processes of the plasma to understand the onset, the evolution, and the result of the magnetic reconnection. Cross-Scale must follow changes of current sheet (thickness, orientation, internal structures) which lead to reconnection. This requires timing analysis on current sheet features observed at the ion scale and to resolve any embedded electron-scale current layer, both with 4-point measurements at 5-50 km studying the electron scale and 100-500 km studying the ion scale. The corresponding cadence for these observations is ~0.01s for electron-scale observations and ~0.02s for ion-scale observations. In addition, a requirement for low-cadence (~1s) measurements on larger scales of the context for the reconnection process implies the need to measure plasma moments and composition, e.g., to determine prevailing Alfven wave speeds or whether the plasma is oxygen-rich.

### 3.2.2 Signatures of Reconnection

In order to identify enhanced parallel electric fields associated with reconnection, simultaneous 4-point measurements of the DC magnetic field and at least 3-point measurements of the DC electric field at electron scale (ion-scale) are required, typically with cadence ~0.01s (~0.02s) to resolve thin structures. Moreover, identification of the enhanced wave activity that may be associated with occurrence of anomalous resistivity and plasma instabilities requires 2D AC electric and 3D magnetic field measurements at ~0.01ms and 1 ms respectively for these ion and electron-scale multi-point observations. Cross-scale must also identify accelerated ion flows and ion beams in the reconnection outflow region, and unmagnetised ions within the ion diffusion region. These typically require multi-point measurements of the velocity distribution function of the thermal ion (< 30keV) population, and/or its moments, at a cadence of order ~1s on spacecraft separated by a typical ion scale (100-500 km). In addition, at least one spacecraft should monitor the presence of higher energy ions and electrons (< 1 MeV and < 200 keV) accelerated by reconnection with a similar time cadence. On the smaller scales, we also need to identify processes occurring within the electron diffusion region, by making fast, multi-point measurements on scales of 5-50 km. The existence of accelerated meandering electrons, electron non-gyrotropy and/or temperature anisotropy $Te_{perp}/Te_{para}$ in the electron scale current layer must be resolved since they affect the growth of instabilities leading to reconnection. Measurements of 3D distributions of electrons < 30 keV with cadence of up to 50 Hz are required to provide several complete velocity distribution functions during a typical crossing time of an electron scale current layer, i.e. ~0.1 s, and to allow comparison with wave activity detected by the fields instruments.

## 3.3 Turbulence science requirements

### 3.3.1 Context and parameters

The recently developed multi-spacecraft techniques and the unique capability of Cross-scale are offering a revolution in our ability to study plasma turbulence. With Cross-scale it will become possible to answer the remaining key questions concerning the nature of the turbulent cascade, particularly near the ion and electron kinetic scales. To achieve this goal fields and particles will have to be measured simultaneously at the various physical scales of the plasma. In a minimal 7 spacecraft configuration, ion and electron moments (density and velocity) and full 3D high resolution distribution function of ions and 3D distribution functions of electrons are required from 4 spacecraft at ion scale, and at least 2 spacecraft at electron scale (or at fluid scale). These should be at a temporal resolution of ~1s at ion scales, and ~0.1s at electron scales (while ~ 4sec (spin) resolution is required in case of measurements at fluid scale). Similarly, AC and DC magnetic (3D) and electric (2D) fields measurements at 4-spacecraft for the available scales (ion+electron, or fluid+ion) are required in the frequency range up to 200Hz. The optimal 12 spacecraft fleet would provide measurements at the fluid, ion and electron scales simultaneously, with the requirements already indicated.

### 3.3.2 The turbulent energy cascade and partition between particles and waves

The description of how energy is transferred and dissipated along the turbulent cascade is important to understand the phenomena occurring at different scales, and the transition between plasma regimes (MHD, Hall-MHD, kinetic…). Cross-Scale must determine, through the k-filtering technique, the 3D full spectra of the field energy and magnetic helicity in frequency and wave vector domain for the dispersive/dissipative scales, as well as the kinetic energy distribution of both ions and electrons. The temporal resolution of these measurements are dependent on the scale and are indicated in Section 3.3.1. Moreover, in order to study the partition of energy between particles and waves, it must provide the relation dispersion for wave modes present in plasma, responsible for the dissipation (or storage) of part of the energy cascading toward both electron and ion scales.

### 3.3.3 Identification, imaging and characterization of coherent structures

Coherent structures, as tangential or rotational discontinuities, shocks and current sheets, are formed along the cascade in solar wind turbulence. For example, the problem of intermittency is strictly related to the scale dependent appearance of structures. Those need to be identified in order to understand how energy is transported to smaller scales, and eventually dissipated into heat or waves. To better understand the origin of such structures, the way they are formed, their orientation in the space, their relationship with anisotropies, and the consequences of their presence in the flow, the correct identification, 3D imaging and characterization are needed, from both field and particle measurements. Since structures are scale dependent, 4 spacecraft measurements are required along all the scales available, with more focus on the small scales of the flow where dissipation occurs.

## 3.4 Orbit Requirements

As mentioned previously, the near Earth environment provides a perfect laboratory in which to investigate the cross-scale coupling of shocks, reconnection and turbulence. The Cross-Scale apogee is of order 25 $R_e$ (up to 30 $R_e$) and the perigee around 10 $R_e$ geocentric distance. This roughly equatorial (14° inclination) orbit enables Cross-Scale to reach the regions where the targeted physics is occurring, including the dayside bowshock (12– 22 $R_e$), reconnection at the magnetopause (8– 12 $R_e$) and in the “tailbox” region of the geomagnetic tail (20–30 $R_e$), and turbulence in the solar wind (12-25 $R_e$), magnetosheath, and geomagnetic tail/plasma sheet.

Cross-Scale’s primary location for studying collisionless shocks is at the Earth’s bow shock, which is at ~14 $R_e$ at the sub-solar point flaring to ~22 $R_e$ at the dawn-dusk terminator. This full range, and tailward of the terminator, is required to efficiently study a range of field geometries and Mach numbers. The primary factors in studying the cross-scale coupling at shocks are the number of shock crossings and the speed of the crossing. This latter is dominated by the bow shock response to even small variations in solar wind pressure, resulting in motions at 10's of km/s. Multiple traversals then occur while the spacecraft are roughly in the appropriate location, and are increased by an orbit apogee that is relatively close to the bow shock, e.g. ~20 $R_e$. Further parameter regimes will be accessible through interplanetary shocks. This requires long periods in the solar wind, facilitated by an orbit apogee > 25 $R_e$.

Cross-Scale's primary location for studying reconnection is at the Earth's magnetopause and Earth's magnetotail current sheet, although reconnection regions also exist in the solar wind and magnetosheath. Typically, the magnetopause is at ~10 $R_e$ distance at the sub-solar point flaring to about ~15 $R_e$ at the dawn-dusk terminator. The most likely location of the near-tail reconnection is between 20 $R_e$ and 30 $R_e$ downtail, centred in the pre-midnight sector. It is required that Cross-Scale cross these reconnection regions. The primary factors in studying the cross-scale coupling are the number of current sheet crossings and the speed of the crossing. In the magnetotail current sheet, natural current sheet oscillations allow multiple crossings around the reconnection region, provided that the spacecraft is close to the neutral sheet in the "tailbox" region. The tailbox location relative to the ecliptic plane depends on the season due to the tilt of the Earth's rotation axis and magnetic field. This requires that spacecraft have an apogee at the neutral sheet at downtail distance > 25 Re with particular constraints on the inclination and other orbital elements. An orbit with this apogee is expected to cover all the key reconnection regions, i.e., magnetopause, magnetotail current sheet, solar wind, and magnetosheath.

Cross-Scale's primary location for studying turbulent cascade is within the solar wind, which requires an orbit apogee >= 25 Re. The low solar activity period during the years 2017-2020 will correspond to steady and stationary fast and slow wind streams, so that turbulent cascade is not affected by local phenomena such as stream interactions or high solar activity (CMEs, solar flares…). Also, the upstream region near the quasi-parallel bow-shock presents a great interest for the enhanced presence of waves. The regions where turbulence anisotropy and coherent structures can be observed spans from the solar wind to the bow shock boundary layer, the magnetosheath and magnetotail. All these regions should therefore be crossed during the course of the mission.

# 4 Payload

This chapter provides an overview of the model scientific model payload of the Cross-Scale mission assessment study, which is presented in more detail in the Payload Definition Document or PDD [51].

## 4.1 Introduction

The model payload for the Cross-Scale mission in the assessment phase industrial studies consisted of a range of different configurations of 11 scientific instruments on the seven science spacecraft. It should be noted that this model payload was defined to be used only in the parallel industrial studies to advance knowledge on the accommodation and interface aspects of a typical payload. Pending down selection to the Definition Phase, the final payload selection will be made during the Announcement of Opportunity for Payload for Cross-Scale in the fourth quarter of 2010.

In parallel to the mission studies by industry, instrument study teams were formed on the basis of the Declaration of Interest for Science Instrumentation for the Cross-Scale mission as part of the Cosmic Vision programme. These studies, funded by ESA member states, aimed at looking into the critical aspects of potential key instruments for the Cross-Scale mission. These instrument study teams will have prepared a final or interim report on their findings and these reports are part of the documentation in the review process. As the instrument studies were running in parallel with the mission studies it is inevitable that some instrument reports have a slightly different instrument conceptual design compared to that used in the industrial studies, in particular as the final model payload for the industrial studies was more or less frozen in January 2009 before the end of the instrument studies.

| Instrument | | | Spacecraft type | | | | |
|---|---|---|---|---|---|---|---|
| **Acronym** | **Measurement** | **Nom. Mass [kg]** | **E1 &E2** | **E3** | **E4/I1 & I3** | **I2** | **I4** |
| MAG | Boom mounted DC vector magnetic field | 0.94 (no boom) | **0.94** | **0.94** | **0.94** | **0.94** | **0.94** |
| ACB | Boom-mounted AC vector magnetic field (1Hz-2kHz, spectra and waveform) | 0.8 (no boom and harness) | **0.8** | **0.8** | **0.8** | **0.8** | **0.8** |
| E2D | 30-50m wire double probe 2D electric field (DC & spectra), (0-100 kHz, DC & spectra) | 8 | **8** | **8** | **8** | **8** | **8** |
| EDEN | Electron density sounder | inside ACDPU | **0** | **0** | **0** | **0** | **0** |
| ACDPU | Common processor & electronics for ACB, E2D & EDEN & MAG) | 3.8 | **3.8** | **3.8** | **3.8** | **3.8** | **3.8** |
| EESA | Dual head thermal 3D electron electrostatic analyzer (3 eV – 30 keV) | 3 | **12** | **0** | **6** | **6** | **6** |
| IESA | Ion Electrostatic Analyzer | 1.5 | **0** | **0** | **3** | **6** | **3** |
| ICA | 3D ion composition < 40 keV | 3.5 | **0** | **0** | **0** | **0** | **3.5** |
| HEP | Solid State high energy particle detector > 30 keV | 1.0 | **0** | **0** | **1** | **0** | **0** |
| CPP | Centralised payload processor | 1.2 | **1.2** | **0** | **1.2** | **1.2** | **1.2** |
| ASP | Active spacecraft potential control | 2 | **2** | | | | |
| | **Total** | | **28.7** | **13.5** | **24.7** | **26.7** | **27.2** |

*Table 1. Instrument based description of all spacecraft types and their payload masses per spacecraft configuration (E1-E4 are electron scale or small scale spacecraft and I1-I4 are the ion scale or medium scale spacecraft). The MAG and ACB resources do not include the booms as these items are to be provided by the spacecraft contractor.*

The model payload comprises basically three categories of instruments; particle, field and their respective digital payload processing units. An overview of the payload is given in Table 1 with their masses. Their place in each of the five different configurations onboard the seven science spacecraft is given in Table 2 with power consumption shown in Table 3. Building on the heritage of recent space plasma mission such as Cluster, the instruments will have a centralised digital processer, in this case one for the particle instruments (CPP) and one for the field instruments (ACDPU). The ACDPU includes a connection to the CPP to provide magnetic field data to the particle instrument data processing.

## 4.2 Payload configurations

Prior to the initiation of the industrial studies, the Cross-Scale SST worked on the definition of the payload on all seven science spacecraft based on the science requirement in the SciRD. Due to mass constraints it was not possible to implement an identical payload configuration on all seven spacecraft and therefore an optimisation was performed which resulted in the configuration depicted in Table 2. The most resource demanding instrument is the fast electron instrument EESA and therefore the full set of sensors is only implemented on two of the electron scale spacecraft. The ion scale spacecraft have a reduced set of sensors of EESA and also a reduced measurement frequency. The ion electrostatic analyzer instrument (IESA) is only carried on the ion scale spacecraft, while the high energy particle instrument (HEP) is located on two spacecraft and the ion composition instrument (ICA) is located on one ion scale spacecraft. The AC and DC magnetic field instruments (ACB and MAG) and the electric field wire booms (E2D) are onboard all seven spacecraft to provide field measurements on all scales. The E2D instrument is identical on all seven spacecraft. By inclining the spin axes of a neighbouring spacecraft, it is possible to combine the E2D data from 2 spacecraft to provide full 3D electric field vectors for frequencies consistent with the propagation and convective time scales and length scales larger than the spacecraft separation. The extent of the relative inclination was set at +/- 20 degrees during the industrial study (spacecraft E1 and E3, Section 5.5). This configuration is referred to in Table 2 as E2Dincl (E2D inclined). This approach removes the need for a rigid axial boom, which would lack the corresponding length while greatly impacting the spacecraft design.

| No | 1 | | 2 | | 3 | | 4 | | 5 | | 6 | | 7 | |
|---|---|---|---|---|---|---|---|---|---|---|---|---|---|---|
| S/C | E1 | # | E2 | # | E3 | # | I1/E4 | # | I2 | # | I3 | # | I4 | # |
| Instr. | ACB | 1 | ACB | 1 | ACB | 1 | ACB | 1 | ACB | 1 | ACB | 1 | ACB | 1 |
| | ASP | 1 | ASP | 1 | | | | | | | | | | |
| | | | E2D | 4 | | | E2D | 4 | E2D | 4 | E2D | 4 | E2D | 4 |
| | E2Dincl[1] | 4 | | | E2Dincl[1] | 4 | | | | | | | | |
| | EDEN[2] | 1 | EDEN[2] | 1 | EDEN[2] | 1 | EDEN[2] | 1 | EDEN[2] | 1 | EDEN[2] | 1 | EDEN[2] | 1 |
| | EESA | 4 | EESA | 4 | | | EESA | 2 | EESA | 2 | EESA | 2 | EESA | 2 |
| | | | | | | | HEP | 1 | | | HEP | 1 | | |
| | | | | | | | | | | | | | ICA | 1 |
| | | | | | | | IESA | 2 | IESA | 4 | IESA | 2 | IESA | 2 |
| | MAG | 1 | MAG | 1 | MAG | 1 | MAG | 1 | MAG | 1 | MAG | 1 | MAG | 1 |
| | ACDPU | 1 | ACDPU | 1 | ACDPU | 1 | ACDPU | 1 | ACDPU | 1 | ACDPU | 1 | ACDPU | 1 |
| | CPP | 1 | CPP | 1 | | | CPP | 1 | CPP | 1 | CPP | 1 | CPP | 1 |

*Table 2. Cross Scale Payload composition and number of units/packages per spacecraft defined by the Science Study Team; [1] E2D inclined is the same instrument as E2D only the spacecraft tilt is in this case 20 degrees with respect to the orbital plane, while for E2D the spacecraft are only 2-5 degrees inclined; [2] the EDEN instrument is incorporated into the ACDPU instrument. Note; E1-E4 are the electron or small-scale spacecraft, I1-I4 are the ion or medium scale spacecraft and I1 and E4 are at the apex of the shared corner.*

To ensure that the probes of the E2D wire antennas are always in sunlight the other five spacecraft will also have spin-axis inclined by a few degrees. The EDEN instrument makes use of the wire antennas of E2D to measure the electron density. EDEN is completely integrated into the ACDPU processor. The ASPOC instrument is part of the payload on two spacecraft where also the maximum number of EESA sensor heads are placed to perform the fast (50 Hz) electron measurements. ASPOC will ensure that the spacecraft charging is kept to a minimum to ensure that the lower range of electron energies can be accurately measured.

| **Instrument acronym** | **Nom. Power per unit [W]** | **E1/ E2 [W]** | **E3 [W]** | **E4/I1 & I3 [W]** | **I2 [W]** | **I4 [W]** |
|---|---|---|---|---|---|---|
| MAG | 1.5 | 1.5 | 1.5 | 1.5 | 1.5 | 1.5 |
| ACB | 0.1 | 0.1 | 0.1 | 0.1 | 0.1 | 0.1 |
| E2D | Inside ACDPU | 0 | 0 | 0 | 0 | 0 |
| EDEN | Inside ACDPU | 0 | 0 | 0 | 0 | 0 |
| ACDPU | 11 | 11 | 11 | 11 | 11 | 11 |
| EESA | 7 | 28 | 0 | 14 | 14 | 14 |
| IESA | 2.0 | 0 | 0 | 4 | 8 | 4 |
| ICA | 6 | 0 | 0 | 0 | 0 | 6 |
| HEP | 1.3 | 0 | 0 | 1.3 | 0 | 0 |
| CPP | 5.3 | 5.3 | | 5.3 | 5.3 | 5.3 |
| ASPOC | 4 | 4 | | | | |
| | Total [W] | 49.9 | 12.6 | 37.2 | 39.9 | 41.9 |

*Table 3. Instrument power consumption for all the five different payload configurations. Note; E1-E4 are the electron or small-scale spacecraft and the I1-I4 are the ion or medium scale spacecraft.*

## 4.3 DC Magnetometer (MAG)

The MAG instrument will measure the DC magnetic field on all spacecraft, sampling field vectors up to 128 Hz. In the near-Earth space, this data will provide a robust global roadmap of 3D space, i.e. the orientation within the solar wind-magnetosphere-system, as well as the more local orientation of the regions where the key science phenomena (shocks, reconnection and turbulence) occur. Simultaneous with this "position" information, multi-point magnetic field measurements at multiple scales will provide current density (local gradient information) or wave properties that are essential for studying interaction with particles, relevant to all the three main science topics.

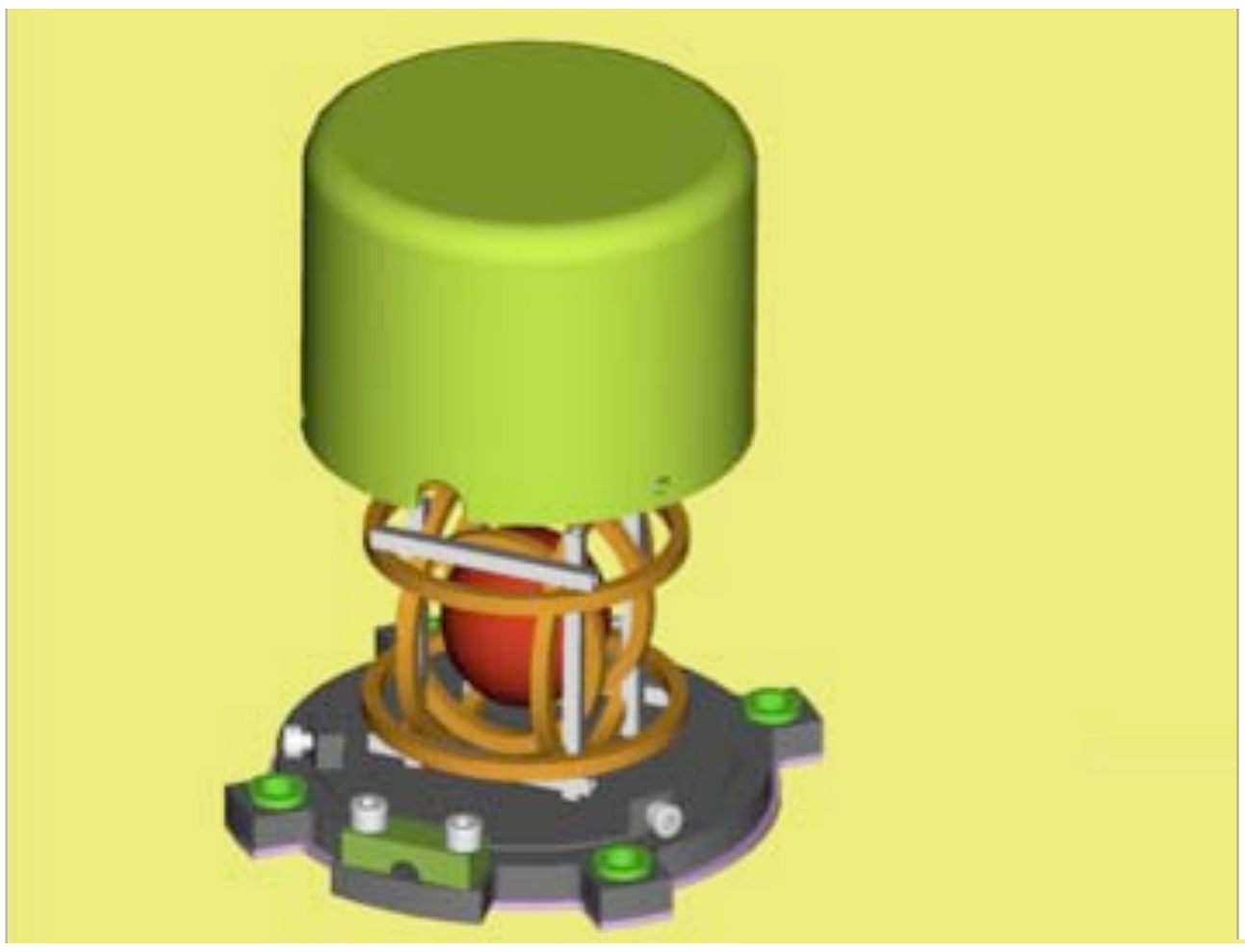

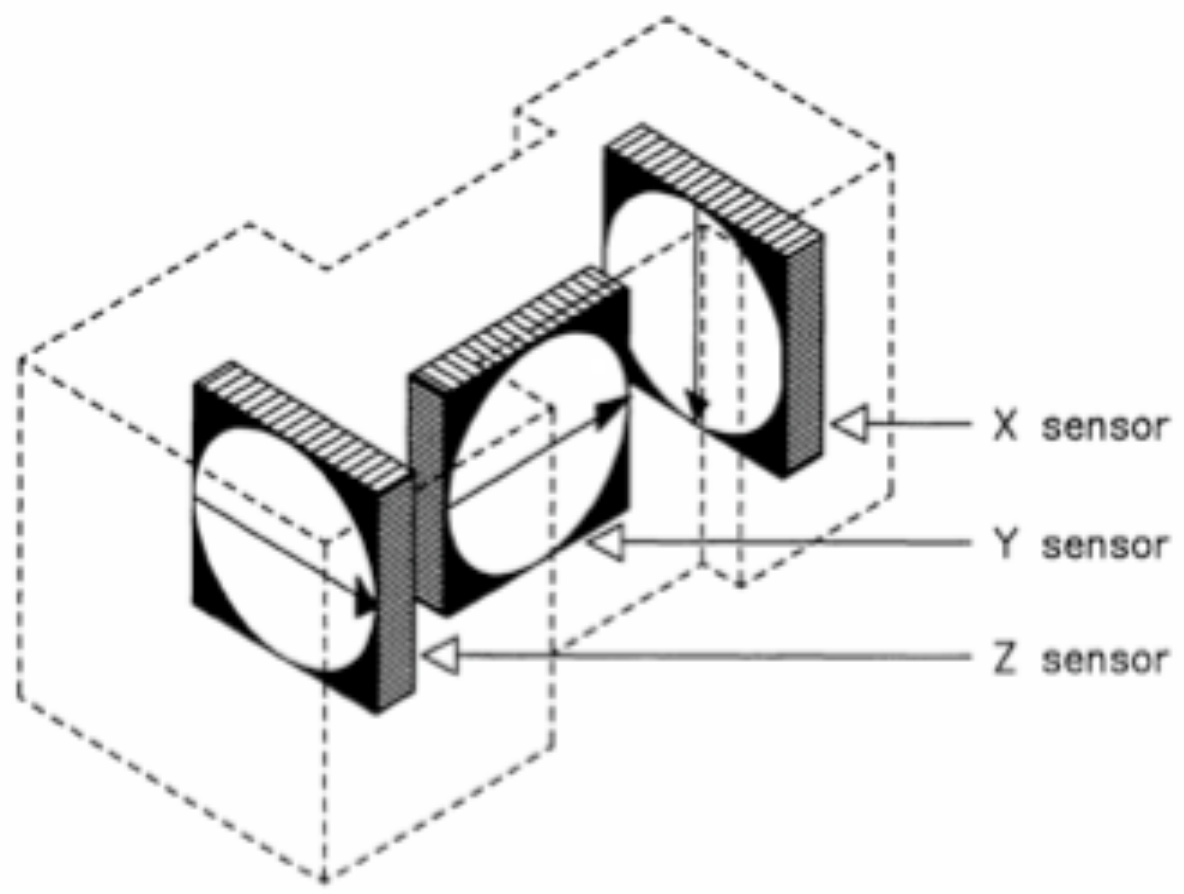

*Figure 18. Left: Fluxgate sensor with opened housing. Right: Schematic of a fluxgate sensor (rectangular prism shape)*

The simultaneous multi-point magnetic field measurements at different scales will allow the variability across the shock to be followed, topological changes in the dynamic current sheet in the reconnection region to be identified and provide the overall turbulence spectrum across the important spatial ranges.

The design of the dual sensor fluxgate magnetometer consists of two sensors, each comprised of a sensor and near-sensor electronics. The sensor itself can have different shapes depending on the design, one solution is shown in the left panel of Figure 18. Two entwined ring-cores are used to measure the magnetic field in three directions in the vector compensated sensor set-up. The magnetic field is measured in X and Z direction via the smaller ring-core, while the larger one is used for the Y and Z components. The ring-cores are equipped with two 3-D coil systems: an inner one to collect (pick-up) the magnetic field dependent second harmonic of the fundamental excitation frequency and an outer Helmholtz coil system to compensate the external field at the ring-core position. The left panel of Figure 18 shows the sensor core with housing and mounting plate made from aluminium. Depending on the design, the ring-cores can be mounted differently, so that the external envelope of the sensor is more rectangular or prism-shaped, as shown in the right panel of Figure 18.

The sensors of the MAG instrument will be located on a spacecraft-provided ~2.8 metre rigid boom. One sensor will be located at the tip of the boom, while the second sensor will be also be positioned on the boom, but at a distance of approximately 1/3 the boom length from the outboard sensor and not less that 50 cm from the spacecraft.

## 4.4 AC Magnetometer (ACB)

AC search coil magnetometers are intended to measure the three components of the magnetic field from near DC to about 2 kHz. It will extend waveform measurements up to 500 Hz and will provide spectral information up to several kHz. These measurements will complete the characterisation of waves responsible for particle scattering and extend the turbulence analysis into the electron dissipation range.

The ACB instrument is a tri-axial search coil based on designs for the Cluster, DEMETER and THEMIS missions. It is composed of 3 ELF/VLF antennas associated with a miniaturized preamplifier built in 3D technology. The 3 orthogonal magnetic antennas are assembled in the most compact way as possible and the supporting structure for these antennae is made of nonmagnetic material (PEEK KETRON).

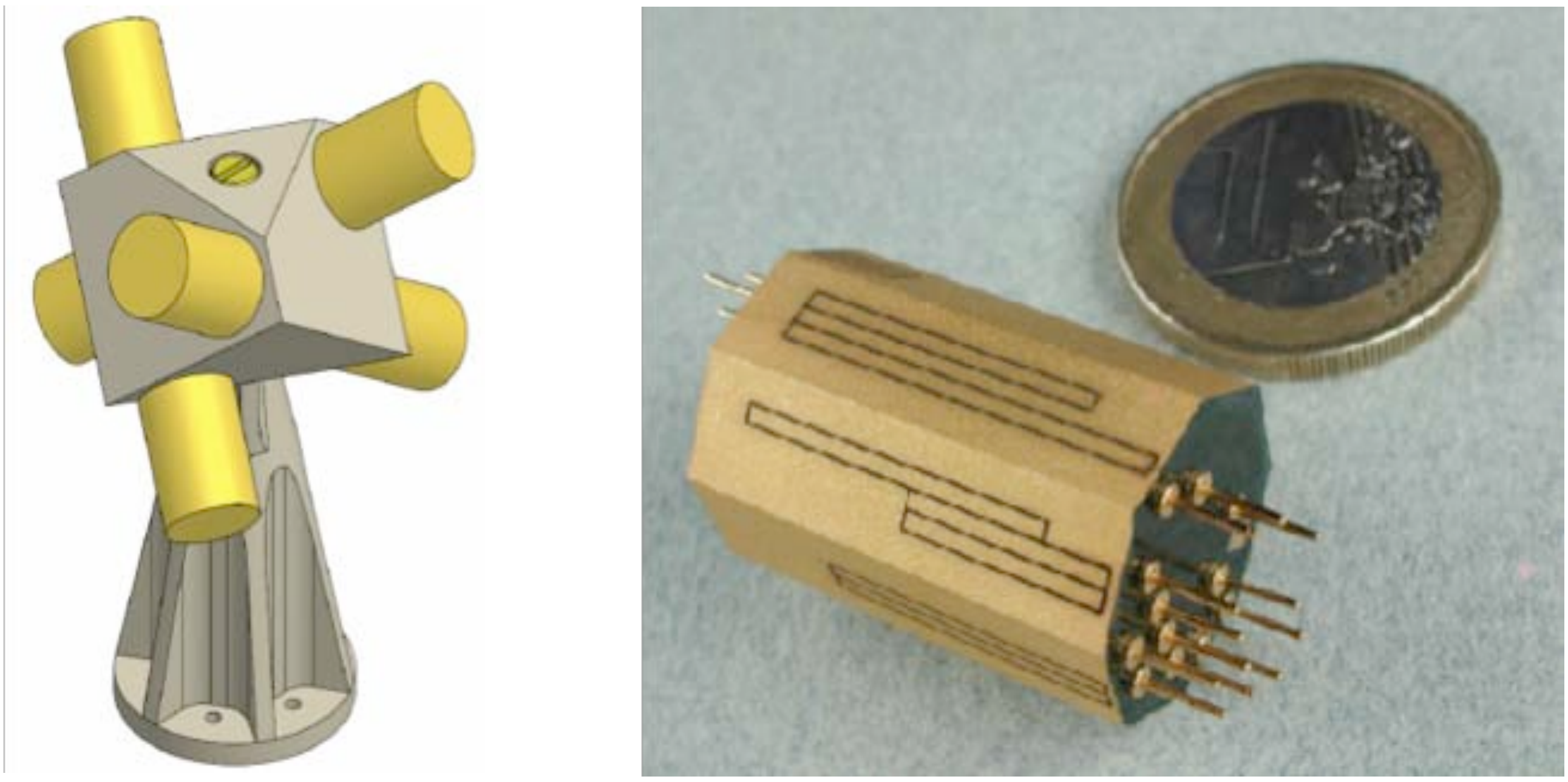

*Figure 19. Left: Sensor representation; Right: Miniaturized preamplifier and its pins.*

The amplification electronic circuit is based on 3D technology and will be mounted in the sensors foot (close to the antennas) to reduce the signal-to-noise ratio. However if the thermal stresses are too critical, the preamplifier can also be installed onboard the satellite.

To minimize noise interference from the spacecraft and other instrument subsystems, the ACB instrument needs to be mounted on a boom extended away from the body of the spacecraft by at least 2 meters. To minimize the S/C interference on the sensor, the 3 antennas must be mounted such that none of the antennas point directly to the body of the satellite. Moreover, the antennas should be mounted at least 1 meter away from any other sensors including any active electronics or magnetic materials. The boom must be of a nonmagnetic material. These requirements are valid for the spacecraft at all scales.

## 4.5 Electromagnetic Field 2D (E2D)

The main goal of the E2D instrument is to measure the electric fields surrounding the spacecraft. By using four wire booms the full electric field in the spin-plane is measured.

A vector component of the electric field is obtained by measuring the observed voltage between two spherical probes at the ends of two wire booms, extending radially from the spacecraft in opposite directions. By combining two such probe pairs (four wire booms) the 2 dimensional electric field components in the spin plane are obtained. Each probe is kept close to the local potential in the space plasma by application of a steady bias current. The potential of any of the sensors with respect to the spacecraft will depend on plasma density, solar illumination and the emission and exchange of photoelectrons by all parts of the spacecraft. The voltage difference between opposite probes is however dominated by the electric field due to the symmetry of the boom assembly. The long wire booms (~ 100 m tip-to-tip) minimize perturbations from inevitable asymmetries on the spacecraft and maximize signal strength. The single-probe voltages provide the spacecraft potential, which in turn depends on the plasma density and is also necessary to monitor for the correct interpretation of particle data. To measure weak signals it is important to mount preamplifiers as close as possible to the sensors, i.e. near the boom ends.

To ensure that perturbations of the potential are minimized and the spacecraft potential well defined, the spacecraft surface should be conductive. The spherical sensors at the boom tips must all be illuminated for the technique to work, meaning that the spacecraft spin axis cannot be exactly perpendicular to the direction to the sun as this would put each sphere in shadow once per spin. This requires the spin axis to be tilted by a few degrees. However, as this relates to measurement quality, not to instrument safety, it is permissible to violate this attitude rule e.g. during s/c manoeuvres. The spacecraft must have good electrostatic cleanliness, preferentially using grounding of all units directly to the s/c chassis, which is then in contact with the conductive surface.

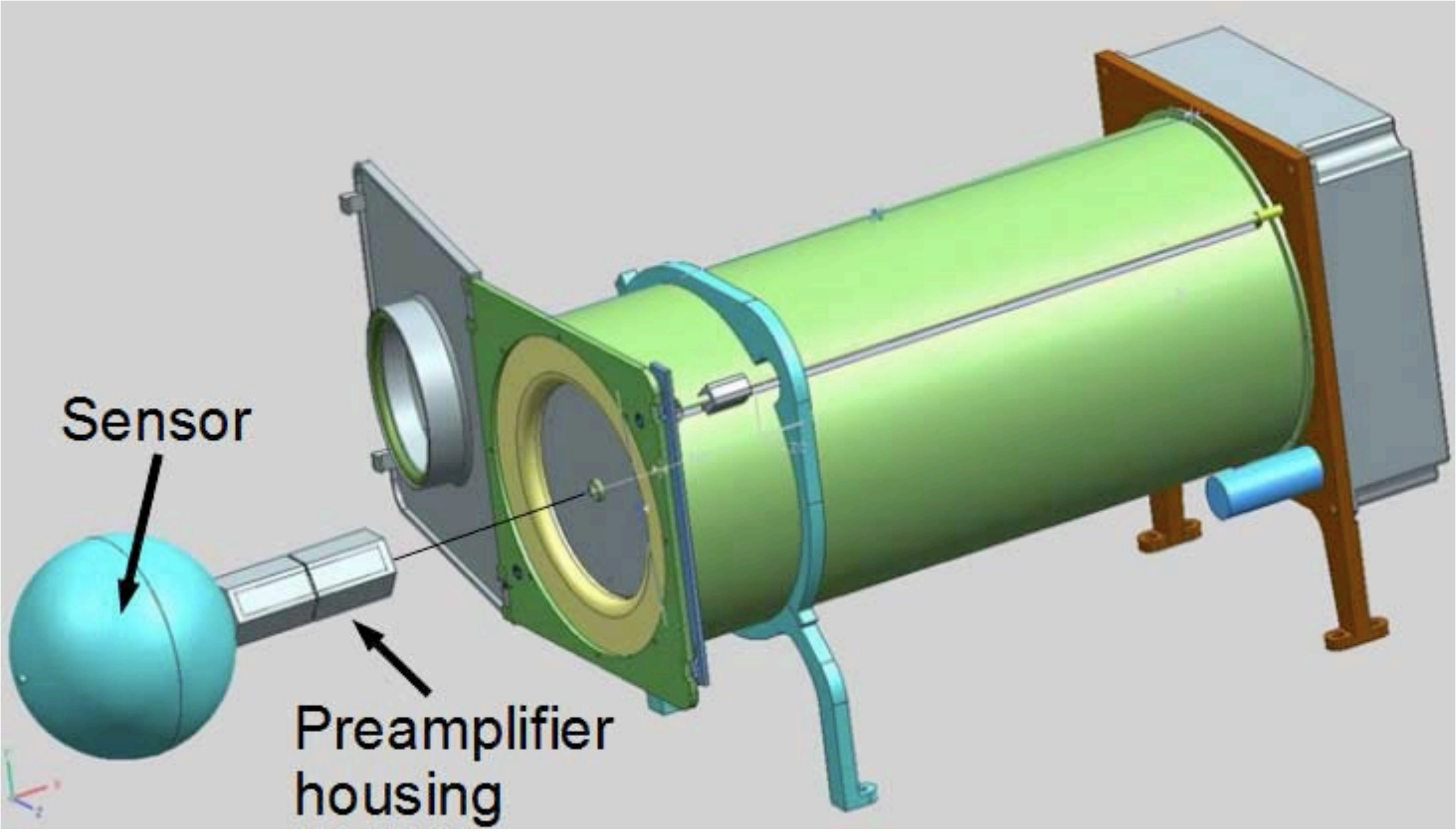

*Figure 20. E2D wire boom unit.*

## 4.6 Electron Density Sounder (EDEN)

The EDEN instrument, as a resonance sounder, can measure the total electron density in the range [0.2 $cm^{-3}$ – 80 $cm^{-3}$] via active stimulation and subsequent detection of the resonances of the local plasma. It will provide a high time resolution, high quality total electron density measurement that is central to the science objectives, and that underpins the calibration of the particle instruments. The EDEN instrument is linked to the E2D antennas and, accordingly, measurements will be performed on board all 7 spacecraft.

The EDEN relaxation sounder is linked to the electric field antennas of Cross-Scale (part of the E2D instrument) and consists of a transmitter and a receiver. The data acquisition and data processing will be performed by ACDPU. The transmitter is connected to the braid of one pair of electric field antenna through the E2D experiment module. When operating in N mode, the transmitter is inactive. When operating in S mode the transmitter is active a signal synthesiser delivers a pulse of sine waves, of either 1ms or 0.5ms duration. The successive frequency values in a sweep follow a table chosen by telecommand. The receiver is connected to the second pair of E2D electric antennas through the E2D experiment module. It is operated in each 13.3 ms listening step, after the transmission of the pulse (the receiver is inhibited during the pulse), during the acquisition slot.

## 4.7 Fields Digital Processing Unit (ACDPU)

The ACDPU lies at the heart of the suite of instruments that measure the electromagnetic wave environment of the spacecraft. It is responsible for the commanding, synchronization, mode control, and the processing of data from the other fields and waves instruments (E2D, ACB and MAG) as indicated in Figure 21. The common electronics box also contains WAXS (High Frequency Receiver, Low Frequency Receiver, Thermal Noise Receiver), Electron density sounder (EDEN), bias electronics for E2D, and a central power distribution unit.

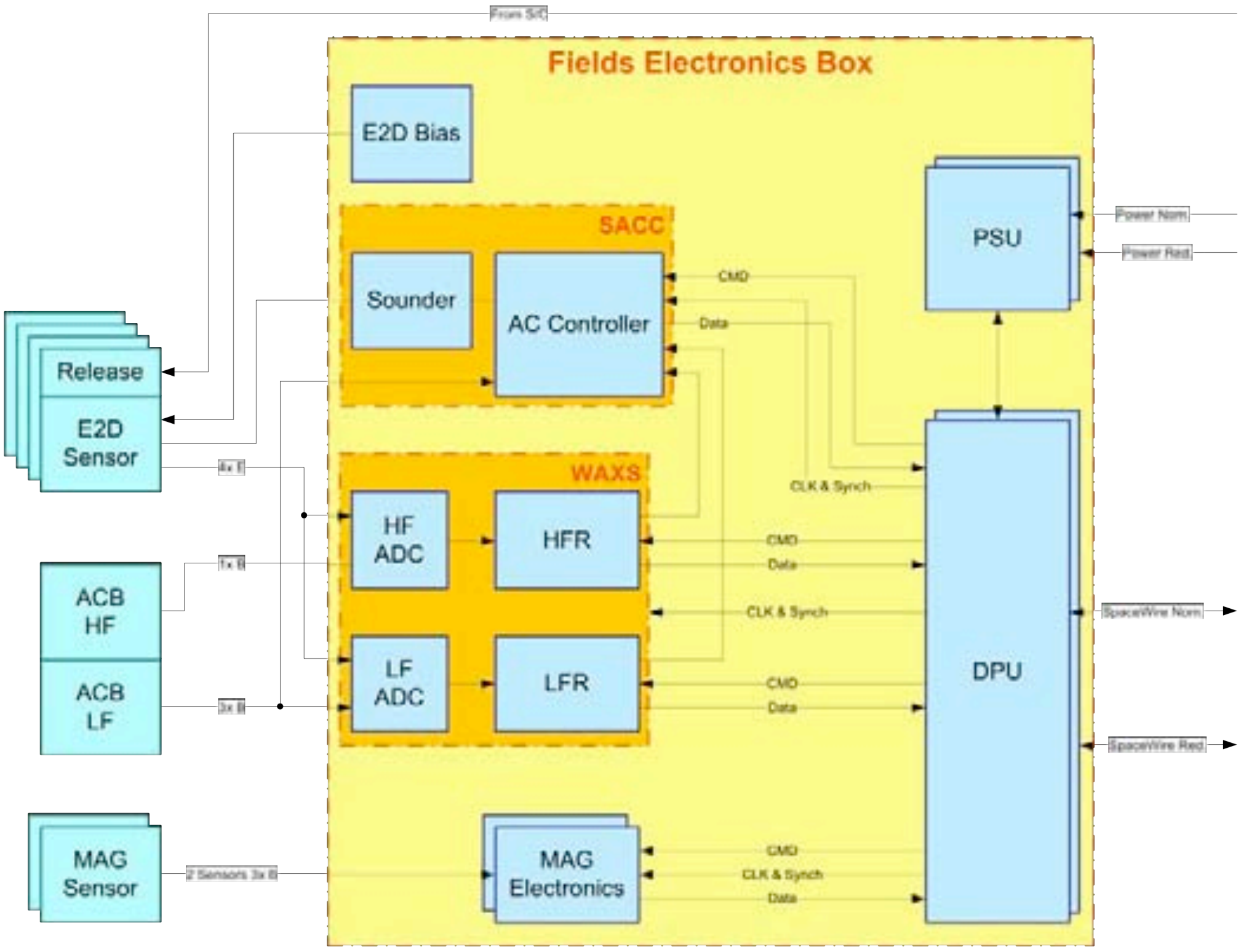

*Figure 21. Block diagram of the Fields DPU (ACDPU) showing connections to the sensors (E2D, ACB, MAG), internal components (wave analyser (WAXS), electron density sounder and AC controller, sensor electronics (MAG and E2D bias), data processing unit, and power supply unit (PSU)).*

This consortium of instruments will provide the following fields measurements to enable Cross-Scale to achieve its scientific objectives:

- Continuous electric and magnetic field waveforms at a sampling rate of at least 500 Hz.
- Multi-component electric and magnetic field spectra up to several kilohertz.
- 2D electric field measurements up to 100 kHz.
- Absolute plasma density using a resonance sounder at low time resolution (~1 second).
- DC magnetic measurements

The instrument performs digital processing on signals acquired by the electric field sensors (E2D), AC magnetic field sensors (ACB), and data pre-processed by the wave analyser (WAXS). Data processing may include: data compression of waveform data as well as spectrogram data (eg JPEG) to reduce the overall data volume (all instruments), transformation of coordinates and forwarding of the data to other instruments, the onboard combination of MAG and ACB data sets and onboard calculation of wave parameters (polarization, wave normal direction). The composition of the ACDPU is given in Table 4 and includes parts of the E2D and EDEN instruments and includes its own power distribution unit.

| Notes | Sub unit | Mass | Mean Power | Peak Power | Voltages |
|---|---|---|---|---|---|
| 0 | E2D & ACB pre-amplifiers | 0 | 0.9 W | 0.9 W | +/-12 |
| 1 | E2D Bias | 0.3 kg | 1.7 W | 2.7 W | +/-12, +5, +3.3 |
| 2 | Sounder | 0.2 kg | 0.5 W | 1.5 W | +28, -6 (TBC) |
| 3 | WAXS High Frequency Receiver | 0.3 kg | 1.5 W | 1.5 W | +/-12, +5, +3.3 |
| 4 | WAXS Low Frequency Receiver | 0.3 kg (TBC) | 1.5 W (TBC) | 1.5 W (TBC) | TBC |
| 5 | Thermal Noise Receiver | 0.5 kg | 1 W | 1 W | TBC |
| 6 | ACDPU Controller | 0.3 kg | 0.75 W | 1.1 W | +3.3 |
| 7 | ACDPU High level data processing | 0.3 kg | 0.75 W | 1.1 W | +3.3 |
| 8 | DCDC & Power distribution | 0.6 kg | 2.4 W | 3.1 W | +28 |
| 9 | Box, backplane, wiring | 1.0 kg | | | |
| | Totals | 3.8 kg | 11 W | 14.4 W | |

*Table 4. Breakdown of mass and power resources for the ACDPU as accommodated by the industrial studies. MAG electronics are incorporated in the Fields DPU that emerged from the parallel instrument studies.*

## 4.8 Electrostatic Analysers

Electrostatic analysers measure the 3D distribution functions of ions and electrons, at energies from a few eV to a few 10's keV. These measurements are essential to determine the fluid parameters (density, flow velocity, temperature, composition...) as well as the kinetic properties (anisotropies, fluctuations with respect to maxwellians, particle beams...) of the plasma. They thus play a central role in any studies regarding the dynamics of collisionless plasmas and their fundamental processes.

Figure 22 shows the principle of a very common type of analyser: the ‘top-hat’, a symmetrical electrostatic analyser which has a uniform 360° field of view (FOV). It consists of three concentric spherical elements: an inner hemisphere, an outer hemisphere, which contains a circular opening and a small circular top cap, which defines the entrance aperture. In the analyser, a potential is applied between the inner and outer plates and only charged particles with a limited range of energy and an initial polar angle are transmitted. The particle exit position, usually identified by the use of a ring-shaped MCP, is a measure of the incident azimuth angle.

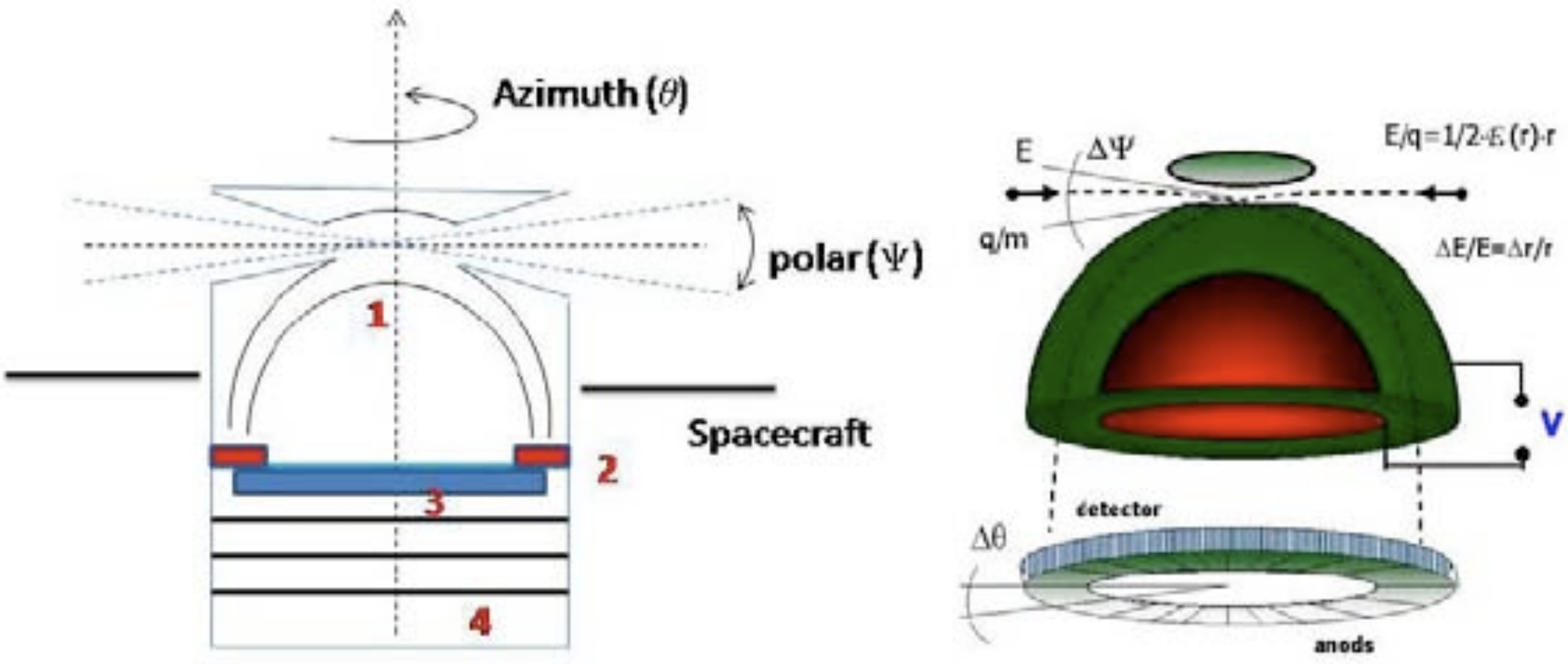

*Figure 22. Principle of a top hat analyser. (1) Optics element: top-hat analyser system. (2) Detector and readout element consisting of an MCP mounted onto an anode and readout board, including HV coupling capacitors and readout electronics. (3) LV and HV boards (4) FPGA-based electronics board/boards for instrument control, position processing, interfaces, counters etc.*

Such sensors are usually are mounted on the spacecraft where the aperture of the optics is parallel to the spin axis and either: (a) tangential to the spacecraft surface, resulting in a 360° FOV; or (b) normal to the spacecraft surface, resulting in a FOV 180°. For a single sensor a full 3D distribution of the particles can then be obtained every (a) ½ spin or (b) 1 spin of the spacecraft.

### 4.8.1 Electron electrostatic Analyser (EESA)

On the electron scale, fast (~50 Hz) measurements of the full 3D electron velocity distribution function need to be made at sufficiently high energy and angular resolution to identify electron beams, temperature anisotropies, multiple populations and other features that may drive the collisionless plasma processes (i.e. reconnection, shocks, turbulence) which are the target of the Cross-Scale mission. Discussed in Section 4.8, tying of the basic instrument time resolution to the spacecraft spin period (four seconds) is adequate at the fluid scale. However, higher time resolutions are needed on the ion and electron scale spacecraft to increase the resolution and provide full 3D capability at the corresponding ion or electron time scales demanded by the science objectives. Multiple sensors, as either dual-heads within a single unit or multiple units, must therefore be deployed. Additionally, on the electron scale, deflector plates need to be employed to increase the field of view for sub-spin resolution and reduce the number of sensors required. Thus the electron electrostatic analyser system described here for use on the electron scale is base-lined as a set of 4 'dual-head' sensors, with each of the 8 heads covering 180° in elevation, and through the use of the Aperture Deflection System (ADS), ±22.5° in azimuth. The combined heads thus provide 360° coverage in azimuth, or complete 4π steradians of the sky. Electrons from a chosen azimuth are deflected into the hemispherical analyser section (which has 180° acceptance in elevation) of each head by applying a voltage across the ADS system.

Electrons are selected in energy by the applied voltage across the concentric hemispheres and are recorded by the MCP (MicroChannel Plate) or CEM (Channel Electron Multiplier) detectors. Sweeping of the ADS and hemispherical voltages allows electron fluxes in the full angular and energy range for each sensor to be obtained. Note that fewer units could be deployed per electron scale spacecraft if larger acceptance angles for the ADS system are used. However, this has drawbacks in the form of irregular angular coverage, potentially large gaps in that coverage and also means the instrument will not be able to cover the full energy range since higher energies cannot be deflected through the larger angles.

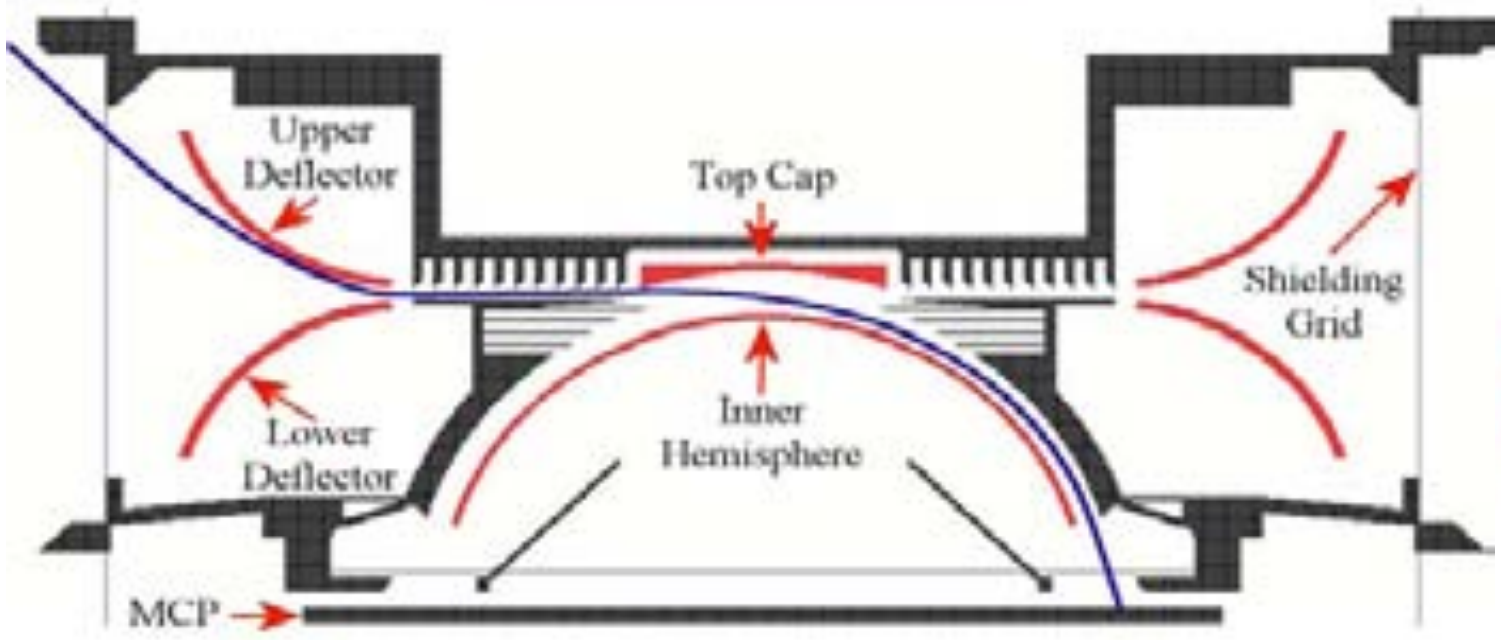

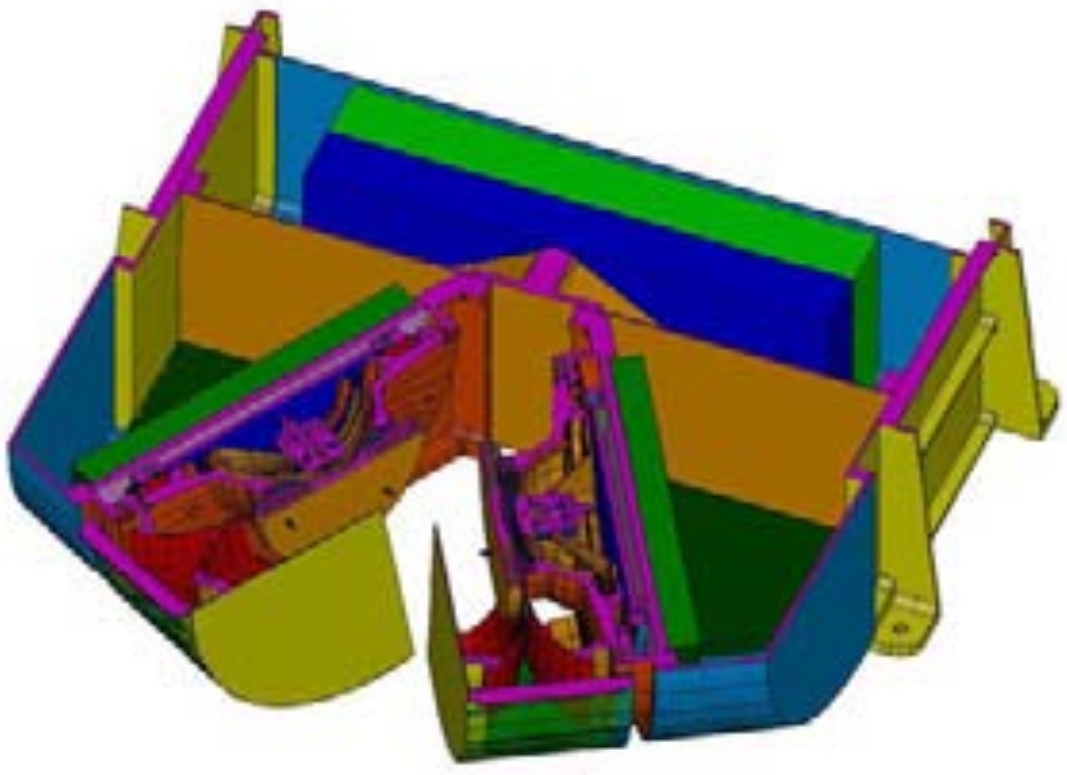

*Figure 23. Left: Electro-optical model of one of the sensor heads of the Dual Head system, right; Dual head unit as used in the design for the MMS mission.*

The EESA is based on a long line of similar sensors flown on Cluster, Cassini and the upcoming Magnetospheric Multi-Scale (MMS) mission. EESA is capable of measuring electrons in the range 3 eV to 30 keV, typically over 30 energy bins with 12 elevation bins and 32 azimuthal bins and the required 50 Hz, with $\Delta E/E \sim 30\%$ if 30 energy bins are used to cover full range.

## 4.8.2 Ion Composition Analyser (ICA)

To measure the distribution functions of various ion species, ICA will utilise the normal electrostatic analyser design presented above but includes an electrostatic time-of-flight (TOF) component as shown in Figure 24. Such a design has a great deal of heritage (Cluster and Cassini). After their direction/energy selection the particles are focussed onto an ultra-thin carbon foil (~0.5 μg $cm^{-2}$) polarized at a voltage of ~ -15 kV, at the entrance of the TOF section. Upon ion impact, the carbon foils emit one or several secondary electrons that are deflected and focused by a dedicated electrostatic optics toward a ring MCP. This provides a START pulse, while the position of the electron impact indicates the azimuthal sector of the incoming ion. When the incoming particles are detected on the MCP at the bottom end of the TOF section this provides the STOP signal.

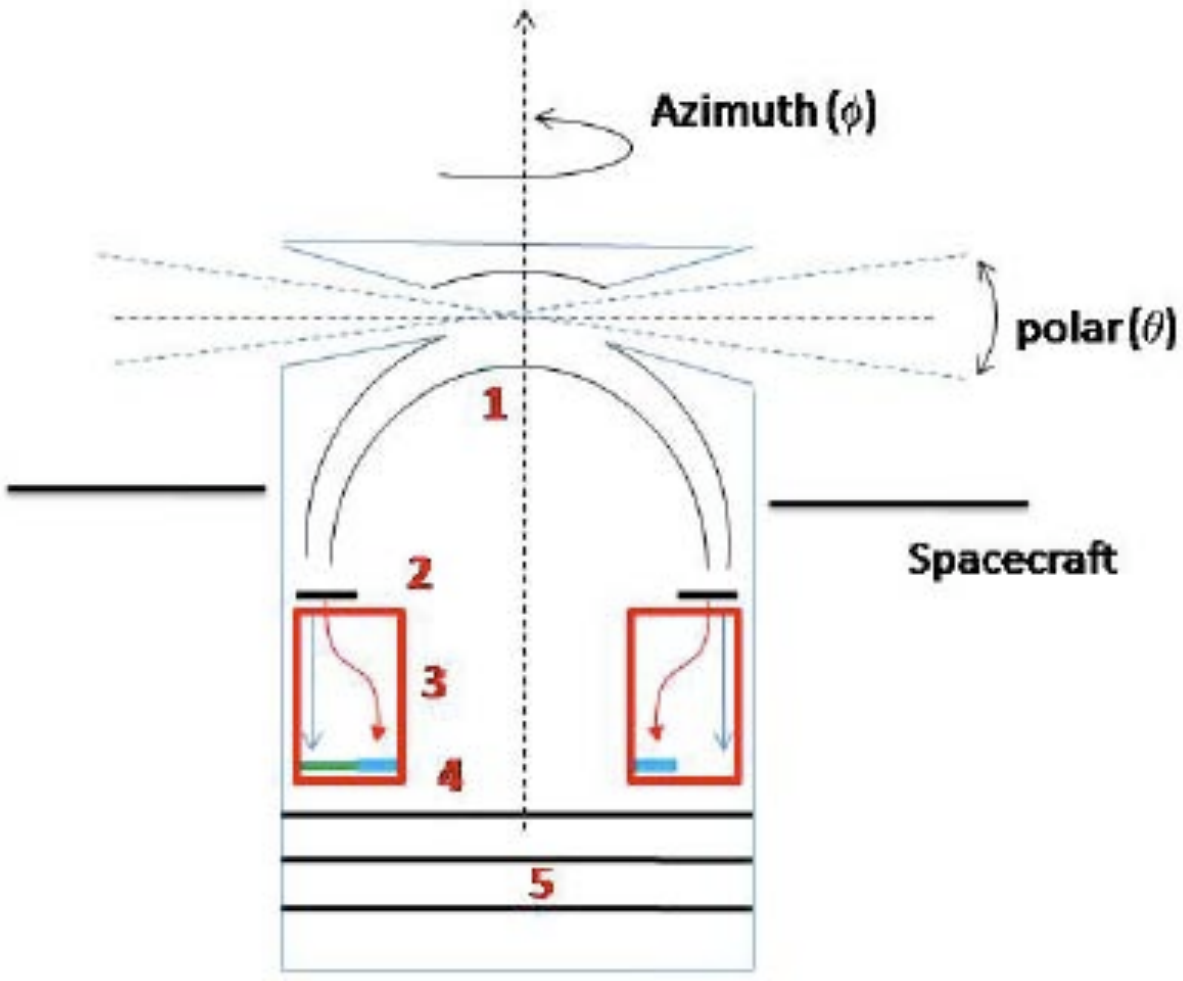

*Figure 24. Principle of a mass composition analyser. Electron Optics element: (1) top-hat analyser system. (2) Carbon foils. (3) TOF section. (4) Start and Stop MCP. (5) LV and HV boards.*

The time-of-flight measurement along the flight path length is used to determine the velocity *v* of the incident ion. Knowledge of the E/q and the velocity of an individual ion will allow the determination of its mass/charge (m/q) and hence separate the dominant ion species: protons (m/q = 1), alpha particles (m/q = 2), and heavier ions (m/q > 2.5).

ICA will be capable of measuring the distribution functions of the major ion species, H, He+, He++, and O+. It will have a 8°x360° FOV and hence able to measure a full distribution every half spin, with an energy range of 10 ev/q – 40 keV/q with 10% resolution.

### 4.8.3 Ion ElectroStatic Analyser (IESA)

The main aspects of IESA have been covered in the general introduction to electrostatic analysers (Section 4.8). IESA is a single head design, covering ions from 3eV to 40 keV, with heritage from Cluster and THEMIS. The FOV can be either 360°x5° or 180°x5° depending on the mounting requirement (FOV normal or tangential to the surface of the spacecraft) resulting in the full 3D distribution being measured in either a half or whole spin respectively. A typical measurement of a 3D distribution will consist of 32 azimuth x 16 polar x 32 energy flux measurements.

## 4.9 High Energy Particle (HEP)

The supra-thermal component of particle distributions is a ubiquitous feature of non-equilibrium, collisionless plasmas including those observed in the near-Earth environment. These populations are most readily described in terms of a kappa function representing a combination of a thermal Maxwellian distribution and a power-law tail. The non-thermal population and rapid field-aligned transport provide the unique capability to remotely sample the acceleration processes and mechanisms taking place within boundaries, shocks and regions of magnetic reconnection. Within the near-Earth plasma environment the non-thermal tail of the distribution is most commonly observed from a few tens of keV and above. The HEP instrument on Cross Scale will measure the full 3D ion and electron particle distributions, at high temporal resolution (~16Hz) in the energy range from ~20 to 1000 keV.

A prospective instrument concept consists of a simple 'pin-hole' design utilising ion implanted silicon solid state detectors to measure supra-thermal ion and electron distributions in the energy range from ~20-1000 keV. The sensor unit comprises a set of detector modules that are similar in design to those adopted for the Imaging Electron Spectrometer (IES) instruments flown on the NASA/Polar and ESA/Cluster spacecraft both of which are still operating successfully after ten and six years respectively. Each energetic electron detection module incorporates a 'pin-hole' aperture, foil and Silicon Solid State Detector (SSD) segmented to provide the desired angular resolution within each modules overall field of view. The baseline configuration of the instrument position on the spacecraft is shown in Figure 25.

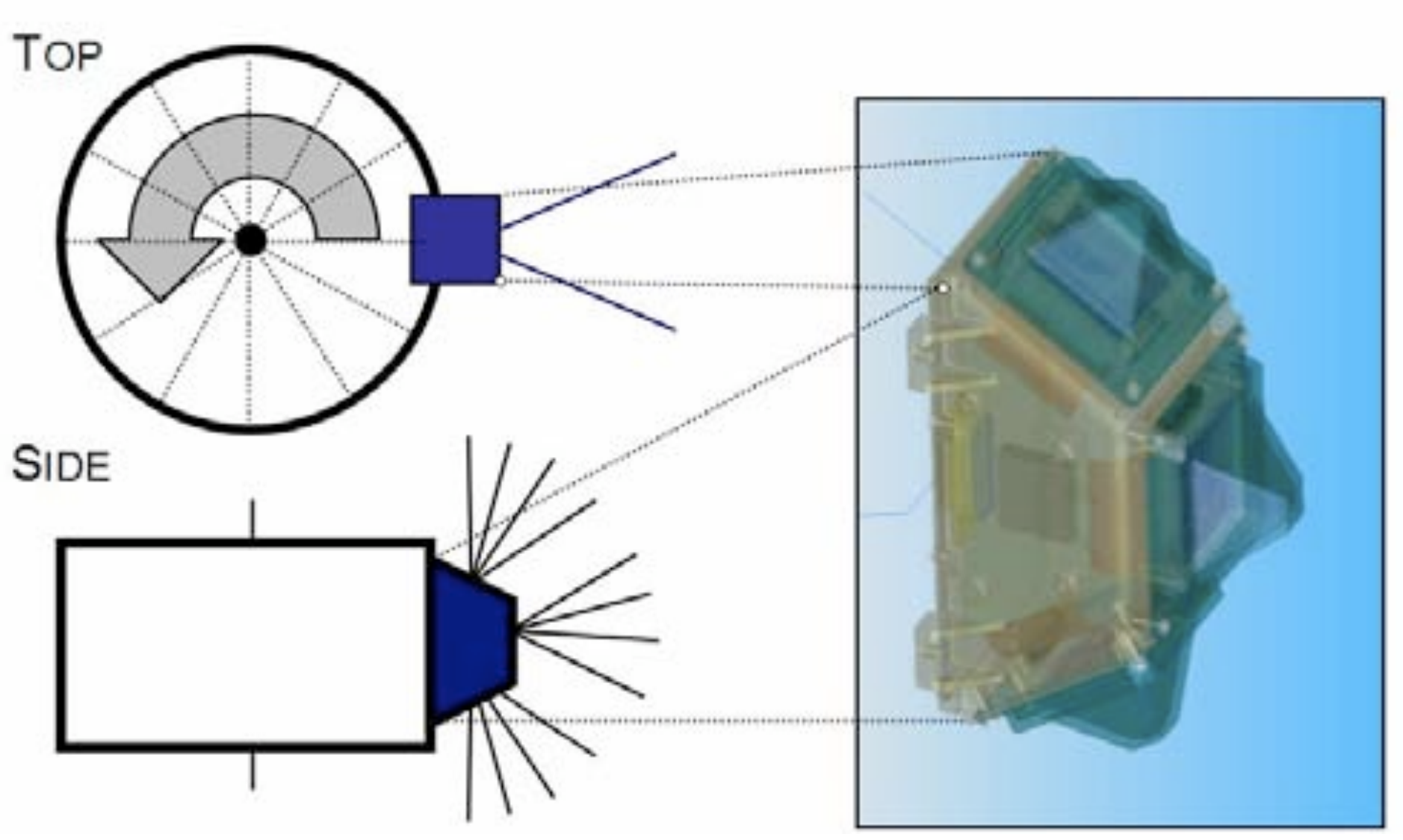

*Figure 25. Baseline configuration of the single HEP sensor unit mounted on the side of the Cross-Scale spacecraft. The HEP unit on the right views in at least 9 (3x3) polar directions. The full 3-D field of view is built up over a spin by sub-dividing the azimuthal directions into a series of sectors with resolution comparable to the polar angles.*

The signals from each SSD element are fed to a front-end, low noise, multi-channel pre-amp with multiplexed digitisation and data readout Application Specific Integrated Circuit (ASIC). This requires minimal external components and is fundamental to the miniaturisation and simplified modular design of the instrument. The remaining sensor unit electronics consist of data buffering, interface to a dedicated or central data processing unit, and power converters for the SSD bias voltage.

# 4.10 Common Payload Processor (CPP)

The CPP will be provide instrument functionality control, memory and computational capability in order to perform the following functions:

- Receive and decode commands from the spacecraft via the Main and Redundant SpaceWire dedicated link.
- Control and management of all the EESA, IESA, ICA and HEP functions.
- Buffer instrument data for the spacecraft On Board Data Handling (OBDH)
- Compress/scale, provide the capability to arrange in pitch-angle (hence requiring the link to ACDPU to obtain information of the MAG measurement), calculate onboard moments, format and transmit sensor measurements according to the available spacecraft telemetry allocation
- For non-lossless compression, verify the full functionality of the local Digital Signal Processor (DSP) compressor, send / retrieve from the DSP raw data / compressed blocks.
- Provide power supplies for CPP itself, and distribute power by latching current limiters to EESA, IESA, HEP and ICA units.
- Verify continuously the health status of CPP both in terms of hardware (HW) (power supplies lines, critical temperatures e.g. DC-DC converter) and in terms of software (SW) (checking of consistency of the checksum appended to all critical tables: operational tables, context tables, Real Time Operating System (RTOS) parameters, etc).

The Data Processing Unit (DPU) will be equipped with a Field Programmable Gate Array (FPGA) based processor, in this case a Leon 3 Fault Tolerant derived processor, able to provide instrument operations control and perform loss-less data compression. The Leon 3 FT represents, at present, the top DPU reference design for FPGA based space applications. The ancillary customized resources block will be developed to guarantee a similar security standard as for example the embedded EDAC (Error Detection Automatic Correction) already developed in the frame of the DPU design in other space missions. Apart for the nominal operation, it will be possible as an option to switch to non-lossless controlled DSP (Digital Signal Processor) based compression operation by a dedicated configurable error parameter.

The onboard management and processing of data is performed by a combination of the HW (FPGA) resident in the detectors' peripherals and the SW and the HW processing units (FPGA, DSP Compressor), resident in the CPP. The Data Handling System (DHS) consists of a Virtual VHDL (VHSIC (Very High Speed Integrated Circuits) hardware description language) microprocessor (custom LEON 3 CPU based) with its related instrument application firmware. This is the main high-reliability processor, which will work as main particle suite interface. The second part is a DSP data compressor and its related DSP compressor application firmware for demanding data rate compressions. In general, the software will be divided into an execution-software and application software.

# 4.11 Active Spacecraft Potential Control (ASPOC)

In order to perform precise plasma observations at energies close to the satellite potential, a satellite potential control device is needed. ASPOC reduces the positive spacecraft potential by emitting an ion beam of a few keV. The beam current sets an upper limit to the potential, which corresponds to the lower end of the energy band measured by the plasma instrumentation.

ASPOC emits a beam of positive ions (indium) to control the electrical potential of the spacecraft. The emission of positive charges from the spacecraft balances the excess of charge accumulating on the vehicle from interactions with the environment in the presence of photo-emission. By adjusting the positive emission current, the spacecraft potential can be adjusted to single-digit positive values. The resulting potential is a function of spacecraft size and shape, the photo-emission properties of the surfaces, the emitted ion beam current, and the characteristics (mainly the density) of the ambient plasma. For a spacecraft of half the sunlit area of the Cluster spacecraft and similar surface parameters, a beam current of 10 μA will be sufficient to set an upper limit to the spacecraft potential of about 5 V. Using up to 50 μA beam current, potentials as low as 2 V are feasible.

The ion emitters are "solid needle" - type liquid metal ion sources using indium as charge material. A solid needle, made of tungsten, is mounted in a heated reservoir with the charge material. A potential of 4 to 10 kV is applied between the needle and an extractor electrode. If the needle is well wetted by the liquid metal, the electrostatic stress at the needle tip pulls the liquid metal towards the extractor electrode and a cone with a tip diameter of a few nm is formed. Field evaporation takes place at the tip and and ion beam is formed. Indium has extremely low vapour pressure, preventing contamination of the source insulators and ambient spacecraft surfaces.

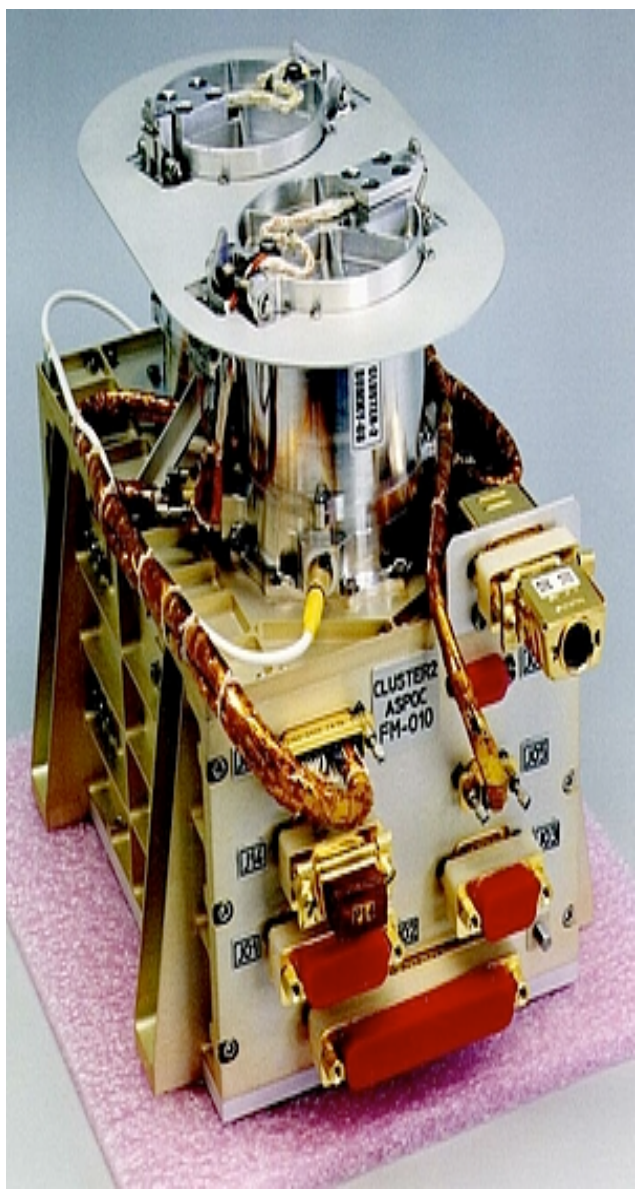

*Figure 26. ASPOC flown on Cluster.*

The instrument comprises two pairs (total 4) of individual ion emitters. Each pair is connected to a dedicated high voltage supply. The ion emitters have heater elements (typical power 0.5 W) to bring the indium in the small reservoir (about 0.5 g) of the active emitter into liquid state (melting point at 156 C). Only one emitter is active at a time. Four emitters are present for lifetime and redundancy reasons. Heritage emitters have demonstrated an achievement of up to 100,000 μAh. A low voltage power converter provides the necessary secondary voltages and power to the heater element of the active ion emitter. Two high voltage converters provide the extraction voltages of typically 4 to 10 kV. A controller based on a microprocessor core running in an FPGA provides the interfaces to the spacecraft data and command system and controls the power supplies.

## 4.12 Payload Data Rates

The SST in collaboration with ESA defined the payload data rates to be used during the industrial mission studies. The industrial contractors used this information to size the memory of the ODBH and the data link to the ground including the number of ground stations to be used. This analysis also included the definition of the ground station contact times per spacecraft since each spacecraft has its own downlink to the ground. The total uncompressed data rate formulated by the SST was 24.19 Mbits/s, while the assumed data compression factor used to reduce the total data volume was 10. In Table 5 the data rates are given per spacecraft and per one and two orbits. This data rate translates to a mass storage capability of 1 Tbit for 2 of the electron spacecraft and 256 Gbit for the other spacecraft, sized to store data for at least 2 orbits leading to a total memory allocation of the constellation with 3.33 Tbit. Of this onboard data volume, around 25% to 33% can be actually downloaded to Earth at the next orbit pass, where 30% corresponds to an average download data rate of 800 kbit/s for the full spacecraft constellation, as required in the Mission Requirements Document or MRD [52]. The method of selection of this fraction of the onboard data is discussed in Chapter 6.

| S/C | Onboard Data rate uncompr. | Onboard Data rate compr. | Onboard Data Volume per Orbit (compr.) | Onboard Data Volume for 2 Orbit (compr.) | Memory accommo-dated |
|---|---|---|---|---|---|
| | [Mbit/s] | [Mbit/s] | [Gbit] | [Gbit] | [Gbit] |
| E1 | 10.47 | 1.047 | 391.83 | 783.66 | 1024 |
| E2 | 10.47 | 1.047 | 391.83 | 783.66 | 1024 |
| E3 | 1.25 | 0.125 | 46.67 | 93.33 | 256 |
| E4/I1 | 0.63 | 0.063 | 23.42 | 46.84 | 256 |
| I2 | 0.60 | 0.060 | 22.58 | 45.17 | 256 |
| I3 | 0.34 | 0.034 | 12.88 | 25.76 | 256 |
| I4 | 0.44 | 0.044 | 16.35 | 32.71 | 256 |
| Total | 24.19 | 2.419 | 905.51 | 1811.02 | 3328 |

*Table 5. Payload data rates (uncompressed and compressed) and memory allocation per spacecraft as baselined during the industrial studies.*

| S/C | Onboard compressed data rate (kbps) | Data Volume per orbit | Data Volume 2 Orbits | Memory | % usage |
|---|---|---|---|---|---|
| | Total | (Gbits) | (Gbits) | (Gbits) | |
| E1 | 1297.6 | 485.8 | 971.6 | 1024 | 95 |
| E2 | 1297.6 | 485.8 | 971.6 | 1024 | 95 |
| E3 | 142.5 | 53.3 | 106.7 | 256 | 42 |
| E4/I1 | 92.1 | 34.5 | 69.0 | 256 | 27 |
| I2 | 85.7 | 32.1 | 64.2 | 256 | 25 |
| I3 | 65.3 | 24.5 | 48.9 | 256 | 19 |
| I4 | 76.8 | 28.8 | 57.5 | 256 | 22 |
| **Totals** | **3057.7** | **1144.8** | **2289.6** | **3328** | |

*Table 6. Updated data-rate and memory usage.*

Parallel to the industrial studies, the instrument teams worked together in consortia, one for fields instruments and one for particles, and studied in more detail the data-rates associated with updated requirements from the instrument teams. The results of these analyses are given in Table 6. It can be seen that although the total data volume per spacecraft is slightly higher compared to that used in the mission studies, these numbers still fit within the allocated memory size per spacecraft with reduced margins. The most important changes with the previous information in Table 5 are the compression factors. The factor 10 compression ratio was modified for the different instruments, becoming 3 for waveforms and other low-dimension time series, 3 for particle data time series and 6 and 8 for omni-directional averaged and 3-D spectra respectively. In addition the number of bits per science–words was modified in certain cases (reflecting a refinement of data resolution). This activity demonstrated that there is scope for flexibility and fine-tuning in the scheme within the current constraints of the 800 kbit/s downlink rate.

## 4.13 Payload Accommodation Issues

The key drivers of the instruments for the design of the spacecraft were the thermal and power characteristics, EMC issues and the position on the payload platform to accommodate each instrument such that the Field of Views were blocked minimally by other parts, i.e. booms and wire antennas. The thermal characteristics of the instruments (shown in Table 8) are especially limiting for the EESA instrument as its thermal operating range is rather small. E2D has a peak power during deployment of the antenna wires, which would only take a small amount of time and during this time other instruments would not operate. The power requirements of the instruments are given in Table 7 dominated by the EESA instrument average power and peak power requirements. The peak power levels are to be handled by the battery of the spacecraft such that the solar arrays are sized to average power levels.

| | DC/DC Conversion | Average Power (W) | Peak power (W) | Duration of peak power (milliseconds) | Instrument power in standby mode (W) |
|---|---|---|---|---|---|
| ASPOC | Inside instrument | 4 | 4.5 | Depends on S/C potential | 0 |
| E2D | DC/DC conversion internally | ACDPU | ACDPU | 0 | 0 |
| EESA | DC/DC internally: 28 V from S/C | 28 for 4 packages | 40 | 0.3 every 5 milliseconds | 8 |
| IESA | DC/DC from S/C: required voltages: +2.5, 5, +/- 12 V | 2 | 2.7 | 10 | 1 |
| ICA | DC/DC from S/C: required voltages: +2.5, 5, +/- 12 V | 6 | 7 | 10 | 2 |
| ACB | DC/DC from S/C: required voltages: +/- 12 V | 0.1 | 0.11 | 1000 | 0 |
| EDEN | DC/DC from S/C: required voltages: +27, -6 and +/- 6, +5.5, +8 V | ACDPU | ACDPU | 1 | 0 |
| MAG (option 1) | DC/DC conversion internally | 1 | 2 | 1000 | 0 |
| HEP | DC/DC conversion internally | 1.3 | TBC | TBC | 0 |
| ACDPU | 28 V (DC/DC internally) | 11 | 14.4 | 10000 | 0 |
| CPP | 28 V (DC/DC internally) | 5.3 | 6.5 | 3000 | 3 |

*Table 7. Power characteristics of Cross-Scale model payload instruments.*

Most instruments have either the DC/DC conversion internally or have this arranged by one of the DPU units ACDPU or CPP. As the payload is divided into five different payload configurations on seven spacecraft the DC/DC conversion performed by instruments or DPUs ensure a simpler interface with fewer different voltage levels to the spacecraft.

| Instrument | Subsystem | Operational temperature range | Non-operational temperature range | Thermal dissipation (estimate) | Require temperature stability; operational (non-operational |
|---|---|---|---|---|---|
| **E2D** | Boom units | -40 °C -- +100 °C [0 °C -- +50 °C during wire boom deployment] | -40 °C -- +100 °C | 0 W and 4 W per unit during wire boom deployment | 1 °C / 1 min |
| | Bias electronics | -25 °C -- +60 °C | -35 °C -- +70 °C | 1.7 W (peak 2.7 W) | 1 °C / 1 min |
| **EESA** | Detector | Maximum temp= +25 °C (ideally +15 °C +/- 10 °C) | -20 °C -- +50 °C | < 1.5 W from each sensor | +/- 10 °C |
| | Electronics | Ideally +15 °C +/- 20 °C | -20 °C -- +50 °C | Included in sensor dissipation | +/- 20 °C |
| **IESA** | | -20 °C -- +50 °C | -40 °C -- +80 °C | 2 W | +/- 20 °C (not critical) |
| **ICA** | | -20 °C -- +50 °C | -40 °C -- +80 °C | 6 W | +/- 20 °C (not critical) |
| **ACB** | Magnetic antennas | -100 °C -- +150 °C | -100 °C -- +150 °C | (TBD) | The temperature gradient along the antennas has to be less than 30 °C |
| | Pre-amplifier | -40 °C -- +80 °C | -50 °C -- +80 °C | < 0.1 W | A thermal blanket shall be provided |
| **EDEN (inside ACDPU)** | Receiver | -20 °C -- +40 °C | -50 °C -- +60 °C | 1.8 W | TBD |
| | transmitter | -20 °C -- +40 °C | -50 °C -- +60 °C | 0.2 W | TBD |
| **MAG** | Sensors (2 per S/C) | -60 °C -- +30 °C | -100 °C -- +80 °C | 50-100 mW (per sensor) | 1°C / 1 min (N/A for non-operational) |
| | electronics | -30 °C -- +60 °C | -40 °C -- +70 °C | 1 (1.5) W without (with) DPU and DC/DC | 1°C / 1 min (N/A for non-operational) |
| **HEP** | | -20 °C -- +40 °C | -30 °C -- +50 °C | 2.3 W + incident solar flux, TBC | < 10°C / 1hour (TBC) |
| **ACDPU** | | -20 °C -- +50 °C | -30 °C -- +60 °C | 2 W | < 10°C / 1 hour |
| **CPP** | | -30 °C -- +60 °C | -50 °C -- +100 °C | 5 W | +/- 5°C |
| **ASP** | | -20 °C -- +40 °C | -30 °C -- +50 °C | 4 W | Uncritical |

*Table 8. Thermal characteristics of Cross Scale model payload instruments.*

## 4.14 Electromagnetic Compatibility (EMC)

For a multi-spacecraft in situ plasma mission the electromagnetic cleanliness and compatibility are rather stringent. The magnetometer instrument is the key instrument regarding the stability of the magnetic field induced by the spacecraft. The value of 0.1 nT during 4000 seconds for the outer sensor is required to ensure the full sensitivity measurement range of the MAG instrument. The inner sensor will measure the time-independent spacecraft induced magnetic field to be used to calibrate the outer sensor aimed at measuring the environment magnetic field.

The electric field cleanliness requirement stems from the E2D instrument. This requires the radiated noise to be less than 15 dBμV in the range 100-500 kHz and conductivity noise to be less than 40 dBμA between 30 Hz and 500 kHz. These values are similar to the Japanese Mercury Magnetospheric Orbiter (MMO). Both the electric and the magnetic cleanliness requirements have an impact on the spacecraft design process (placement of EM interfering components) and on the Assembly Integration and Verification process. The spacecraft will have to be tested in a special facility to measure the total spacecraft induced electromagnetic field.

## 4.15 Radiation Protection

The radiation environment during the complete mission lifetime including launch and transfer has been identified to be ~ 10 krad behind 4 mm of aluminium. However, a number of instruments are currently not shielded by 4 mm of aluminium and therefore shielding analysis as a function of position of the spacecraft is needed to ensure that each instrument has the correct amount of shielding implemented. In the industrial cases a preliminary assessment of required shielding mass was performed and taken into account into the final total space element mass budget.

## 4.16 Conclusions & recommendations

In this chapter the model payload of the Cross Scale mission concept has been described including the key accommodation requirements to the spacecraft. The following conclusions can be made;

- the current model payload fulfils the science requirements for the Cross Scale mission
- the five different payload configurations increases the complexity of the spacecraft design and AIV process
- the accommodation of the instruments on the spacecraft is a multidimensional problem, which needs many iterations and is sensitive to changes in mass of the instruments
- the EMC and radiation issues with respect to the payload are critical for the final scientific performance of the instruments
- the usage of two common payload processors (one for particle instruments and one for field instruments) simplify the interfaces to the spacecraft

A number of potential changes in the composition of the payload could be implemented to reduce the payload complexity. The SST has defined the five different payload configurations to reduce the total mission payload mass as the overall mass budget was identified to be a key mission driver during the Cross Scale CDF study in 2007. Potentially, in the next Cosmic Vision phase a new detailed analysis should be performed into the reduction of the different payload configurations. Continued coordination with the SCOPE mission offers further possibilities for refining and potentially simplifying both the payload configurations and the mission programmatics.

# 5 Mission Design

The baseline for the Cross-Scale Assessment Study involves 7 ESA spacecraft. The benefit of this approach is to have a feasible, stand-alone ESA mission tailored to achieve the minimum scientific objectives described in Chapter 2, by enabling the study of coupling between two scales at a time. This approach facilitates the possibility of further involvement by international partners, in particular SCOPE and hence achieving the optimum coverage of spacecraft across all three scales simultaneously.

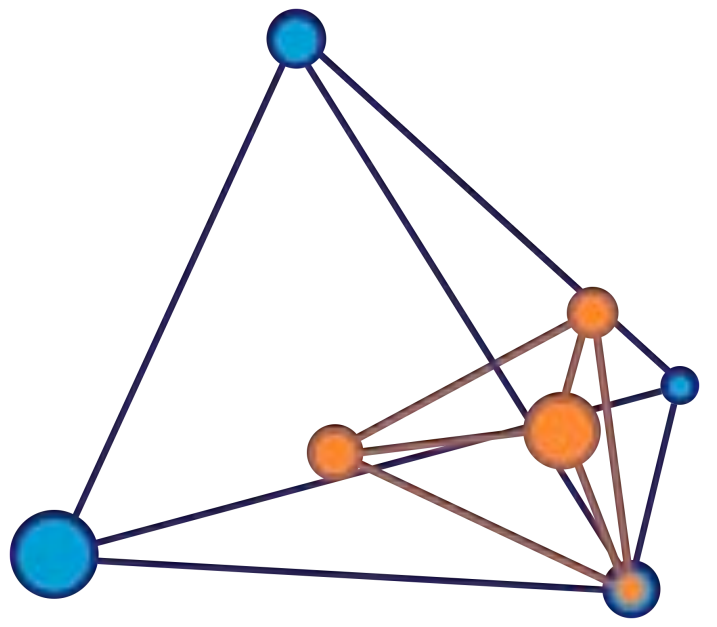

*Figure 27. Baseline ESA constellation of the seven spacecraft at two scales with a shared corner. This configuration is optimized close to apogee at 160° and 200° true anomaly, the angle between the direction of periapsis and the position of the spacecraft along the orbit.*

With the 7 spacecraft baseline, the assessment study has concluded with two technically feasible options for the successful implementation of the mission, referred to as solution 1 and 2 below. This chapter provides an overview of the common mission concepts and where appropriate focuses on the specifics of each of the industry studies.

## 5.1 Mission Requirements and Objectives

For completeness, the top-level requirements used to derive the Cross-Scale mission profile are repeated here. More detailed science and mission requirements can be found in the SciRD and the MRD.

- Cross-Scale shall perform an in-situ multidimensional scientific exploration of universal plasma phenomena occurring in near-Earth space. This requires:
  - at least two length scales to be sampled at the same time
  - a constellation of seven spacecraft
  - the spacecraft to form two nested tetrahedra, with one spacecraft at a shared corner, facilitating the coverage of two scales simultaneously
- The Cross-Scale spacecraft constellation shall visit the following relevant regions in near-Earth space where the most scientifically interesting plasma processes occur
  - Bow shock
  - Magnetosheath
  - Magnetopause and tail current sheet (reconnection regions)
  - Solar wind
- The Cross-Scale constellation shall be in an optimized spatial configuration to measure multiple scale plasma phenomena when visiting the aforementioned regions.
- The relative timing of science data between any two spacecraft shall be retrievable
- The relative position of each spacecraft in the constellation shall be known
- The acquired science data shall be relayed to Earth
- Space debris requirement: the operational orbit remains outside the protected space regions.

## 5.2 Mission Profile

All seven identical Cross-Scale science spacecraft will maintain a 10 $R_E$ x 25 $R_E$, 14º inclined, target orbit. Once deployed, the constellation will not require permanent constellation maintenance, as the spacecraft will follow their individual Keplerian orbits (with slight difference in eccentricity, inclination and radii) around Earth, with an optimized tetrahedral configuration (see Figure 27) aimed to occur close to apogee at around 160° and 200° true anomaly (see Figure 28). Each spacecraft will spin at a rate of 15 rpm to obtain the required observation cadence and also to provide rotational stiffness to the wire booms.

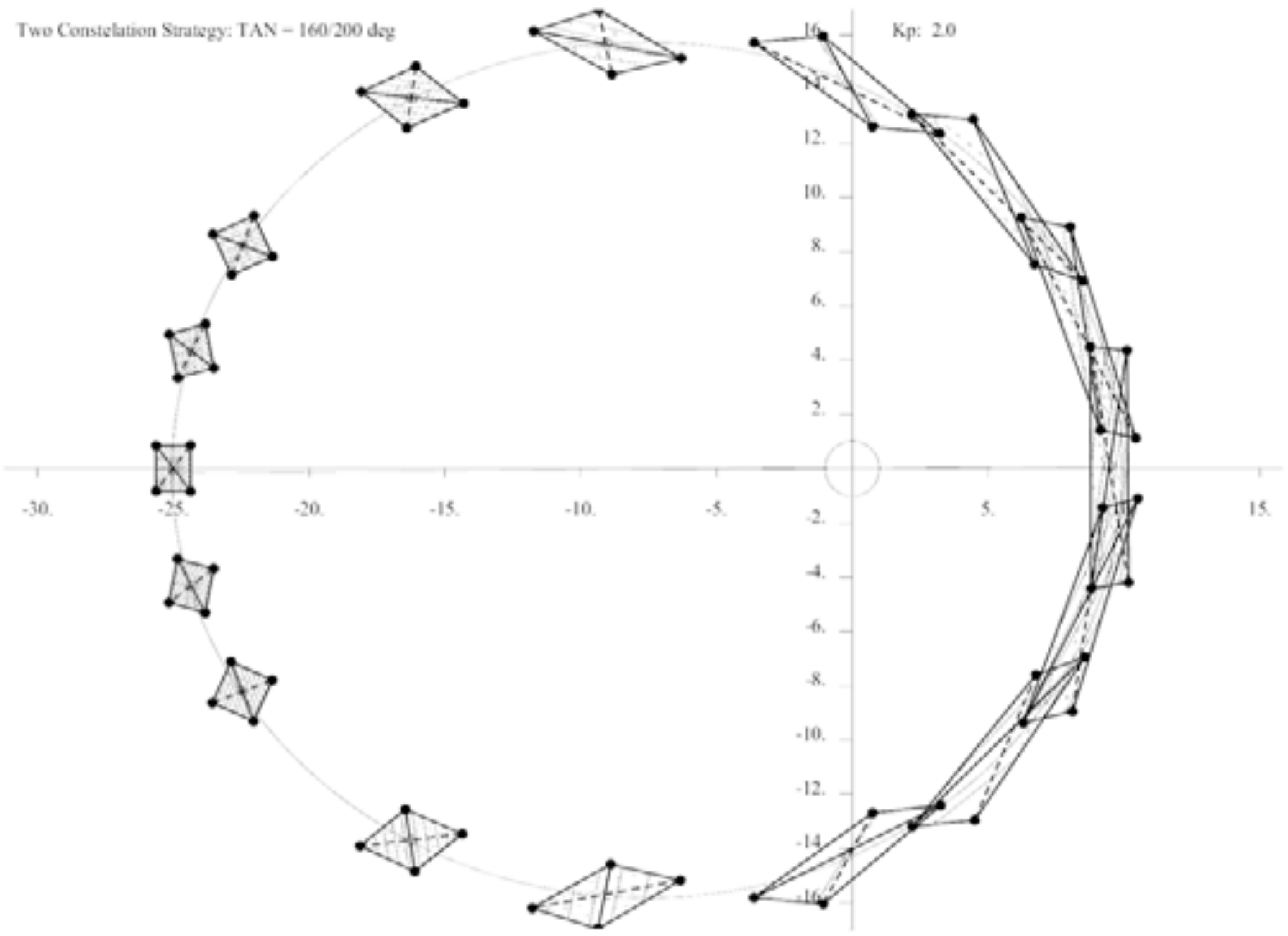

*Figure 28. Evolution of a tetrahedron geometry over one 10x25 $R_E$ orbit revolution, in this case a regular tetrahedron is obtained at 170° and 190° true anomaly.*

The scientific payload will carry out measurements along the orbit with all instruments and the resulting data will be recorded on the mass memory onboard the spacecraft and transferred to Earth during communication windows. After one year of science operations the constellation configuration will be changed and the electron scale spacecraft will transfer from the small e-scale to the fluid scale, again with a shared corner of the ion–scale. Scientific measurements will be performed for another year in this new configuration. No de-orbiting or specific parking orbit is required [53] after the mission lifetime, as for at least 25 years after this time the constellation orbit remains outside the 'protected' zone of the low Earth and geosynchronous orbit regions and is predicted to be collision free within the constellation itself.

## 5.3 Launch and transfer to target orbit

Several launcher types and single versus dual launch have been traded off, leading to cost effective SF-2B as baseline launch vehicle. All seven spacecraft will be launched on a single Soyuz-Fregat 2B from CSG (Centre Spatial Guyanais) into an elliptical orbit (200 km x 5.3 $R_E$ and 3570 kg injected mass for solution 1 or 219 km x 3.8 $R_E$ and 3703 kg injected mass for solution 2). Due to launch restrictions of the SF-2B from Kourou the insertion is limited to below 6° inclination for a reasonable mass performance. From the launcher insertion orbit the spacecraft composite, using its own propulsion stage, will transfer to the target orbit at 10 $R_E$ x 25 $R_E$, with a required inclination of 14° and an argument of perigee of 205° (mid October tailbox crossing). This requires a series of apogee raising manoeuvres and a perigee raising with combined inclination change [54]. The total required $\Delta v$ for the transfer is 1633 m/s and the final mass inserted into the target orbit is 2069 kg. An option for the transfer will use a lunar resonant orbit for the perigee raising phase, with the benefit of increase mass at the target orbit but with the drawback of a 5 month increase in the transfer period due to the additional resonance phase. This option also has an increased radiation dose, as the total duration spent in the radiation belts is increased.

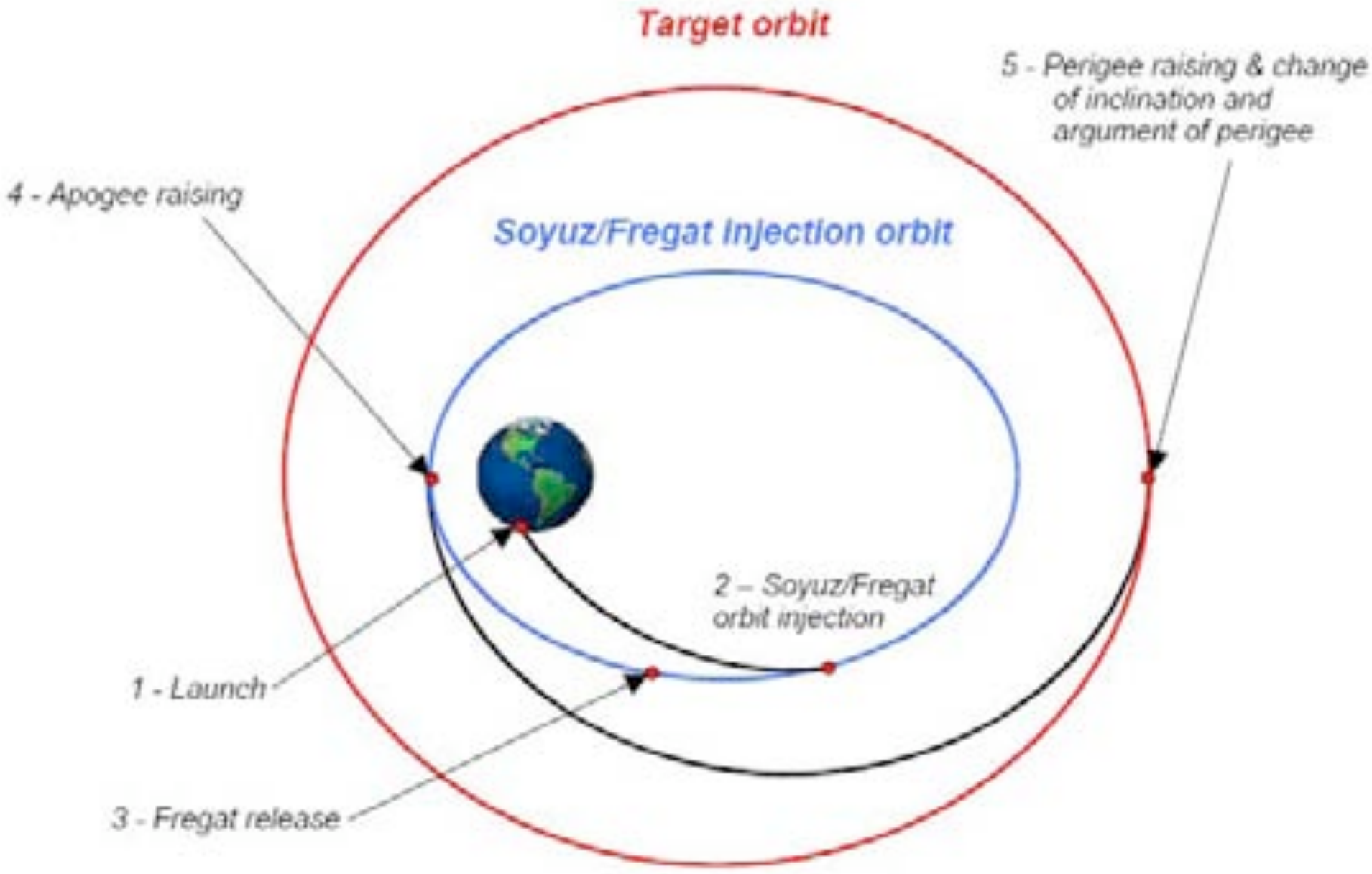

*Figure 29. Launch and transfer scenario into the $10R_E$ x $25R_E$ orbit.*

## 5.4 Transfer module design

The transfer options to bring the spacecraft from the launcher insertion orbit to the target orbit are: solution 1, a composite stacked configuration (Figure 30), where the seven spacecraft are stacked on a propulsion module (PRM) and solution 2, a composite consisting of a PRM with a cylindrical carrier structure to which the seven spacecraft are attached (Figure 31).

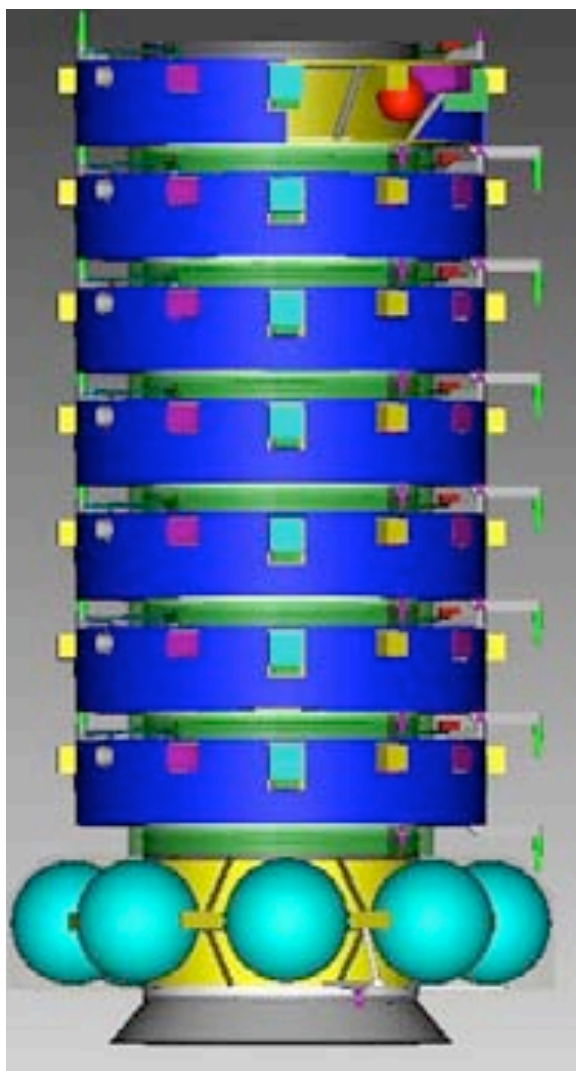

*Figure 30. Solution 1 - seven spacecraft in a stacked configuration.*

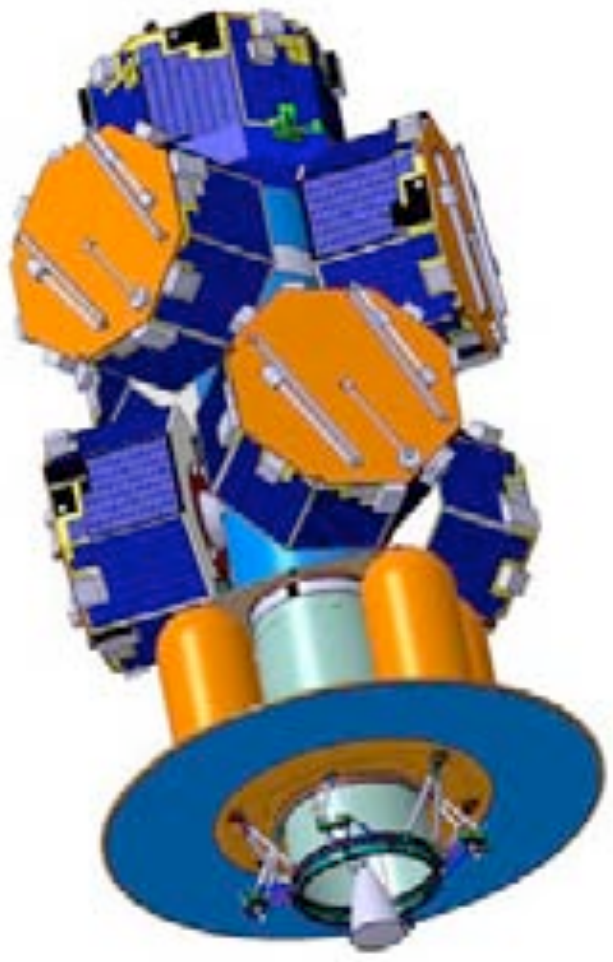

*Figure 31. Solution 2 - seven spacecraft transferred by a dispenser system.*

Both options require an efficient propulsion system with a main engine to perform the transfer (Δv in the order of 1660 m/s, or reduced by 150-320 m/s in case of lunar resonance).

The **total launch mass** of the solution 1 composite is **3547 kg** and for solution 2 composite **3692 kg** including 20% system margin.

### 5.4.1 PRM - Solution 1

This solution drives the structural design of the spacecraft, as the lowest spacecraft has to carry the rest of the pile. The mechanical connection between the spacecraft is a clamp band mechanism. A 1666 mm central tube design has been chosen for the main structure and verified to be stiff enough for the launch environment. The individual spacecraft are larger in diameter compared to solution 2, as the inner diameter drives the radial dimensions, but at the same time the height can be reduced. The complete module concept has the 1666 mm central tube surrounded by 8 propellant tanks, able to hold 1440 kg of propellant. One 400N main engine, based on a high specific impulse MMH/MON propellant pressurised system and 12 RCS (Reaction Control System) thrusters (10N) are also accommodated.

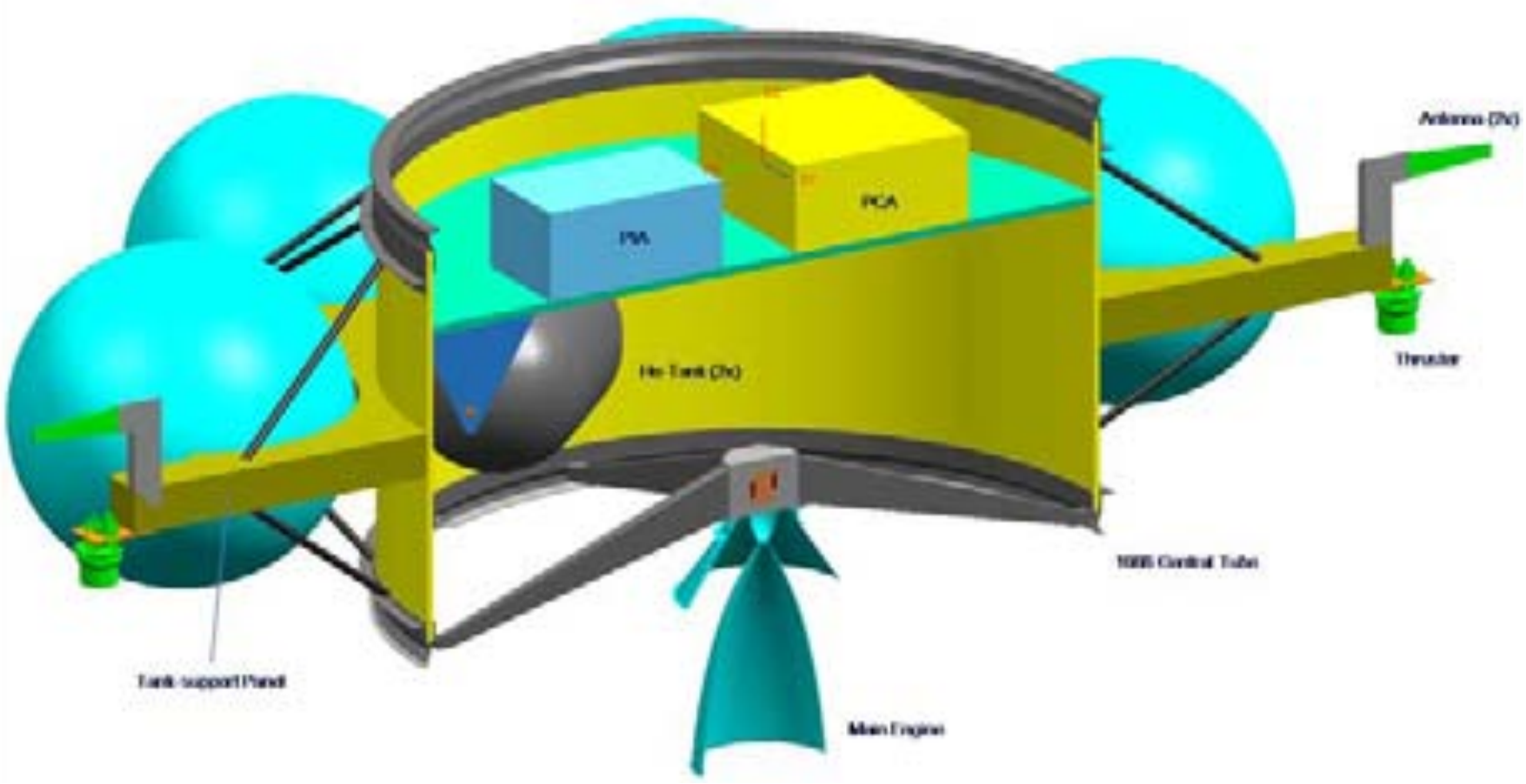

*Figure 32. Propulsion module design for Solution 1 to transfer the seven science spacecraft to the target orbit, showing the 1666 mm central tube, antenna, thrusters, main engine, pressurant control assembly (PCA), propellant isolation assembly (PIA) and propellant tanks.*

Electrical power during the transfer phase will come from the science spacecraft so the propulsion module does not require its own power source (i.e. no additional solar panels). One science spacecraft (the one closest to the PRM) will control the slowly spinning (2-3 rpm) stack during the transfer phase; hence no additional intelligence is required on the propulsion stage. The AOCS (Attitude and Orbit Control Subsystems) thrusters on the science spacecraft are required to stabilize the slow rotation of the composite during transfer, as the long axis of rotation is not stable. Propellant allocation has been included in the budgets for this stabilisation. The introduction of this slow spin introduces a similar spacecraft thermal environment to the science phase, with the exception that the radiators do not see cold space but the neighbouring spacecraft. This is not problematic, as all spacecraft will be in a hibernation mode during transfer, with the exception of the one being used to control the entire stack.

### 5.4.2 PRM - Solution 2

This solution has a more complex design, in particular the thermal aspect of the transfer phase, as each spacecraft has different thermal conditions to accommodate. To solve these issues, during transfer the spacecraft will be in shadow, generated by the PRM solar panel structure (Figure 33). This PRM array will provide power during the transfer and keep the science spacecraft within the allowed non-operational temperature range.

In contrast, the overall size and mass of the individual spacecraft can be kept at a minimum, as the dispenser is the load carrying structure. The size of the spacecraft is in this case then primarily driven by the required solar panel size.

The basic design of the propulsion module is based on the LISA-PF PRM design. The modifications from this baseline have the following characteristics.

- A sun-pointing configuration during cruise phase, with the main engine and long-axis of the composite sun-pointing.
- A modified LISA-PRM cylinder with tapered dimensions (and increased stiffening) for the Cross-Scale PRM.
- A dedicated power system, including a large solar array (6.6 m2) forming an annulus around the base of the PRM tank support structure. The solar array provides a power of 1402 WEOL and an occulting structure to the spacecraft and PRM tanks during transfer. Power is also provided by the composite battery, in eclipse or burn modes, controlled by the PRM Power Control and Distribution Unit (PCDU) (~20kg).
- Intelligence and control for the composite in transfer, provided by the top science spacecraft.

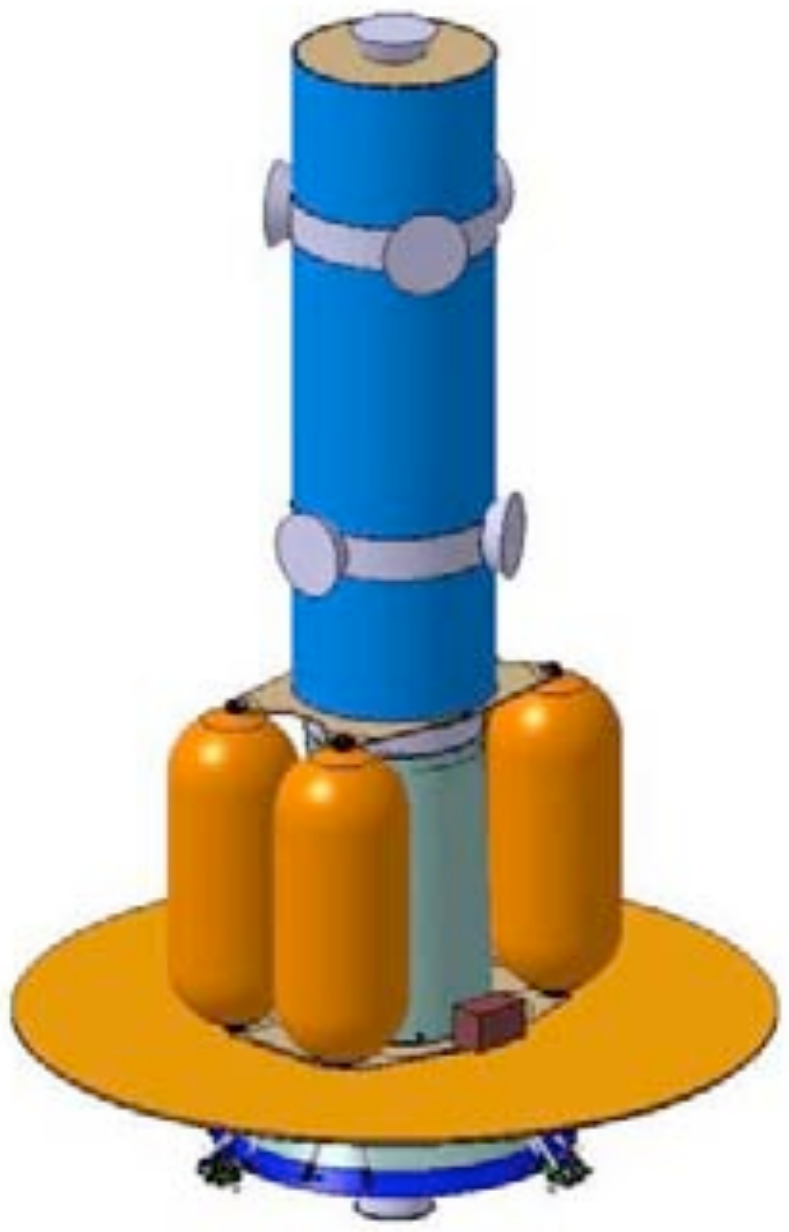

*Figure 33. Propulsion module design for Solution 2 to transfer the seven science spacecraft to the target orbit, showing the modified LISA-PF PRM, central tube assembly and the radial solar panel/sun shield structure.*

## 5.5 Deployment of the constellation

Once in their target orbit the seven spacecraft will be deployed in sequential order from the transfer stack/dispenser. The final science spacecraft to be deployed will be responsible for the stack/dispenser control and after separation the PRM will be of no further use and will be separated from the science spacecraft constellation with a small differential drift to avoid later collision. Each spacecraft will then use RCS thrusters to fine-tune its individual orbit to enable an optimised tetrahedral configuration to be formed near apogee (Figure 27 and Figure 28). The required $\Delta v$ for this transfer is below 10 m/s.

Once deployed, the science spacecraft will spin up to around 15 rpm and the instruments booms will be deployed. A further spin up manoeuvre will occur to ensure the final spin rate of 15 rpm is achieved. To obtain a three dimensional measurement of the electric field, two of the electron spacecraft (e1 and e3) will have their spin axes inclined by 20° with respect to the ecliptic plane. All other spacecraft have a small spin axis tilt (~5°), to avoid shadowing of the wire-booms sensors (See Section 4.5.)

### 5.5.1 Collision avoidance

Once the seven spacecraft are deployed into their operational orbits, collision avoidance must be assured and this is achieved by the careful selection of initial orbital deployment parameters. Detailed analysis (including all known perturbation factors such as lunar-solar and solar radiation pressure) have shown that there will be no collision between the spacecraft within the mission lifetime plus 25 years. Nevertheless a regular tracking of the individual spacecraft during the mission lifetime is foreseen, as the science measurements require knowledge of the constellation geometry. The actual measured spacecraft positions will be incorporated into an orbit propagation model, which will enable the prediction of the future geometry and to verify 'no-collision' risk. In the case of a collision risk arising a small reconfiguration can then be introduced.

As mentioned, the science of the mission requires accurate determination of the relative distances and timing between the spacecraft. On the medium (and large) scale this can be done with tracking from ground. For the small scale spacecraft a distance resolution below 125 m is required along with accuracy of 0.1 ms or less in terms of relative timing. To satisfy these requirements onboard inter-spacecraft RF-link (ISL) equipment will be used on the electron scale spacecraft. A combination of the ISL measurements, tracking from ground and use of the orbit propagation model will be sufficient to derive the spacecraft constellation geometry at any time.

## 5.6 Science Spacecraft Design

Following from the two different transfer module designs, the design of the science spacecraft are also somewhat different. In terms of instrument accommodation, both industrial studies have worked on making modular interfaces such that one instrument interface can host different particle instruments for instance. This modular approach ensures a larger flexibility in instrument delivery to the prime integrator and makes last minute changes in placement on spacecraft possible. In both cases a number of restrictions from the instruments influenced their position in the spacecraft payload compartment;

- angular separation of payload units
- Field of View requirements including avoidance of FoV intrusions for different instruments
- Avoidance of physical contact between spacecraft subsystems (e.g. booms) and instruments
- Centre of Gravity trimming of payload as part of overall spacecraft CoG analysis
- Position sharing of different instruments on different spacecraft to minimise the reduction of solar cells on spacecraft

For all spacecraft the process of accommodating five different payload configurations is an iterative process which has to take into account all the above requirements for each instrument.

### 5.6.1 Spacecraft Design-Solution 1

The spacecraft design for the stacked configuration (Figure 30) is shown in Figure 34. All seven science spacecraft have a basic similar design. The only deviations that exist are on the payload complement (see Table 2 and Chapter 4) and the mass memory allocation. Differences have been kept to a minimum to reduce the non-recurring cost (NRE).

*Structure and Configuration*

As with the PRM, the science spacecraft design is built around a 1666 mm central tube (Figure 34) with the overall spacecraft dimensions being 245 cm in diameter and 50 cm in height. Instruments and spacecraft equipment are accommodated between the central tube and the outer panel, as well on the top and bottom panel. Once in the stacked configuration accessibility to the central tube of any of the seven spacecraft is very limited and so no equipment is accommodated there. Several instruments protrude from the surface of the spacecraft via cutouts in the solar array. Interfaces for the clamp band are located on the upper and lower deck of the spacecraft to facilitate separation of the spacecraft from the stack after transfer. The structural stiffness has been verified to be compliant with the launch environment.

## *AOCS*

Star mapper, sun sensors and accelerometers are foreseen as sensors for AOCS. Currently there is no European star mapper solution available for a 15 rpm spin rate, as the Cluster solution is obsolete and the current star mapper spin rate limit is at 10 rpm. However a solution exists which only requires a software update to cope with the 15 rpm requirements. An additional slit sun sensor is used to give a precise sun-pulse timing signal, as required by the payload and for initial setup of the star mapper. The AOCS is designed to cope with the requirement to control the stack during the transfer phase and to handle the less demanding science phase.

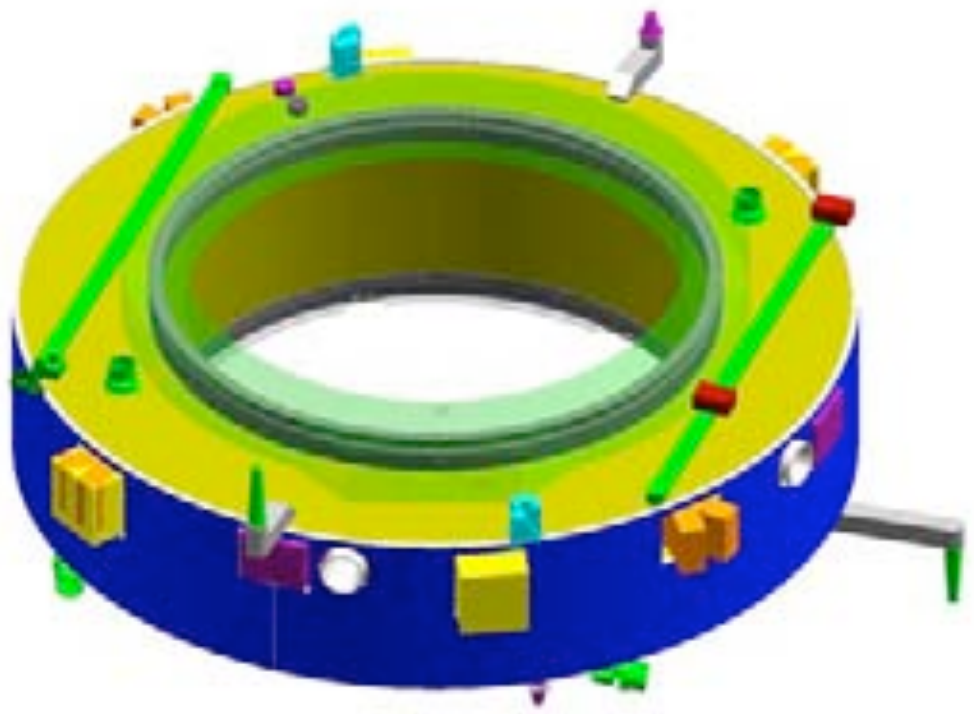

*Figure 34. Design of the science spacecraft for the stacked configuration (solution 1)*

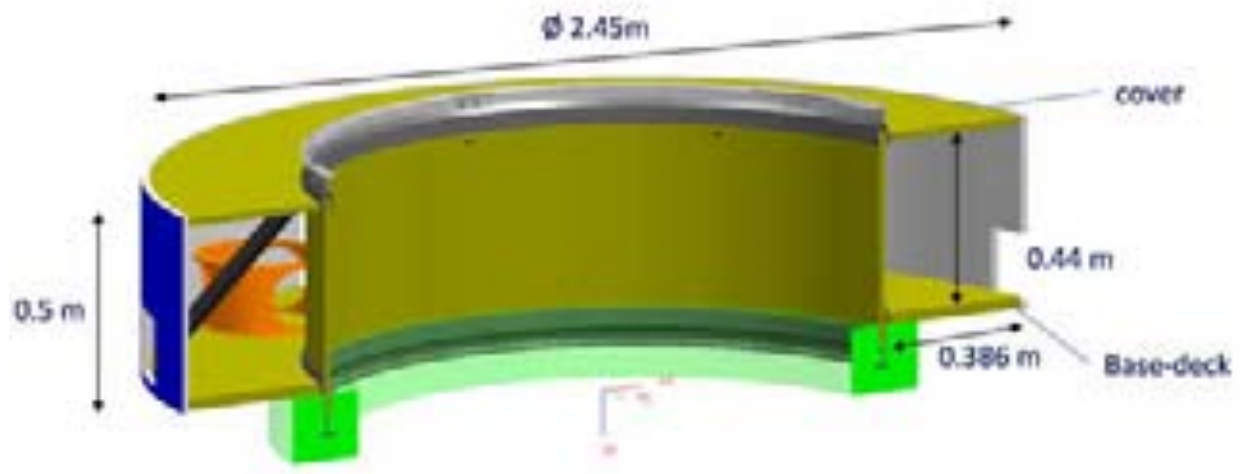

*Figure 35. Main dimension of the science spacecraft (solution 1)*

## *Propulsion*

The propulsion system design for the science spacecraft (Figure 36) is based on a simple mono-propellant system in blow-down mode (where the pressure decreases, as with an emptying balloon, rather than a pressure regulated system).

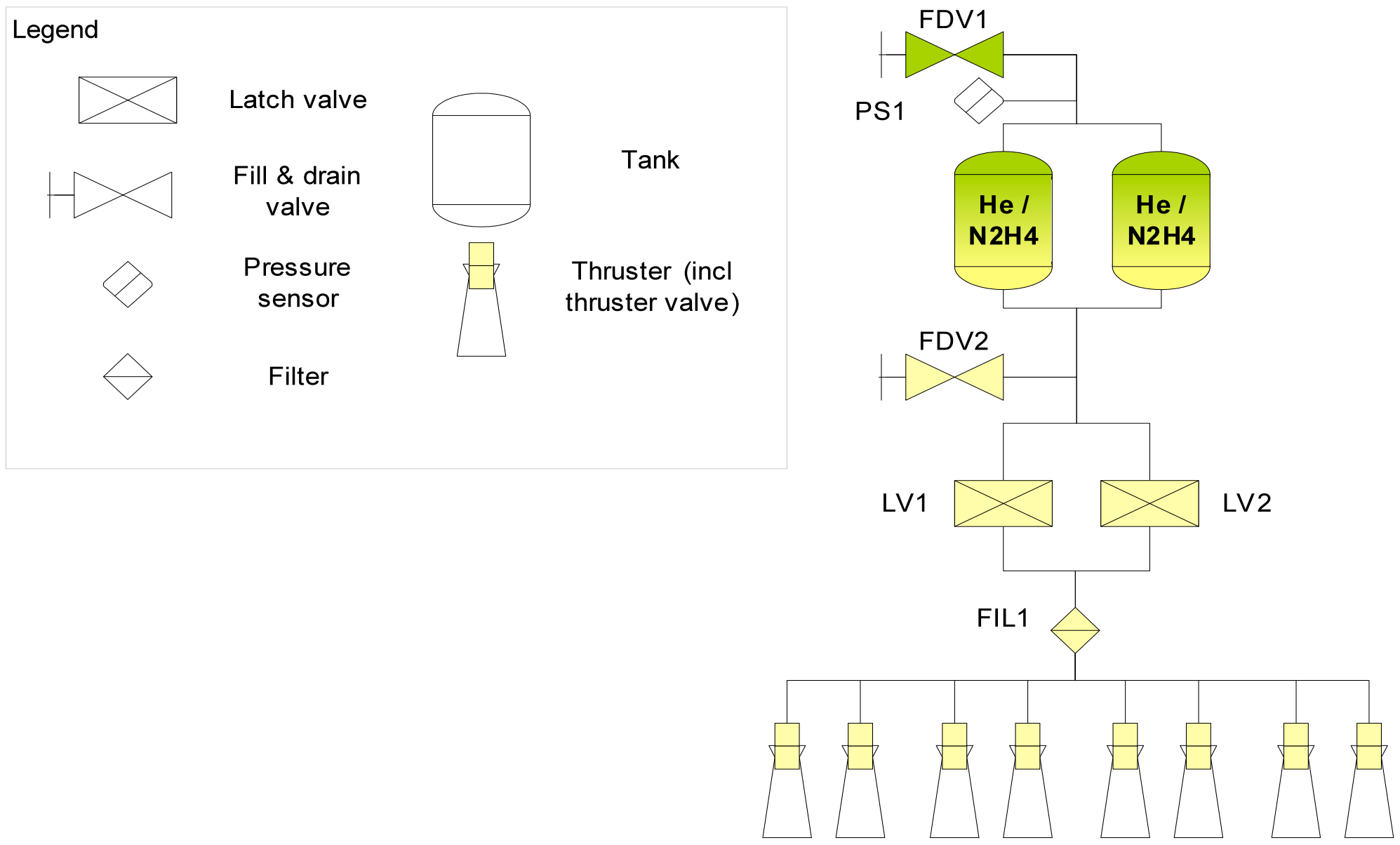

*Figure 36. Propulsion subsystem architecture for the science satellites of solution 1.*

The maximum total Δv required from the system is 125 m/s (variations exist between the spacecraft depending on their location in the configuration), using only 11 kg of propellant, stored in two small tanks. The eight 5N hydrazine RCS thrusters will be used for: (1) spin up; (2) transfer to final tetrahedral science configuration and reconfiguration; (3) spin rate and pointing maintenance and; (4) nutation control. The total dry mass of the propulsion system is 8.74 kg (including margin).

## Power

A PCDU of mass 4.8 kg will handle onboard power, primarily coming from the solar array accommodated around the cylindrical outer surface of the spacecraft. The panels are sized (12 kg, 3.7 $m^2$) to supply the power demand sufficiently, even with the loss of one string (~ 16 cells). The maximum power requirement is 236 W during science operation including simultaneous data download. The minimum power demand occurs during safe mode and is 80 W. Attention has been paid to consider the resulting shadowing by instruments protruding through the solar panel and cell free areas around the panel cut outs have been foreseen to avoid partial shadowing of a cell string. The most demanding case is the E1-spacecraft which along with E2 has with the highest power requirements but in addition has the further constraint of a 20° solar aspect angle due to the spin axis tilt, affecting the power generation capability of the arrays. During eclipse 180W are drawn from the Li-Ion battery (28Ah, 6.5kg) for a maximum duration of 3.3 hours.

## Data handling System

Figure 37 shows the architecture of the data handling subsystem. The two major tasks of the Data Handling System (DHS) are (1) control of the spacecraft subsystems and (2) storage of science data. The figure describes the connection between the satellite management unit (SMU) and the payload components, satellite avionics (including Star Tracker (STR), Accelerometer (ACC) and Coarse Sun Sensor (CSS)), spacecraft-spacecraft and spacecraft-ground telemetry tracking and control (TT&C), electronic power sub-system (EPS), propellant, and temperature and humidity controls (THC) for payload (PL) and platform (PF). The mass memory, input-output (I/O) channels, DC-DC convertors, Telemetry Telecommand reconfiguration module (TMTC-RM), Processor module (PM, A+B for redundancy) comprises the SMU. Connections to the data bus and SC-SC TT&C are via a Radio Frequency system (RFS), to the instrument via SpaceWire (Spw) and to the ground via a virtual channel (VC).

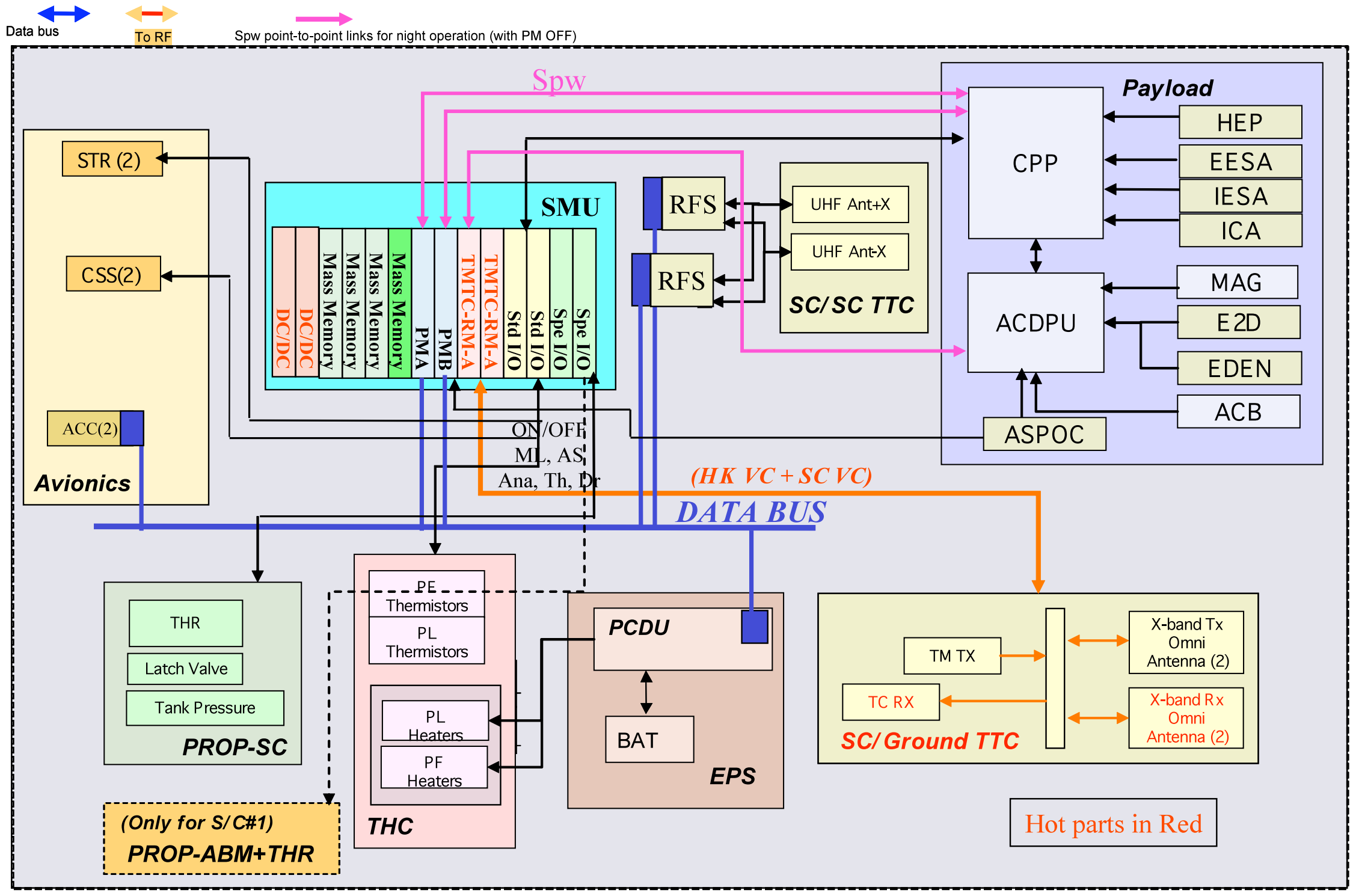

*Figure 37. Data handling system architecture.*

The instrument data are to be controlled by PI-provided Data Processing Units (DPUs): the Central Payload Processor (CPP) for all the particle instruments and the ACDPU for all the field instruments. These DPUs will handle all data compression. All data exchange and instrument commanding will be done via two spacewire links from the DHS. The onboard NAND-flash-memory for science data buffering is sized to either 256 Gbit or 1 Tbit dependent on the payload requirement. The maximum mass (dependent on the memory allocation) for the DHS is 16.6 kg. CPP and ACDPU are considered as part of the instrument mass budget.

## *Communications*

As stated previously an average orbit download data rate of around 800 kbit/s for the full spacecraft constellation was baselined. Onboard RF-equipment has been sized to support higher download rates (up to 6.4 Mbit/s), allowing for much shorter download link duration, actually limited to less than 8 hours/day.

To facilitate the downlink across the constellation, Time Division Multiple Access (TDMA) communication with ground stations is foreseen, with a pro-rata time allocation per spacecraft, corresponding to the fraction of produced data volume (~30% total orbit volume). In addition, the two electron scale spacecraft, E1 and E2, could use different download frequencies to facilitate simultaneous downlink utilising Frequency Division Multiple Access (FDMA), in cases when both spacecraft are in the FoV of the ground station (see also Chapter 6).

X-band has been baselined for the data downlink, although an S-band solution would also be possible. Due to the wide angular range between the ground station and the spinning spacecraft along their orbit, a non-directive Low Gain Quadrifilar Helix Antenna is proposed. X-Band RF power amplification will be via Travelling Wave Tube Amplifiers (TWTA) providing $13W_{RF}$ output power. In the case of S-band, Solid State Power Amplifiers would be used. Overall the proposed Telemetry, Tracking and Command (TT&C) system has a mass of 14.3 kg and a power consumption of 22W in receive and 61 W in transmitting mode.

## *Thermal Design*

The thermal design of the spacecraft must carefully consider the impact of the spacecraft spin axis (20° or 5° tilt). A total radiator surface area of 1 $m^2$ will be spread over the upper panel, close to the excess heat sources. The remaining upper panel and the entire bottom platform is to be covered with Multi-Layer Insulation (MLI). The circular side panel are covered with the solar cells. Thermal analysis confirms that the most stringent temperature constraints, given by the EESA instrument, can be met by this design (Table 8). To cope with internal power dissipation and hence temperature variation associated with various operational modes, 28 heaters are also accommodated on the spacecraft.

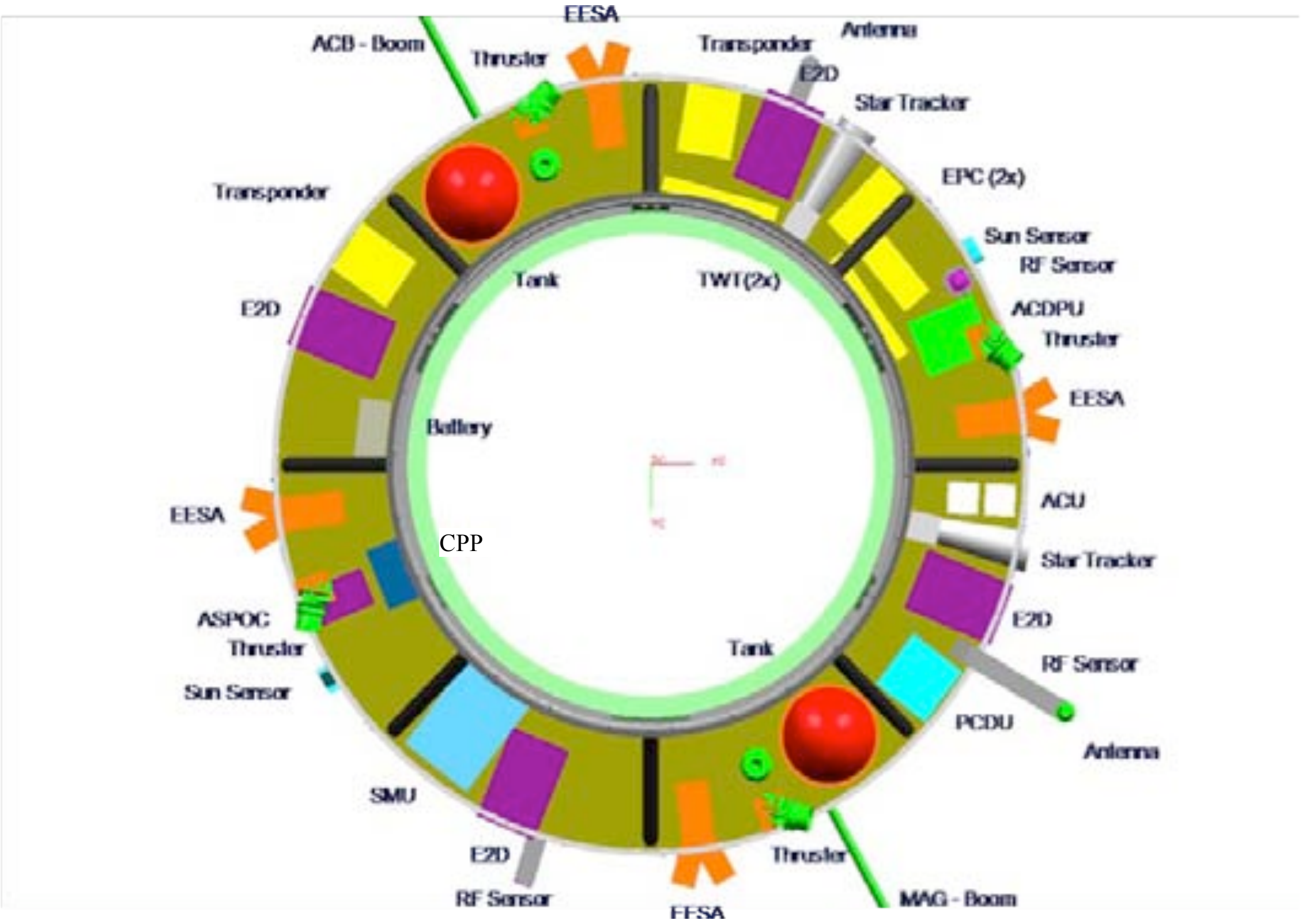

*Figure 38. Accommodation of scientific payload, described in the previous chapter, for the E1 spacecraft from solution 1. Figure also indicate the positions of Travelling Wave Tube (TWT) components of the spacecraft radio frequency (RF) system, thrusters, electric power conditioner (EPC), Power Control and Distribution Unit (PCDU) and Satellite Management Unit (SMU).*

### *Payload Accommodation*

The spacecraft have been sized for the maximum P/L demands; hence every spacecraft can be reconfigured to any P/L configuration in case of Assembly, Integration and Verification (AIV) problems. Details for accommodation (Field of View (FOV), Electro-Magnetic Compatibility (EMC) requirements, deployment sequence, interferences, etc.) have been consolidated with the parallel instrument studies. Due to the tailoring of the payload complement the P/L mass and power ranges between 15kg/15W for the E3 spacecraft to 33.3 kg/60W for the E1 and E2 spacecraft.

## 5.6.2 Spacecraft Design Solution 2

Solution 2 is the design for the dispenser-based configuration (Figure 31) and is shown in Figure 39. As with solution 1, the spacecraft are rather similar with deviations driven by payload complement and the mass memory allocation. To avoid repetition, discussion of solution 2 will focus on the deviations from solution 1, but still aims to provide a complete picture.

### *Structure and Configuration*

The spacecraft is an octagonal prism structure, 1.59 m in diameter and 0.98 in height. This platform structure consists of a top deck (the payload module) and a bottom deck, connected via four vertical walls. The eight plain solar panels are mounted at the edge of the top and bottom decks and carry some loads in order to stiffen the structure. Separation of the spacecraft from the carrier will be implemented via four explosive bolts. The dedicated payload module enables payload-platform AIV simplification and is compatible with the five payload configurations (Table 2).

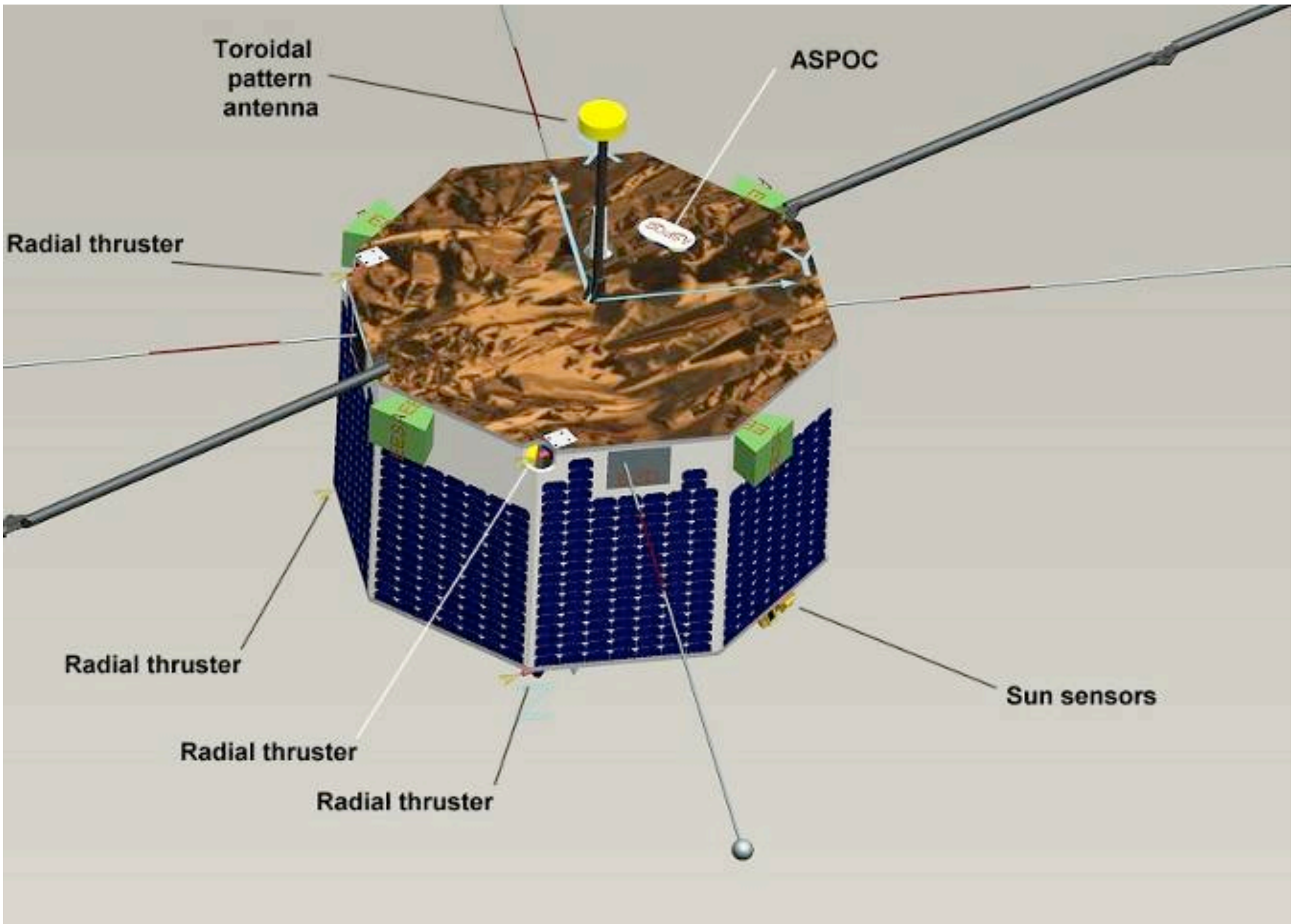

*Figure 39. Design of the science spacecraft for the dispenser-based configuration (solution 2).*

## *AOCS*

The 15 rpm spin rate drives the star tracker design and the Danish Technical University (DTU) Advanced Stellar Compass has been baselined. Sun sensors provide spin phase data for instruments and radial thruster firings. Passive nutation dampers provide a clean spin before boom deployment. Radial 1-N thrusters are located at the edge of top and bottom deck to optimise their capability. Two redundant thrusters provide $\Delta v$ along the spin-axis, i.e. for out-of-plane manoeuvres. The spacecraft will spin up to ≈5 rpm after separation (perpendicular to the orbital plane) and then fire thrusters to take up position in the constellation. The spin axis will be adjusted to the correct attitude for science operations and then the booms will be deployed and the spin increased to 15 rpm. The wire boom deployment out to the full 50 m requires an addition of 3.7 kg to the propellant mass.

## *Propulsion*

The 1-N thrusters provide impulse for both orbit changes and attitude manoeuvres. Two tanks capable of holding 15 litres of Swedish Space Corporation's (SSC) non-toxic High Performance Green Propellant (HPGP) monopropellant are placed on vertical panels on each side of the spin axis and also operate in the blow-down mode.

## *Power*

Electrical power design is driven by the spacecraft dimensions (surface available to solar arrays) and by battery sizing to cope with the ≤ 3.3 hours eclipses. The transponder is the highest power consumer (68 W) followed by the payload (≤59 W). The eight solar panels feed ≤ 237 Watts to a 28V bus system.

| **Subsystem** | **Power (Watts)** | |
|---|---|---|
| | **Science (sunlit)** | **Eclipse (3.3 hrs)** |
| Avionics | 22.2 | 22.2 |
| RF system | 68.0 | 8.0 |
| AOCS | 4.3 | 4.3 |
| Propulsion actuators | 5.5 | 5.5 |
| Thermal | 0.0 | 20.0 |
| Power | 16.8 | 16.8 |
| Platform | 116.8 | 76.78 |
| Payload | 59.0 | 59.0 |
| Total | 175.8 | 135.8 |
| Total including 20% ESA margin | 210.8 | 162.7 |

*Table 9. Power budget for science spacecraft (solution 2)*

All spacecraft are equipped with a 7.1 kg battery. The long orbital period and very few eclipses enable a high depth-of-discharge, ≈75%, while maintaining reliability. An example of the power budget breakdown is given in Table 9.

## *Data handling system*

The DHS system is based on the Swedish Prisma satellite solution with a strong heritage from ESA's SMART-1 Moon probe. The mass memory uses flash technology, 3D plus 4GB modules, the largest flash devices available.

## *Communications*

This solution meets the data downlink requirements of ~30% of a single orbit data volume. The main downlink antenna has a toroidal gain pattern for which the gain in the spin plane is +2.2 dBi and about -2 dBi at ±30° from the spin plane. This antenna is mounted on a 70 cm long boom extending along the spin axis from the top deck. A hemispherical pattern antenna is located on each of the top and bottom deck. One of these antennas is connected to the back-up transponder, and the other is connected to the inter-satellite-link system. By using a transponder with 10 Watts RF output power a downlink data rate of about 700 kbps can be achieved at apogee if the spacecraft spin axis is near the ecliptic north pole. If pointed 20º away from the north-pole the date rate at apogee that can be supported is about 200 kbps.

### *Thermal Design*

The thermal control system is driven by the requirements to keep the spacecraft sufficiently warm during the long eclipses and during the transfer phase when the spacecraft are in shadow. To avoid losing too much heat through the solar panels during the transfer phase the panels will be covered by multi-layer insulation on their backsides. During other mission phases the excess needs to be radiated. This takes place via a radiator on the bottom deck. As this radiator will face the carrier during the transfer phase, there is a reduced need for power from the composite to maintain spacecraft temperature.

### *Payload Accommodation*

The upper spacecraft deck is used as standardized payload deck to accommodate the five different P/L configurations. A general overview of the accommodation availability on the platform for the instruments is shown in Figure 40. This case shows all instruments accommodated, where tailoring to a particular scale would result in the removal or replacement of the relevant payload across the deck.

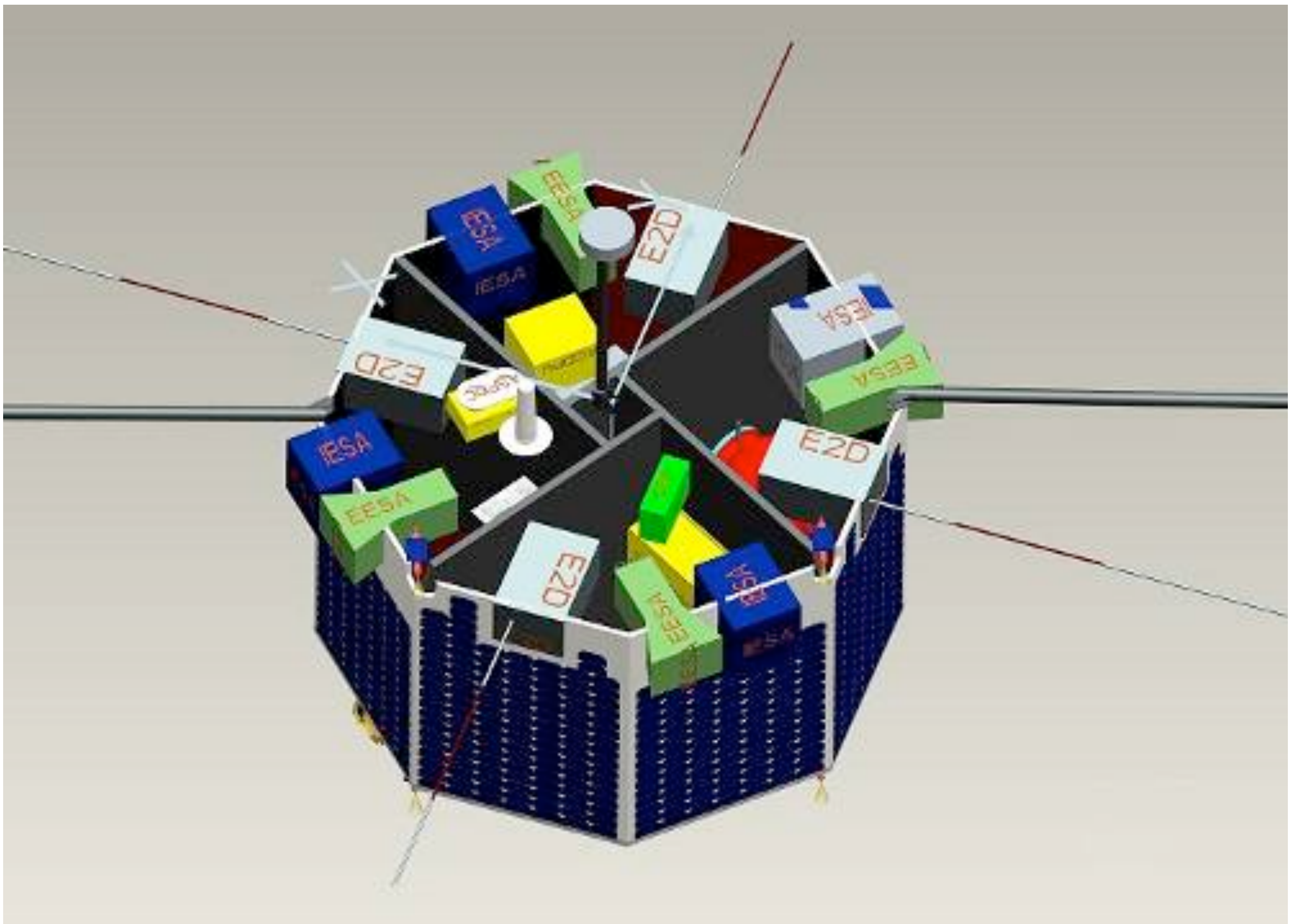

*Figure 40. Payload accommodation for science spacecraft (solution 2).*

## 5.7 Electromagnetic Compatibility

The MRD lists the EMC requirements of the spacecraft and payload. A similar approach as for Cluster will be adopted to keep EMC issues within the required limits. Surface charging of the entire spacecraft needs to be limited to 1V maximum difference, requiring a conductive spacecraft surface. Electrostatic cleanliness is achieved by using conducting MLI, paint and solar cell cover glass. The DC magnetic requirement is satisfied with a stiff boom of 2.9 m length enabling mounting of the sensors away from the main spacecraft body.

## 5.8 Mass budget

Table 10 summarizes the mass budget for both solutions for the science spacecraft (note: solution 1 uses average p/l mass whereas solution 2 uses the maximum p/l mass). Table 11 summarizes the mass budget for the propulsion module/dispenser and Table 12 gives the overall mission mass budget. On subsystem level, mass margins have been applied according to Technology Readiness Level (TRL) status and a 20 % system margin has been applied in addition (following [54]). Both solutions fit into the launcher performance and provide a few kg of spare mass.

## 5.9 Mission Environment

Cross Scale will be deployed into a highly elliptical Earth orbit by the launcher. The spacecraft composite will then transferred to the target orbit by a sequence of apogee and perigee raising manoeuvres. During this phase the composite orbit will cross through the Earth's radiation belts on several occasions, but the crossings will be of relatively short duration and once the spacecraft reach the target orbit of $10R_E$ x $25R_E$, no further radiation belt crossings will occur. Hence the overall dose is limited. In the case of lunar resonance being used to aid the transfer, the phase duration will be increased from a 1-2 weeks to 5 - 6 months. A detailed analysis is presented in [55]. In solution 2 (with a lunar resonance strategy) 21 kg of radiation shielding are included to keep the Total Ionizing Dose or TID within acceptable limits.

| **Science spacecraft Item** | **Solution1 [kg]** | **Solution2 [kg]** |
|---|---|---|
| Structure | 71.0 | 41.8 |
| Thermal Control | 7.9 | 4.6 |
| Mechanism | 7.8 | 6.6 |
| Communication | 14.3 | 11.7 |
| Data Handling | 16.6 | 6.6 |
| AOCS | 9.7 | 3.6 |
| Propulsion | 8.7 | 11.5 |
| Power | 24.2 | 38.3 |
| Harness | 13.0 | 6.0 |
| RF sensors | 2.1 | 4.0 |
| Instruments | 27.5 | 33.3 |
| **Total Dry** | **202.8** | **168.0** |
| 20% System margin | 40.56 | 33.6 |
| Propellant | 11.9 | 12 |
| **Total Wet** | **255.26** | **213.6** |

*Table 10. Mass budget for science spacecraft for both solutions.*

| **Carrier Item** | **Solution1 [kg]** | **Solution2 [kg]** |
|---|---|---|
| Structure | 104.1 | 115.0 |
| Thermal Control | 13.2 | 11.0 |
| Mechanism | 0.0 | 0.0 |
| Communication | 1.2 | 1.7 |
| Data Handling | 0.0 | 0.0 |
| AOCS | 0.0 | 9.2 |
| Propulsion | 94.0 | 116.0 |
| Power | 0.0 | 191.4 |
| Harness | 5.0 | 20.3 |
| | | |
| **Total Dry** | **217.5** | **464.6** |
| 20% System margin | 43.5 | 92.92 |
| Propellant | 1426 | 1375.2 |
| **Total Wet** | **1687.0** | **1932.7** |

*Table 11. Mass budget for carrier spacecraft for both solutions.*

| **Overall Mission Mass Breakdown** | **Solution 1 [kg]** | **Solution 2 [kg]** |
|---|---|---|
| 7 Science spacecraft | 1786.8 | 1668.8 |
| Carrier | 1687.0 | 1932.7 |
| launch adapter (modified) | 72.9 | 90 |
| **Total Wet incl. 20% SM** | 3546.72 | 3691.5 |
| SF-2B insertion mass | 3570.0 | 3703.0 |
| **Launcher excess margin** | 23.3 | 11.5 |

*Table 12. Total System Mass budget (note: Solution 1 is inserted into a 200 km x 5.3 $R_E$ orbit and Solution 2 into a 219 km x 3.8 $R_E$ orbit).*

## 5.10 Critical Elements and Drivers

Significant heritage exists from Cluster, Double Star, THEMIS and MMS. The main driver and critical elements for Cross Scale are:

- Number of spacecraft launched simultaneously
- Data volume and downlink driving power and hence size of solar panel arrays
- Number of P/L items to be integrated on time
- Mission AIV for seven spacecraft and associated schedule

No new technology is needed for Cross Scale. However, upgrade of existing technology is required for the following items:

- Interspacecraft link (ISL)
- Star mapper
- Mass memory
- Structures

## 5.11 Cross-Scale and SCOPE

The ESA 7 spacecraft assessment study resulted in two viable mission concepts. In parallel to these studies, JAXA have been studying the SCOPE [1] ('cross Scale COupling in Plasma universE') mission concept in cooperation with CSA. The 5 spacecraft SCOPE constellation is made up of a mother spacecraft and near daughter, to be provided by JAXA, with payloads well suited to studying electron scale physics. The 3 far daughters provide contextual information on larger scales and will be provided by CSA. Using the H2-A launch vehicle, the SCOPE constellation will be inserted into an initial orbit of 250km x 24$R_E$ and using a propulsion module will together attain a final operational orbit of 10 $R_E$ x 25 $R_E$.

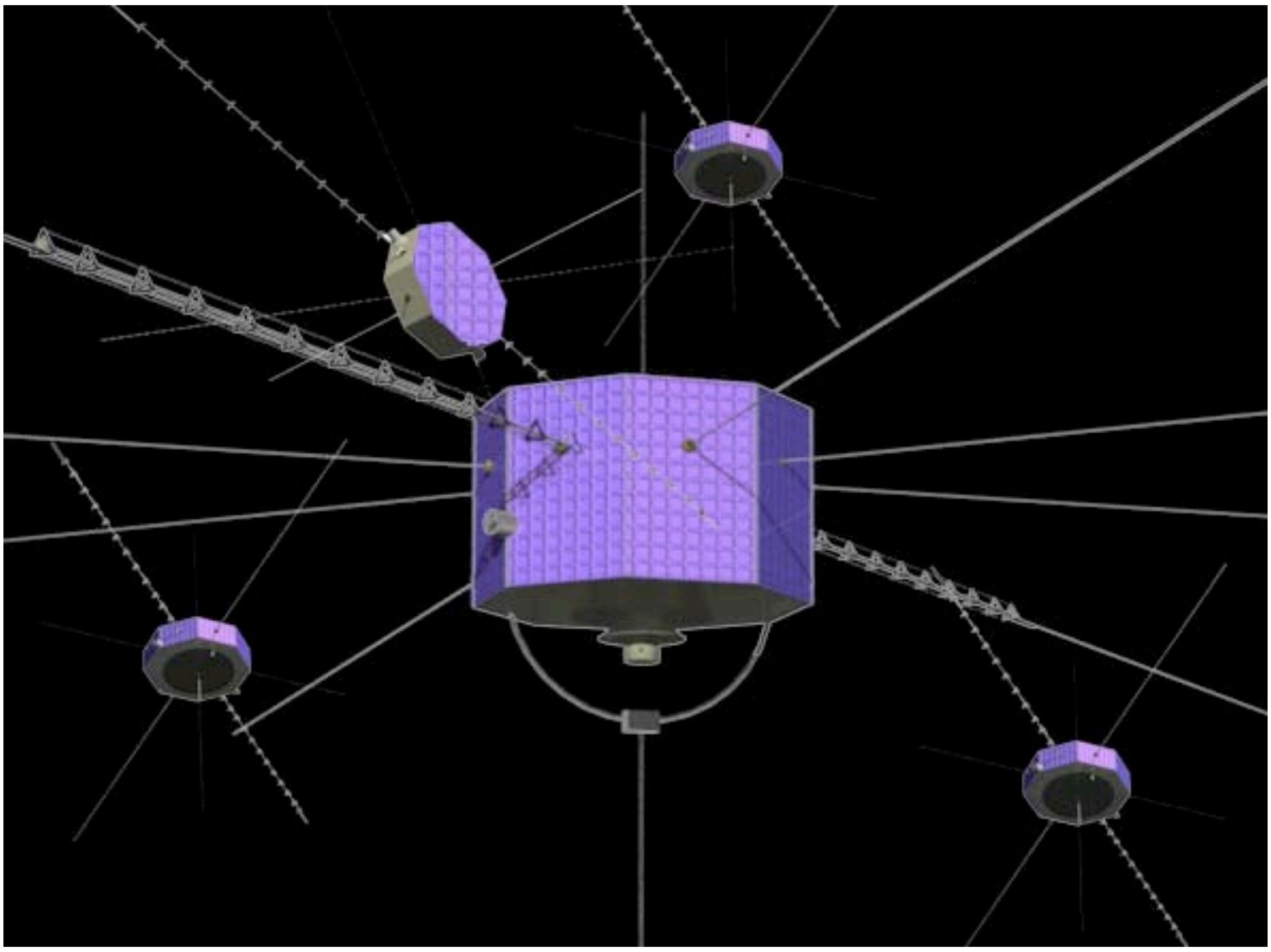

*Figure 41. SCOPE spacecraft: mother and near daughter, with 3 far daughters.*

The close alignment of science goals, in addition to similar spacecraft concepts (e.g. Cross-Scale's E1 and E3 and SCOPE's Mother-Daughter pairing) highlight the significant benefits a collaborative or merged mission would represent. In this case the combination of SCOPE spacecraft and the Cross Scale constellation would bring about the optimum situation for investigations of Cross-Scale coupling, namely the realisation of three full concentric tetrahedra without corner sharing across all three scales (electron, ion and fluid scale) simultaneously. Simplification of the payload configuration might be possible in a joint mission scenario. As indicated the proposed target orbits are identical. Only a harmonisation on the inter spacecraft link (ISL) technical solution and harmonisation of spacecraft operation for collision avoidance is required in this case. No other technical changes on the ESA science spacecraft or transfer modules are necessary to realise this opportunity. In addition, it should be noted that the H2-A launcher with the full SCOPE constellation has spare capacity of around 1.5 tonnes providing ample opportunity for a dual launch scenario with other international partners. In conclusion, the synergies of a combined SCOPE-Cross-Scale enterprise are compelling and scientifically highly desirable but have not been fully investigated at this stage, only to the extent of revealing exciting possibilities. Such a merging would be investigated more comprehensively in the next phase.

# 6 Mission Operations

This chapter provides a brief overview of the proposed Mission Operations Centre (MOC) and Science Ground Segment (SGS) components of the Cross-Scale mission. Further details of these subjects can be found in the Industrial study reports, the Mission Assumption Document (MAD) [56] and the Science Operations Assumptions Document (SOAD) [57]. A section on aspects of the proposed Science Management Plan (SMP) and International collaboration is also included.

An overview of the baseline infrastructure for the Cross-Scale mission operations is presented in Figure 42 and includes the SGS, the MOC, the instrument teams and the community. The Science Working Team (SWT) will be made up of PIs and project Scientist with supportive members from each of the other components displayed.

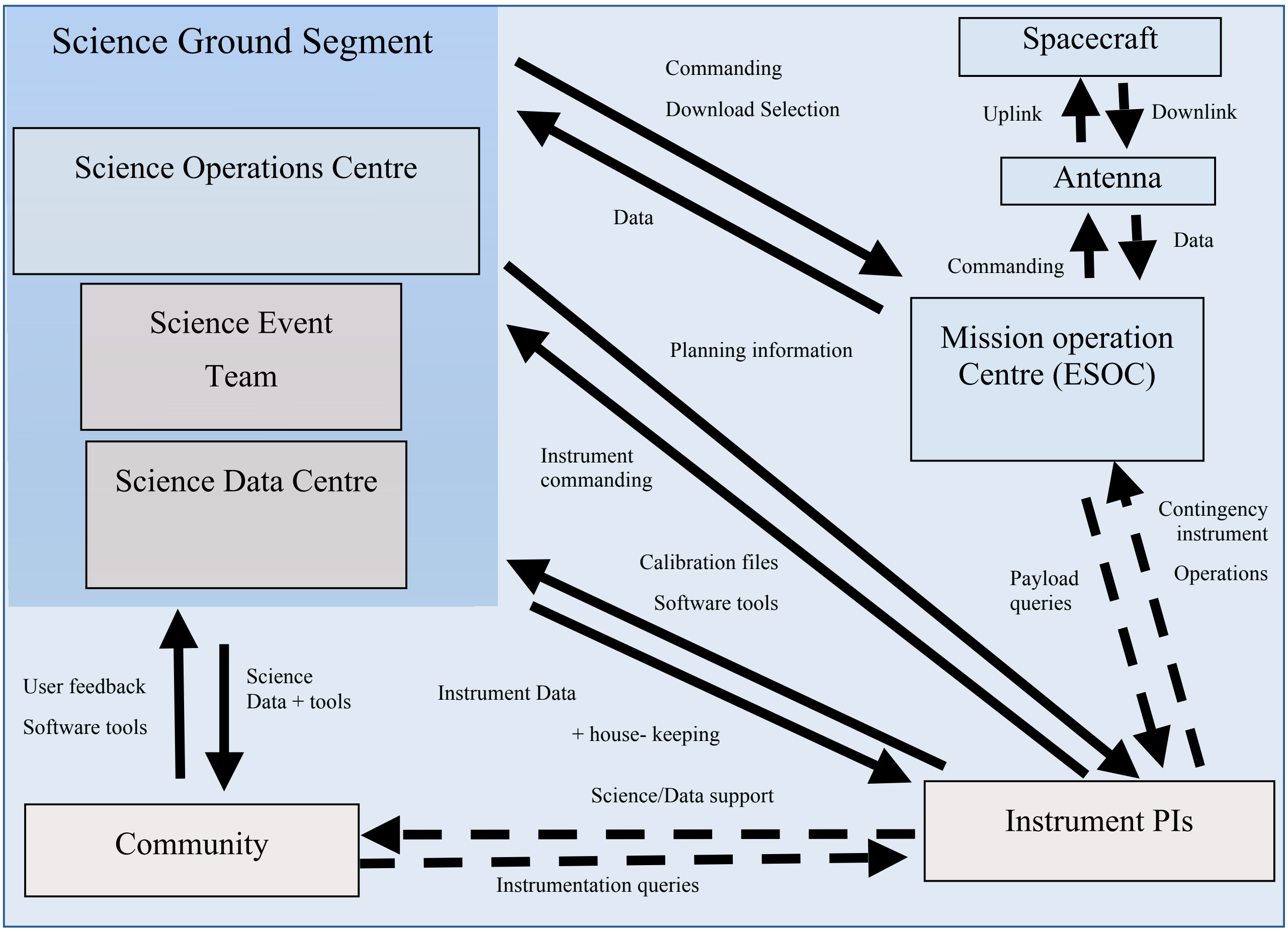

*Figure 42. Cross-Scale Operations infrastructure concept. Arrows indicate information flow, solid lines being routine and dashed being less frequent.*

After commissioning, the baseline during the nominal mission is that all instruments on all spacecraft will operate at 100% duty cycle. It is assumed that instruments are on also during eclipses. The Science Operations Centre (SOC) will implement constraints and rules regarding each instrument. These will be provided by the SWT (including the instrument Principle Investigators or PIs), with the aim to make commanding as autonomous as possible. Constellation changes during the mission, currently expected at the end of the first year of the mission, will require instrument switch-on and switch-off to prevent possible damage from thruster firings. The timing of such manoeuvres and the extent of inter-spacecraft separation re-configurations will be decided by the SWT and along with the instrument rules will constitute the long-term plan. It is envisaged that this automation will require a strict set of guidelines and policies to be implemented under the guidance of the SWT. Short-term modifications should be considered in the case of instrument and payload issues and would most likely be dealt with within the scope of the medium/short term plan.

The underlying expectation/assumption is that orbit-by-orbit 'real-time' payload commanding will NOT be required, but instead that instrument commanding will be carried out via a medium/short-term plan. This would be broken into planning periods (PPs) made up of a number of orbits (1-4 weeks cycle ~ 2-7 orbits), during which time payload operations are consolidated and uploaded to the MOC. The aim will be to balance SGS (and MOC) manpower requirements with the ability to react to problems on the short term. This includes the ability to fine tune instrument performance and therefore deliver the highest quality data. The operations architecture will be implemented to allow seamless transition between such autonomous commanding and shorter timescale payload operations, for example in the event of instrument safety concerns. This is also to facilitate the orbit-by-orbit modification of Science Event List (SEL) discussed in detail below.

# 6.1 Mission Operations Centre

The European Space Operations Centre (ESOC) will provide the MOC for the Cross-Scale mission and will prepare a ground segment including all facilities, hardware, software, documentation, the respective validation, and trained staff, which are required to conduct the mission operations. The MOC will use operational concepts proven with Solar and Planetary Science missions (Cluster, Rosetta, VEX, MEX, BepiColombo) and will adapt them to the Cross-Scale mission. The concept for the establishment of the Cross-Scale ground segment shall be the maximum sharing and reuse of manpower, facilities and tools from the Solar and Planetary Science family of missions. Sharing/reuse depends on the time phasing of the Cross-Scale mission with the corresponding on-going projects, which could be in the exploitation or in the preparation phase. All operations will be conducted by ESOC according to procedures in the long-term plan, contained in the FOP (Flight Operations Plan).

## 6.1.1 Nominal operations

Nominal spacecraft control during the commissioning and nominal operations phases shall be "offline". In particular manned operational interfaces to other entities shall require nominal working hours only, with possible exceptions for selected operations during critical phases. The contacts between the mission control centre and the spacecraft, except for collecting payload and housekeeping telemetry, will therefore primarily be used for pre-programming of autonomous operation functions on the spacecraft, and for data collection for off-line status assessment. This will be conducted by uplinking of a master schedule of commands for later execution on the S/C. It should be noted that any spacecraft commanding is planned under the presence of a spacecraft controller, or SPACON, with an on-call engineer.

The ground reaction time will be within 48h after detection of an anomaly and any anomalies dealt with in the next coverage slot. Therefore the use of any near real time reactions is limited to exceptional cases. No required real time reaction below 12 hours is assumed during any mission phase. The need for any short-term reaction is clearly defined in the FOP and unambiguously identified in the spacecraft telemetry. It is assumed that any problems are detected in the House Keeping (HK) telemetry and that flight control/contingency recovery procedures are available.

## 6.1.2 Communications

The ground station coverage for Cross-Scale must satisfy four basic requirements

- sufficient data downlink capacity;
- sufficient command and spacecraft maintenance uplink capability;
- support quick recovery of on board problems;
- support of Full Resolution science data selection cycle.

As indicated in Chapter 5, during the assessment study, a typical link budget of 800 kb/s has been assumed, such that up to ~ 33% of the total orbit full resolution science data is retrievable during an orbit. This is based on sufficient on board storage being available for a 2-orbit storage cycle for this full resolution science data.

The selection of science data, discussed in more detail below, will be supported by ESOC but will be based on off-line operations. To achieve the downlink and uplink requirements, two 15m ground stations (G/S) with coverage of up to 24h per day have been assumed (time is assumed to include link setup and antenna movement time between sessions with different Cross-Scale satellites). Both X and S band communications are considered, depending on the mission profile. A viable option to utilize a single 35m deep space antenna has also been examined, for example Cebreros or New Norcia, with extension to a second station during critical phases. In this case 8-hour passes every day (or ¼ orbit) were sufficient to cover the downlink budget, utilizing X-band communication. An example of the requirements on ground station (G/S) access during the different mission phases for a 35m antenna is given in Table 13.

| **Mission Phase** | **Coverage** |
|---|---|
| LEOP (4 weeks duration) | Dual G/S, > 16 hour per ¼ orbit |
| Moon assisted perigee raising phase option | Single/Dual G/S, dependent on criticality |
| Separation and spacecraft deployment | Single G/S, 10 hour per ¼ orbit |
| Commissioning | Single G/S,10 hour per ¼ orbit |
| Nominal Phase and Extended Phase (including maintenance and reconfiguration manoeuvres) | Single G/S, 8 hours per ¼ orbit |

*Table 13. Mission Phase and G/S coverage (35m case).*

For the operational access to the satellites, TDMA (Time Division Multiple Access) is chosen. All satellites may have the same frequency. A strategy for safe mode has to be defined that minimizes the mutual interference between the satellites. In addition a dual beam reception, i.e. the reception of the RF signals of two satellites in parallel in a single station has also been considered to optimize downlink time for the E1 and E2 spacecraft (i.e. those with the largest data generation capability). However, with all satellites having the same frequency there is likely to be an interference constraint for the scheduling of passes with 2 antennas for parallel downlinks from 2 satellites. A proposed solution and a manner to increase the operational flexibility would be to divide the constellation into two groups with different frequencies. In this case access would also be implemented via (FDMA) Frequency Division Multiple Access. In the case of the 35m X-band stations, such dual beam reception is not a valid option because of the much smaller beamwidth.

The ground station schemes considered facilitate contact periods available for maintenance / trouble-shooting / anomaly recovery. For example for the 35m case, one spare ground station slot would be available every two days. This contributes to minimise the risk of a spacecraft contingency impacting the science return of the other spacecraft. The duration of the spare slot of 4h is adequate for a first level recovery attempt. This strategy also has the advantage that the respective required engineers have a large enough prewarning time to prepare for the recovery.

### 6.1.3 Science Data Retrieval

As noted in the previous section, the downlink capacity cannot deliver 100% of the full resolution science data. Accordingly, two types of science data products are envisaged. Summary Data are comprised of reduced –resolution and onboard processed products together with Housekeeping data. This will be telemetred at the earliest opportunity and provides a 100% orbital coverage, science-quality product. The Summary data fills less than 10% of the downlink budget. Full resolution data must be stored onboard and subsets need to be selected for downlink. Approximately 25-33% of that data can be telemetred in the remaining link budget, ample to meet the science objectives.

The selection of this full resolution (FR) science data shall be based on off-line operations. It is assumed that sufficient on board storage will be available to enable the integration of the FR data selection into the nominal PP cycle such that planning work (at the SOC and at the MOC) only occurs during nominal working hours. ESA/ESOC will fulfill these requirements by providing an automated interface to the SOC concerning the downlink data selection command requests. The core aspect of this is the uplink of a list of time-tagged commands (known as the Science Event List or SEL) that identify which segments of the FR data accumulated over an orbit should be telemetred to the ground. The SEL is provided for uplink to the MOC by the SOC. To describe the principle of SEL selection we define a representative structuring of the onboard FR stored data. One orbit of FR data on spacecraft # is divided into i time periods or sectors, t, in onboard storage:

| | | | | | | | | | | | |
|---|---|---|---|---|---|---|---|---|---|---|---|
| S/C # | $t_{\#}1$ | $t_{\#}2$ | $t_{\#}3$ | $t_{\#}4$ | $t_{\#}5$ | $t_{\#}6$ | $t_{\#}7$ | $t_{\#}8$ | … | $t_{\#}$ (i-1) | $t_{\#}i$ |

The SEL for spacecraft #, for orbit n, contains the selected time periods or sectors of full resolution data:

| | | |
|---|---|---|
| S/C # | Orbit n | Sector requests: $t_{\#}1$, $t_{\#}2$, $t_{\#}28$, … $t_{\#}$(i-20), $t_{\#}i$ |

The SEL is applied to all FR data from all 7 spacecraft, with the SEL requested time periods or sectors indicated in red:

| | | | | | | | | | | | |
|---|---|---|---|---|---|---|---|---|---|---|---|
| S/C 1 | $t_1 1$ | $t_1 2$ | $t_1 3$ | $t_1 4$ | $t_1 5$ | $t_1 6$ | $t_1 7$ | $t_1 8$ | … | $t_1$(i-1) | $t_1 i$ |
| S/C 2 | $t_2 1$ | $t_2 2$ | $t_2 3$ | $t_2 4$ | $t_2 5$ | $t_2 6$ | $t_2 7$ | $t_2 8$ | … | $t_2$(i-1) | $t_2 i$ |
| S/C 3 | $t_3 1$ | $t_3 2$ | $t_3 3$ | $t_3 4$ | $t_3 5$ | $t_3 6$ | $t_3 7$ | $t_3 8$ | … | $t_3$(i-1) | $t_3 i$ |
| S/C 4 | $t_4 1$ | $t_4 2$ | $t_4 3$ | $t_4 4$ | $t_4 5$ | $t_4 6$ | $t_4 7$ | $t_4 8$ | … | $t_4$(i-1) | $t_4 i$ |
| S/C 5 | $t_5 1$ | $t_5 2$ | $t_5 3$ | $t_5 4$ | $t_5 5$ | $t_5 6$ | $t_5 7$ | $t_5 8$ | … | $t_5$(i-1) | $t_5 i$ |
| S/C 6 | $t_6 1$ | $t_6 2$ | $t_6 3$ | $t_6 4$ | $t_6 5$ | $t_6 6$ | $t_6 7$ | $t_6 8$ | … | $t_6$(i-1) | $t_6 i$ |
| S/C 7 | $t_7 1$ | $t_7 2$ | $t_7 3$ | $t_7 4$ | $t_7 5$ | $t_7 6$ | $t_7 7$ | $t_7 8$ | … | $t_7$(i-1) | $t_7 i$ |

The resultant total data for downlink is then:

| | | | | | | | | | | | | | |
|---|---|---|---|---|---|---|---|---|---|---|---|---|---|
| $t_1 2$ | $t_1 3$ | $t_1 5$ | $t_1 6$ | $t_2 2$ | $t_2 3$ | $t_2 5$ | $t_2 6$ | $t_3 2$ | $t_3 3$ | $t_3 5$ | $t_3 6$ | $t_4 2$ | $t_4 3$ |
| $t_4 5$ | $t_4 6$ | $t_5 2$ | $t_5 3$ | $t_5 5$ | $t_5 6$ | $t_6 2$ | $t_6 3$ | $t_6 5$ | $t_6 6$ | $t_7 2$ | $t_7 3$ | $t_7 5$ | $t_7 6$ |

where the total size of data is less than or equal to the down link budget (~800 kb/s) and the SEL will be strictly checked to ensure this limit is not exceeded (by the SOC and the MOC). Each of these sectors, t, contains a defined set of FR data products from the instruments on that spacecraft. The SEL will be spacecraft specific, indicating the selected sectors from the specific spacecraft, indicated above by subscripts.

The baseline implementation of the SEL occurs on the timescale of the medium term plan (~ a number of weeks) to provide ample time for MOC ground station planning. Moreover, it is assumed that the same time intervals will be selected from each spacecraft. Following from this mode of operation, it should be possible to update the SEL on an orbit-by-orbit basis: a 'Campaign' option. This process is described further in below in Section 6.2.3.

### 6.1.4 Orbit and Attitude control

The trajectory, attitude and coverage analyses required for mission preparation are carried out by the Mission Analysis component of the MOC. For mission operations the Flight Dynamics (FD) support will provide orbit determination of the stack/composite during the LEOP and optionally lunar assisted transfer using two-way ranging and two-way coherent Doppler data. Orbit determination will be carried out using range and Doppler and inter satellite link data. Attitude Control System Monitoring will include monitoring and verification of the on-board functions such as star tracker window and sensitivity setting. FD will support trajectory and manoeuvre optimisation: the manoeuvres performed for LEOP, lunar assisted transfer, tetrahedron maintenance and reconfiguration will be optimised to minimise propellant consumption and to take into account all operational conditions. Manoeuvre command sequence generation will be provided for input to the master schedule updates related to all orbit and attitude manoeuvres.

## 6.2 Science Ground Segment

The Science Ground Segment is made up of the SOC, the Science Data Centre (SDC) and the Science Event Team (SET) (Figure 42). The main activities of these groups are outlined in Figure 43.

### 6.2.1 Science Operations Centre

As with the MOC, the implementation of the SOC will benefit from the experience and heritage of the operation of missions such as the joint ESA-CNSA Double Star program, NASA's THEMIS spacecraft, but most of all ESA's own Cluster mission. The SOC will implement constraints or rules regarding the spacecraft and each instrument, which are provided by the instrument PIs and the SWT, with the aim to make commanding as autonomous as possible. The SOC will have the sole responsibility to provide, via predicted orbit data, time-tagged information for input into the planning files that are the basis for the PI-SOC commanding cycle. This information will include event times of eclipses and model magnetospheric region boundaries, including uncertainties on these positions. The SWT, with the guidance of the SET, will provide further scientific constraints and/or policies to be used in the planning files. A primary output of this activity will be rules related to which science targets are prioritised during particular parts of the mission and will be the basis of the SEL.

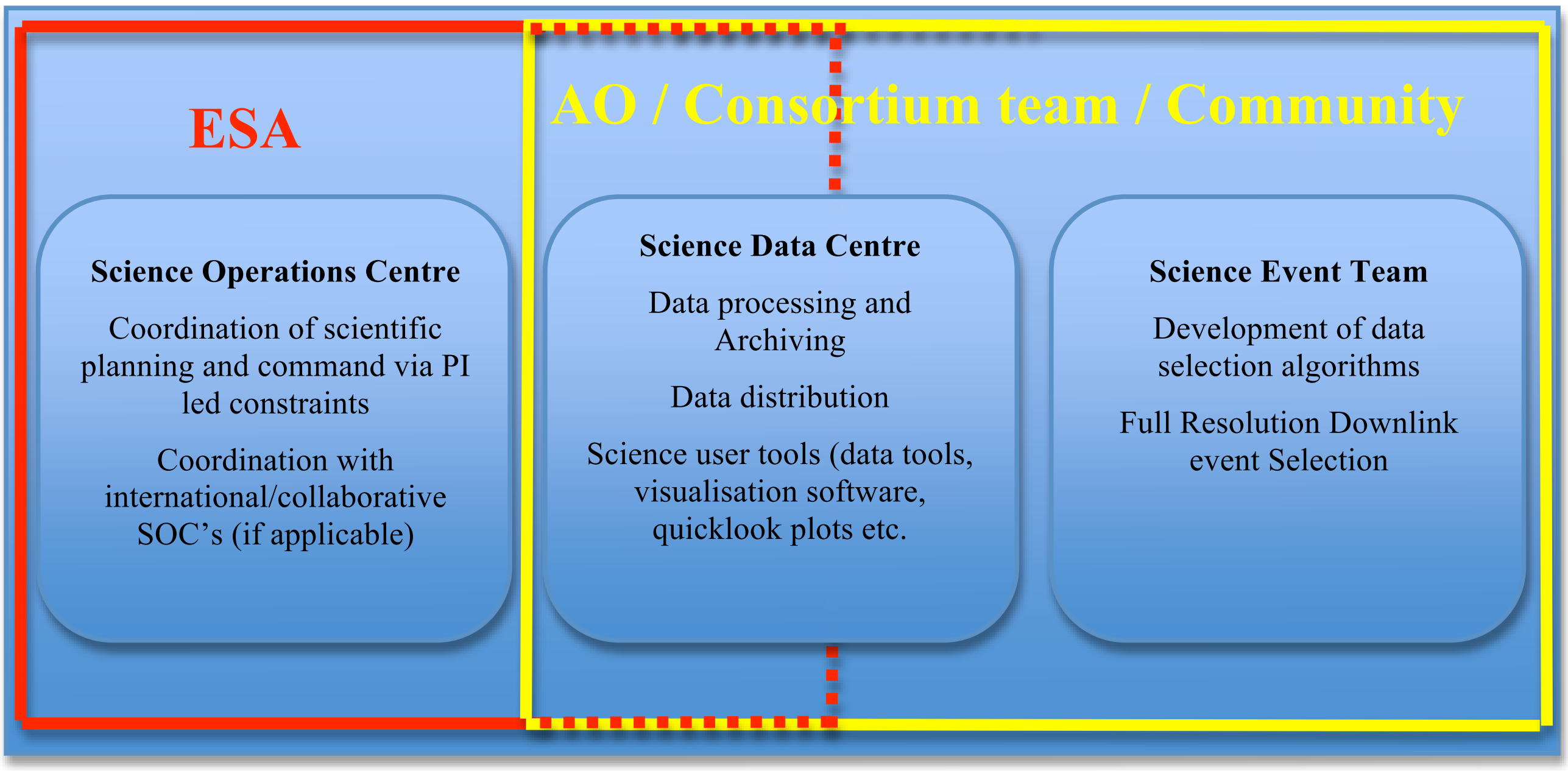

*Figure 43. SGS outline. ESA and AO/Consortium team/Community responsibilities are indicated.*

### 6.2.2 Science Data Centre

The baseline SDC is a consortium team, selected via an AO, with a PI-level status, together with an associated ESA component, which will be responsible for the development and provision of long-term archiving facility at ESA. The prime activities of the SDC are envisaged to be data processing and archiving, data distribution and the provision of user analysis and display tools. Cross-Scale will significantly benefit in this area from the Cluster and Double Star Data Centre heritage and in particular the experience of the implementation and success of ESA's Cluster Active Archive.

Due to the nature of the science, a space plasma physics mission requires a coupled approach to data analysis and exploitation in order to maintain the coupling of wave, field and particle observations as the primary focus. This is even more relevant for Cross-Scale, as the science requirements relate specifically to cross-scale coupling in plasmas, and hence coupling numerous, multi-point and multi-instrument measurements. As indicated in Figure 42, the routine data pipeline at the SDC will involve the combination of the raw data (comprising the orbit summary data and the full resolution subset) with PI-provided calibration files. The SDC will provide the infrastructure to archive and retrieve all the mission data (raw, HK, Science) including the provision of interfaces and tools to visualise and scientifically utilise the data. The archiving of the data is considered to be part of the general data system, in that data that are used by the science community during the active part of the mission should also be the archive data (same format, etc) and are therefore one and the same.

### 6.2.3 Science Event Team

The prime activities of the SET are the creation, validation, verification and modification of the SEL (Section 6.1.2). This validation would be in terms of how well it captured the required science targets, and would include the development of data selection algorithms. It is envisaged that the SET will be made up of an SWT-nominated team drawn from the international science community, with a range of science backgrounds affiliated to the mission science objectives, with the ESA component being an interface to the SOC.

As alluded to above, the primary aspect of the downlink is to return the selected portion(s) of the FR data stored onboard the spacecraft. In addition to this a lower resolution, reduced set of parameters derived from the FR data covering an entire orbit would be downlinked. The relative size of these summary data plus the HK data for the orbit comprise of a small fraction (≤ 10%) of the total download capacity. The SDC provides this summary data to the SET to monitor the validity and accuracy of the SEL and also for modification of the SEL in the 'Campaign' mode.

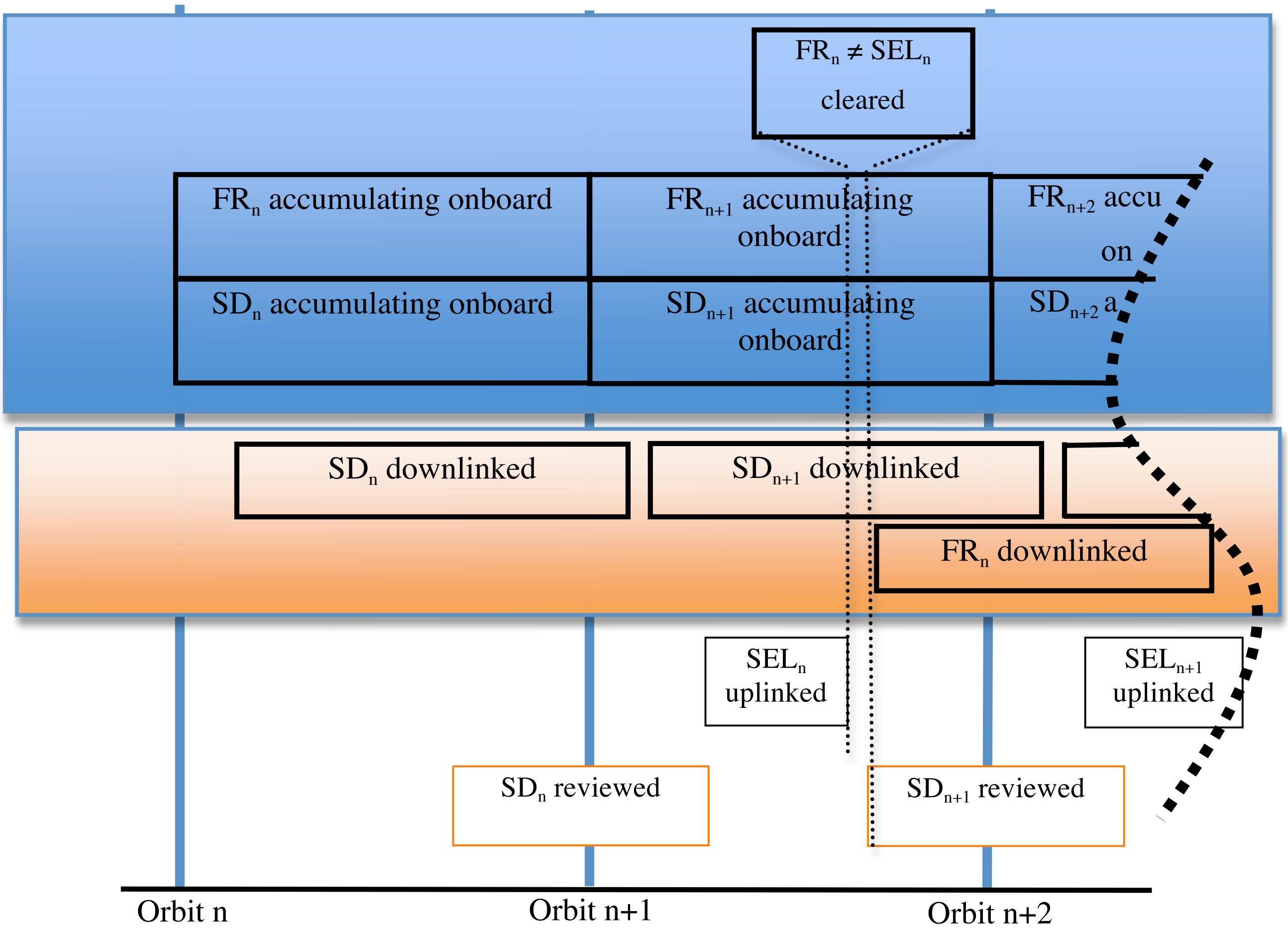

*Figure 44. Example timeline for 'Campaign' operations*

In the nominal case, during the nominal PP cycle, summary data from orbit n+1 and the pre-selected (according to the SEL n) FR data from orbit n are down-linked during orbit n+1. During a 'Campaign' style operation, SEL n is actively updated before uplink during orbit n+1, based on quick analysis of the summary data from orbit n. This process is described briefly below and in Figure 44.

During orbit n, the spacecraft take Full Resolution ($FR_n$) data for the entire orbit and at the same time telemeter the summary data ($SD_n$). During the second half of orbit n, the review of the partial $SD_n$ data begins, to identify periods of interest, where we assume a lag of around 1/2 day between data measurement and conversion to $SD_n$ and subsequent downlink. Following from this, we therefore assume that the final part of $SD_n$ will be downlinked within the first 0.5 days of orbit n+1 and that the review of this data will take up to 1 further day. As previously mentioned, the summary data amounts to <10% of the total downlink budget for the orbit. Assuming total orbit coverage for downlink this equates to around 5 hours downlink time, so 0.5 days should be ample margin for the final part of the $SD_n$ to be downloaded. Based on the review of the $SD_n$, the SET provide the SOC with an updated SEL (via the SET-SOC interface). This information is uplinked to the spacecraft during of orbit n+1 and the non–selected $FR_n$ data (i.e. $FR_n \neq SEL_n$) are cleared from memory and only the $FR_n = SEL_n$ data downloaded in the remaining time of orbit n+1 and the beginning of orbit n+2.

The implementation of this activity during operations is dependent on the time-scale of the planning cycle or planning period (PP). For example if the current PP operates on a three-week time scale, this process would incur additional support from the SOC and the MOC and would not be considered nominal. However, in the case of a 1-week PP (i.e. near orbit by orbit), the timescale for the update of the SEL would be iterated in phase with the nominal uplink timetable. This is indicated schematically in Figure 45, where a time period of uplink during the orbit is associated with the SEL. Before that deadline, a modified SEL could be uplinked (with a sufficient margin for checking etc). Due to the phasing of the orbit period with the working week, there is a requirement to modify the time allocation of a particular PP to keep operations costs down (i.e. to restrict as much as possible the SOC/MOC activities to the normal working week). For example, 3 successive PP may be made of 2 orbit lengths, followed by a single orbit PP. Part of the long term plan would involve the allocation of these 1 orbit PPs, which would be associated with the 'Campaign' style SEL update. Here, a series of such 'Campaign' PPs could be targeted at a specific region of scientific interest.

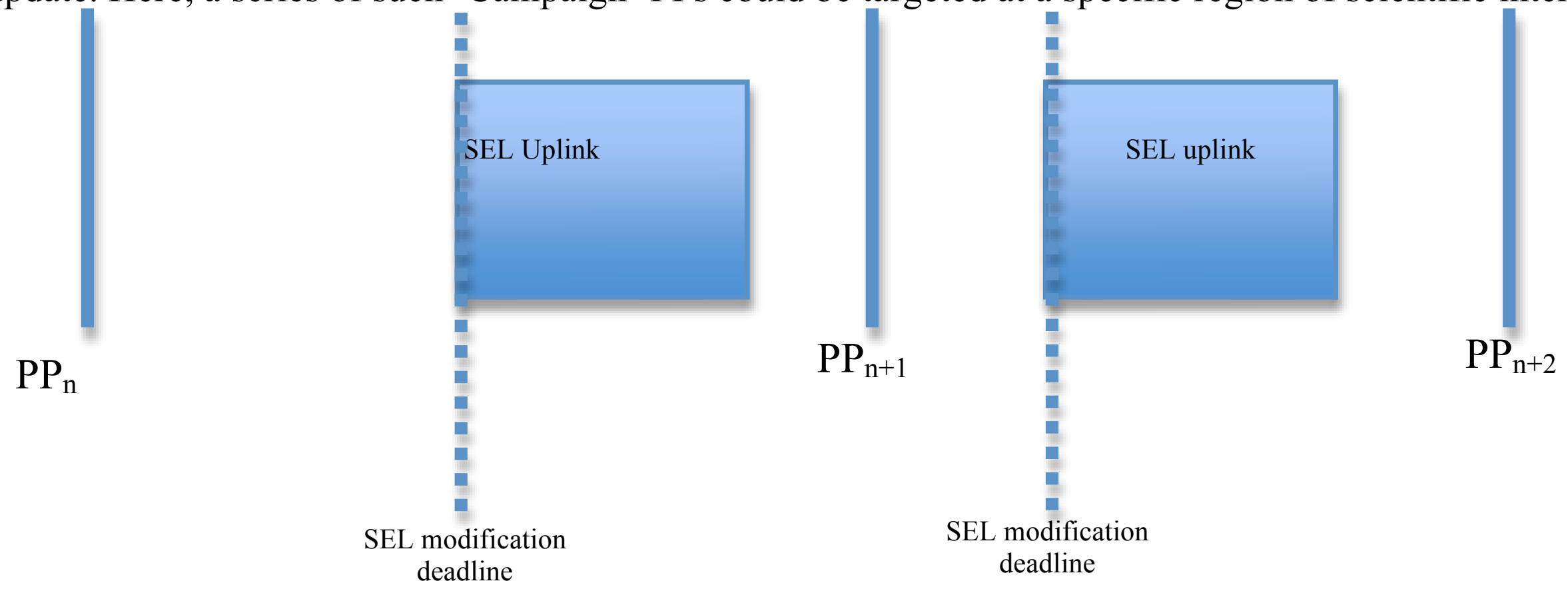

*Figure 45. Schematic of SEL uplink during successive PPs*

An alternative option of data selection could be driven by onboard selection of periods of data. This would necessitate onboard triggers being implemented to identify regions of interest. This would require the addition of more complexity onboard the spacecraft. An additional task of the SET would then be focused on the tuning of such triggers.

The 'Campaign' mode could become a more frequent or even nominal FR download scheme if the onboard data storage volume was increased. Each factor increase would result in additional time to analyse the summary data by increasing, by an orbit period, the time required to return the updated SEL. This would enable such operations to be feasible in PPs greater than one orbit.

## 6.3 International Collaboration

The baseline consideration in this study has been for a 7-spacecraft ESA mission. As mentioned in Section 5.11 it is feasible that this configuration will be augmented with the inclusion of the SCOPE mission (JAXA/CSA) and possibly other agencies (NASA, ROSCOSMOS).

The combination of the SCOPE mission with Cross-Scale can be done in a coordinated yet independent manner, in which each agency will operate their spacecraft independently. The data acquisition and selection schemes are also considered synergetic and the inclusion of SCOPE would simply involve broader membership of the SET. Increased data downlink could be realised by having more ground stations.

In this case we envisage a coordinated approach to spacecraft and science commanding, as has been carried out previously by the Agency (Double Star, Kaguya/Smart -1, Hinode, BepiColombo). The ESA spacecraft will be operated by ESA and the JAXA spacecraft by JAXA, with an open line of communication between the MOC's to exchange orbit data, especially with regard to those spacecraft that are separated by small scales. This could include a dedicated interface person or personnel at each MOC. The same is expected for the SOC, which will be responsible for the interface and coordination with the SCOPE SOC, including the exchange of the SEL and coordination of other data activities.

## 6.4 Towards a Science Management Plan

The Cross-Scale Science Management Plan (SMP) will outline the management scheme that will be used to achieve the scientific objectives of the Cross-Scale mission, up to and including the post-operations phase and the manner in which the community will be involved. Possible modes of participation to the Cross-Scale mission are: Principal Investigator (PI), Co-Principle Investigator (Co-PI), Co-Investigator (Co-I), Guest Investigator (GI) and Inter-Disciplinary Scientist (IDS). PIs will head Instrument consortia from which consortia members can be recognised as Co-Is. In the case of large consortia where major developments are carried out in a country or institute other than the PI, a Co-PI may be appointed. GIs will participate in the data collection and analysis of the mission. Finally, IDSs are considered to be expert in overarching science themes, for example connecting astrophysical or laboratory phenomena to the Cross-Scale observations.

A fundamental aspect of the science involved in Cross-Scale necessitates the rapid dissemination of data. To facilitate this, it is envisaged that an open data policy be imposed, such that the SDC provide the community as soon as possible the best calibrated data from the Cross-Scale mission. The SDC will be responsible for communicating instrument calibration activities and updates to the community and for the long term archiving of the data. Such a single entity data centre/archive is more efficient, reduces overhead and also ensures quality and consistency control. It is also much more efficient for the community to have a 'one-stop shop' for data. Due to the central role of the SDC it is expected that it will also be a PI-level activity in the mission.

The 'Campaign' mode represents an interesting community engagement opportunity, where an AO would be made for proposals for campaign targets during a short number of orbits or PPs. The winning proposals would then conduct the SEL creation with the support of the SET.

In terms of broader communication to the public, outreach activities will be a highly active aspect of the mission. Cross-Scale is an ambitious undertaking targeted at quantifying fundamental, complex processes. The goal of any outreach programme should therefore capture the essence and universality of the key phenomena. The spacecraft will be studying shocks, reconnection and turbulence within the near-Earth environment. Thus solar storms, geomagnetic activity, and their consequences (spectacular aurorae, telecomunications and technical systems outages and bio-hazards) provide visual and tangible vehicles to convey the mission message. Cross-Scale will tap into the experience of previous missions in engaging the public, such as the Cluster top story activities [58] which are regularly percolated into both scientific and mainstream news media and the THEMIS mission education and public outreach activities. Here dedicated education resources are available to the public to engage on mission related activities [59].

# 7 Management

## 7.1 M-Class Procurement Approach

This chapter describes one possible scenario of the management approach for the Cross-Scale mission implementation, which may of course evolve as the mission progresses through the various phases. The procurement schedule and model philosophy described herein is a result of the assessment study. The procurement approach is based on a fixed price contract instead of a cost reimbursement type of contract.

## 7.2 Procurement Philosophy

The proposed procurement scheme for Cross-Scale is based on the concept that the payload (instruments and associated processing, data handling and control components) will be provided by national agency funded Principle Investigators. The booms (except the Electric field instrument wire booms) are to be provided as part of the spacecraft.

ESA will have overall responsibility for:

- The overall spacecraft mission and design (industrial contract)
- Provision and Integration of the spacecraft bus and payload interfaces (industrial contract)
- System testing and payload integration (industrial contract)
- Spacecraft Launch and Operations (Arianespace, ESOC and ESAC)
- Acquisition and distribution of data to the Science Data Centre (ESOC, ESAC)

## 7.3 Scientific Management and Instrument selection

Following the further down selection of the Cross-Scale mission ESA will release an Invitation To Tender (ITT) for the selection of two competitive industrial contractors for the definition phase (Phase A/B1) in March 2010, which will specifically consider the payload design concepts of Phase 0/A but accounting for the design of specific instruments rather than a model payload. To facilitate this an Announcement of Opportunity (AO) will call for proposals for the payload instruments and the Science Data Centre. The AO will be carried out in parallel with the Phase A/B1 ITT.

The AO will solicit proposals of scientific investigation for the Cross-Scale mission and will be explicit in terms of the spacecraft mission, resources, technical interfaces, schedule, deliverable items and associated items. The AO will include a number of Cross-Scale technical documents, including the MRD, PDD, SciRD, SMP and a draft EID-A. In addition to these documents, and of particular relevance, the Science Management Plan will outline how the science community will be associated with the mission and how the scientific objectives of the mission will be fulfilled. Furthermore, the SMP will aim to optimise the scientific return of Cross-Scale, with emphasis on community participation, science operations and data management, access and archiving. When answering the AO, proposers will be expected to clearly identify a Principal Investigator (PI) and a lead funding agency. Details of other funding agencies should include indication of agreements with this lead funding agency. The proposal should describe the scientific objectives and the design and development of the envisaged instrumentation. This should also include the management structure of instrument consortium, detailing the responsibilities for the scientific, technical, operational and analysis aspects.

A Payload Review Committee will perform a full review of the responses to the AO and select the payload for the Definition phase. It should be noted that the selected PIs are responsible for obtaining the necessary funding from the appropriate national authorities, where the national funding agency of the country of the proposing PI is considered as the Lead Funding Agency for the instrument. Moreover, proposals in response to the AO will have to be accompanied by “Letters of Endorsement” (LoE’s) stating the awareness of the funding agencies of the scale of activities expected to be funded, both during Definition Phase and during Implementation Phase activities. The LoE should also state the willingness of the funding agencies to fund the Definition Phase activities should the proposal be selected. Shortly after the mission's adoption in the program by SPC the commitment of the Member States to deliver the payload will be requested. Such a commitment will be formalized through Multi-Lateral Agreements (MLA) between all parties involved.

Following selection, refined resource allocation and instrument interfaces will be negotiated with the PIs prior to the start of the spacecraft Phase A/B1 and will be frozen before commencing the Implementation Phase B2/C/D (i.e. at the end of the Definition Phase A/B1).

At the end of Definition Phase a scientific evaluation by ESA's scientific advisory bodies will provide a recommendation regarding the final selection of missions to go into implementation Phase. This recommendation will be provided to the Science Programme Committee (SPC) for approval and the successful candidates will move into the Implementation Phase (Phase B2/C/D) and a Prime industry contract will be selected via a further ITT.

During the instrument development phase, the project team conducts a preliminary design review, a critical design review and a flight model review. To support the project team in this and other tasks, a Cross-Scale Science Working Team will be formed, made up of PIs and the project scientist along with support from components of the MOC and SOC. This SWT will form the primary scientific voice of Cross-Scale, chaired by the Project Scientist.

## 7.4 Industrial Management

It is proposed that an industrial prime contractor, with the responsibility for the design, manufacturing, integration, testing and assembly of the spacecraft, will carry out the Cross-Scale spacecraft procurement.

Industrial contracts will be funded and placed by ESA. The responsibility for control and monitoring the contracts and provision for liaison between partners, contractors and PI groups will be with the ESA project team.

Ground segment, Launcher procurement and Mission and Science operations are the responsibility of ESA.

## 7.5 Development Philosophy

Being a multi-spacecraft mission, Cross-Scale poses a challenge to AIV, modelling and testing activities at both platform and payload level. To address these challenges, parallel integration and testing philosophies will be implemented. Furthermore, the similarity of the spacecraft introduces the benefit of modularity, recurrence and interchangability among the constituents, in addition to a learning curve factor. In the case of the payload there is also the benefit of instrument heritage (Cluster, THEMIS and upcoming missions such as MMS).

The assessment study has resulted in 2 solutions. The first is centred on the creation of a fully equipped spacecraft, the EQM (Electrical Qualification Model), which is used to perform all qualification testing before embarking on the AIV of the seven flight model spacecraft, which are only subjected to acceptance tests. The EQM spacecraft consists of a fully representative structure and EQM models of platform and instruments and its dedicated payload module can be adapted to different instrument suites. A second approach starts with the creation of 2 PFMs (Proto Flight Model) to cover all payload configurations. These models 'test' the AIV and testing chain and subsequently the other 5 PFMs are initiated at staggered intervals.

Due to the payload heritage (Cluster, THEMIS, Double Star and MMS), extensive experience for payload calibration, AIV and logistics already exist which can serve as a basis for Cross-Scale. The total number of instruments units to be flown adds up to 93 for the seven spacecraft constellation. For comparison, counting the same way for Cluster brings a total number of 72 units for the 4 Cluster spacecraft. Therefore efficient time and cost saving approaches for production procurement, AIV and testing are necessary. Even if the qualification procedure only has to be performed for one unit of a specific instrument all others have to be tested as well. In particular for instruments that are required in large numbers or that require considerable qualification testing an Engineering Model/Qualification Model approach is most practical. However, in the Cross-Scale case the most suitable approach will be to perform the qualification testing already with Qualification Models (QM). These QMs shall already consist of the full flight design and the according flight standards. In contrast to a PFM approach these units are not supposed to fly but can be refurbished to serve as a basic set of flight spares. The test of instrument QM's is recommended in order to analyse the real behaviour in particular of the electronics components. Thus, in case design changes turn out to be necessary this only requires design changes before starting the production of large quantities of the instrument flight models (FM).

## 7.6 Schedule

The Definition Phase (A/B1) system study is expected to start in July 2010 for a period of 16 months, with the objective to enable final adoption of the mission early 2012. It will include two major reviews: the Baseline Critical Review (BCR), to be held by the mid-term of the study, and the System Requirements Review (SRR), which will close the Definition Phase. In parallel Technology Development Activities (TDAs) defined for Cross Scale will be initiated directly after mission down-selection in February 2010. These TDAs will provide input for the system study and in the next mission down-selection phase. After potential mission adoption early in 2012 a prime contractor for the mission will be chosen for phase B2/C/D through open competition and by taking into account geographical distribution requirements.

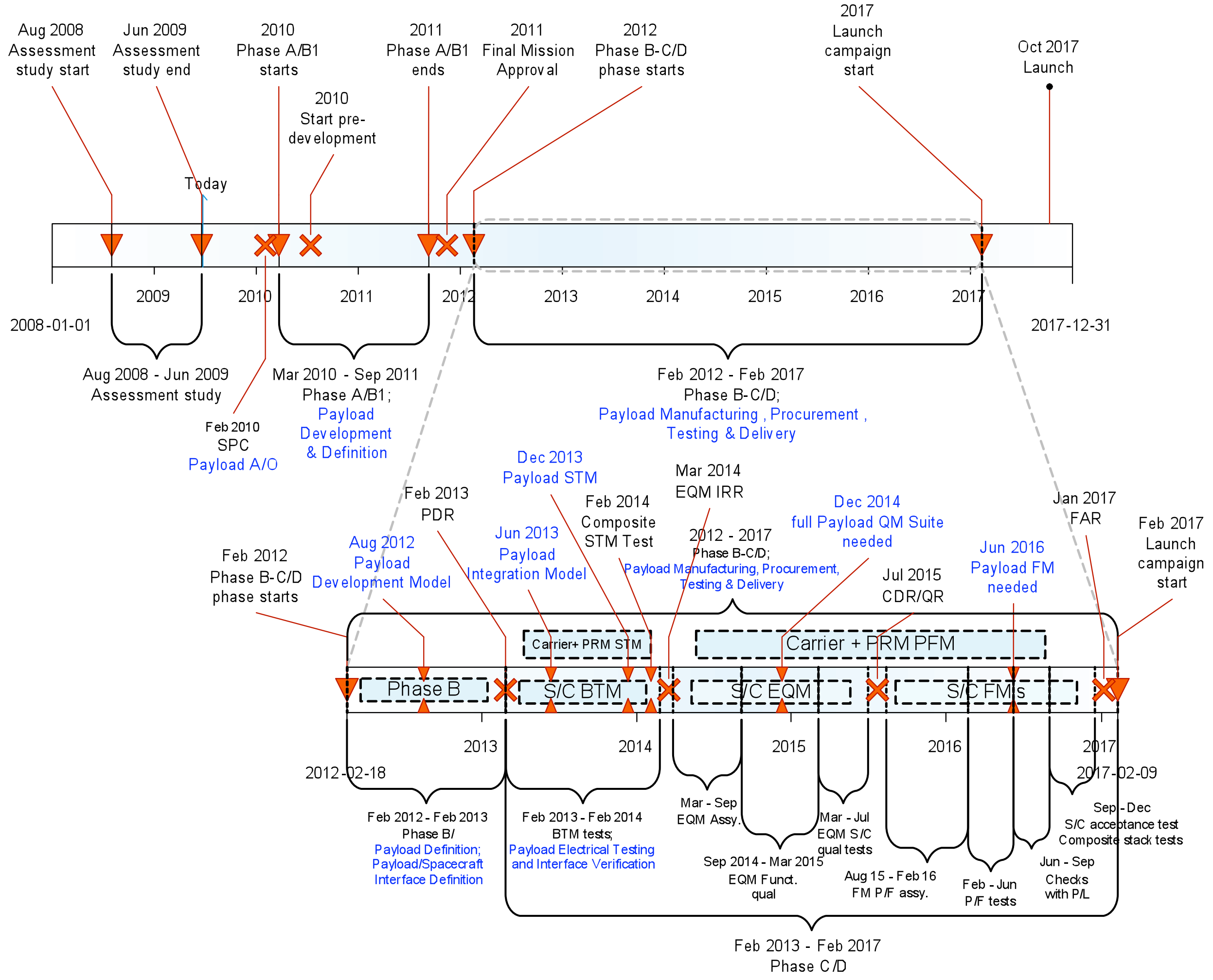

*Figure 46. Summary of the current programmatics timeline for the ESA Cross-Scale mission up until launch, including particular key detail for the payload and platform testing (EQM approach)*

Developing 7 very similar but non-identical, electro-magnetically-clean science spacecraft demands an extremely intensive schedule from start of Phase B2/C/D to launch, a period of only 5 years. An example of such a schedule, related to the EQM+7 approach discussed above is shown in Figure 46, with the critical design review (CDR) in July 2015, the Flight Acceptance Review (FAR) in January 2017 and launch campaign beginning February 2017. Figure 47 gives a more detailed breakdown of the master schedule, in this case from the initial 2 PFM approach.

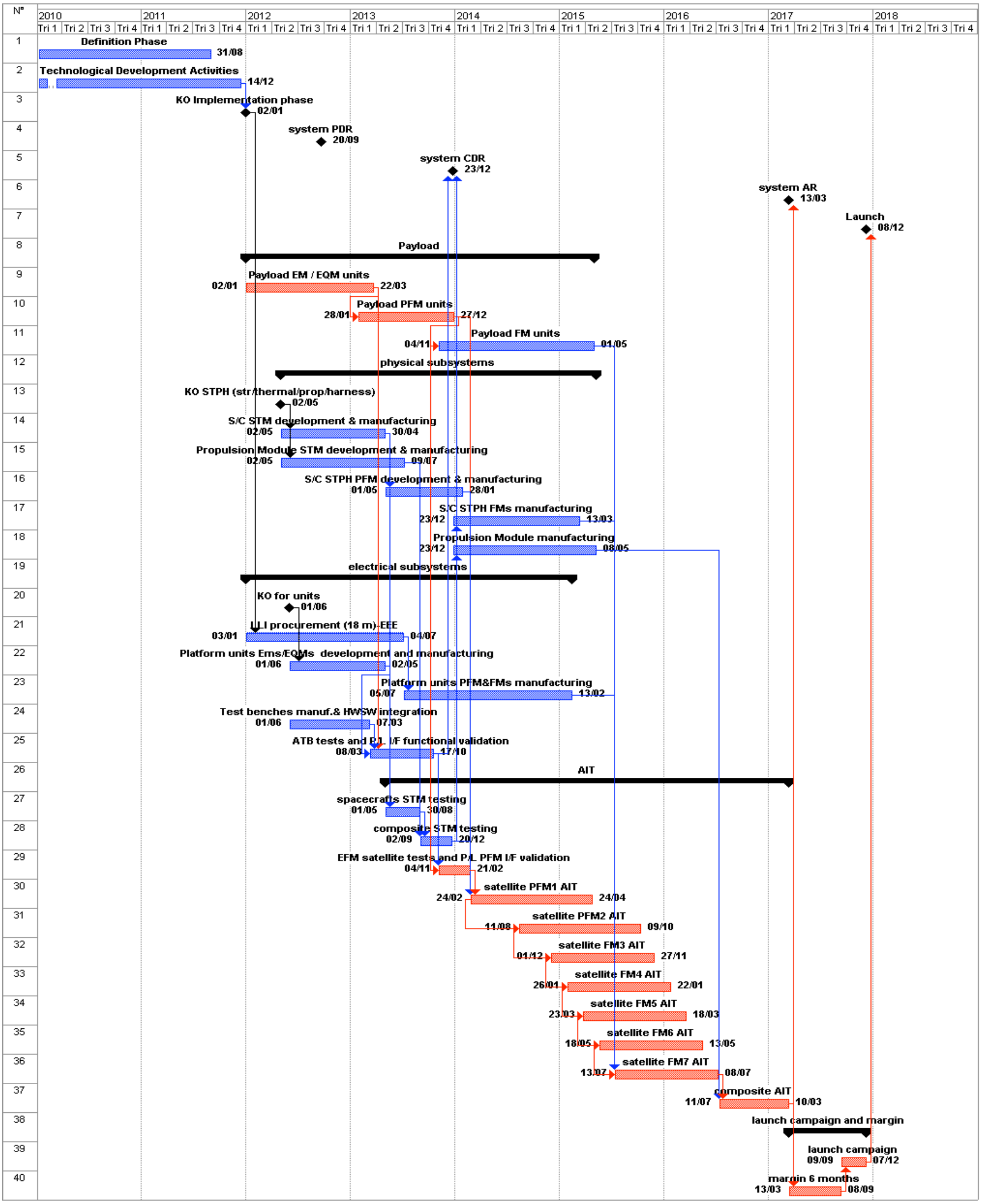

*Figure 47 Cross-Scale master schedule (PFM approach)*

# Acronyms

| | |
|---|---|
| AC | Alternating Current |
| ACB | AC Magnetometer |
| ACDPU | Field Instruments DPU |
| ADC | Analog to Digital Converter |
| ADS | Aperture Deflection System |
| ACC | Accelerometer |
| AIV | Assembly, Integration and Verification |
| AMPTE | Active Magnetospheric Particle Tracer Explorers |
| AO | Announcement of Opportunity |
| AOCS | Attitude and Orbit Control System |
| ASIC | Application Specific Integrated Circuit |
| ASPOC | Active Spacecraft Potential Control |
| BCR | Baseline Critical Review |
| CDR | Critical Design Review |
| CEM | Channel Electron Multiplier |
| CoG | Centre Of Gravity |
| Co-I | Co-Investigator |
| CNRS | Centre national de la recherche scientifique |
| CNSA | China National Space Administration |
| CPP | Central Payload Processor |
| CSA | Canadian Space Agency |
| CSG | Centre Spatial Guyanais |
| CCS | Coarse Sun Sensor |
| DC | Direct Current |
| DEMETER | Detection of ElectroMagnetic Emissions Transmitted from Earthquake Regions) |
| DHS | Data Handling System |
| DPU | Data Processing Unit |
| DSP | Digital Signal Processor |
| DTU | Danish Technical University |
| E2D | Electromagnetic Field 2D instrument |
| EDEN | Electron Density Sounder |
| EESA | Electron Electrostatic Analyser |
| EID-A | Experiment Interface Document –part A |
| EMC | Electromagnetic Compatibility |
| EPC | Electronic Power Conditioner |
| EPS | Electronic Power Sub-system |
| EQM | Electrical Qualification Model |
| ESA | European Space Agency |
| ESAC | European Space Astronomy Centre |
| ESOC | European Space Operations Centre |
| ESTEC | European Space Research and Technology Centre |
| FAR | Flight Acceptance Review |
| FD | Flight Dynamics |
| FDMA | Frequency Division Multiple Access |
| FM | Flight Model |
| FOP | Flight Operation Plan |
| FOV | Field Of View |
| FPGA | Field Programmable Array |
| FR | Full Resolution |
| GI | Guest Investigator |
| GS | Ground Station |
| HEP | High Energy Particle instrument |
| HK | House Keeping |
| HPGP | High Performance Green Propellant |
| HW | HardWare |
| I/O | Input/Output |
| ICA | Ion Composition Analyser |
| IDS | Inter-Disciplinary Scientist |
| IESA | Ion Electrostatic Analyser |
| ISAS | Institute of Space and Astronautical Science |
| ISEE | International Sun-Earth Explorer |
| ISL | Inter Spacecraft Link |
| ITT | Invitation to Tender |
| IWF | Institut für Weltraumforschung |
| JAXA | Japanese Aerospace eXploration Agency |
| KTH | Royal Institute of Technology (Kungliga Tekniska Högskolan), Stockholm, Sweden |
| LICRYL-CNR | Liquid Crystal Laboratory - Consiglio Nazionale delle Ricerche |
| LEOP | Launch and Early Orbit Phase |
| LISA | Laser Interferometer Space Antenna |
| LoE | Letter of Endorsement |
| LPC2E | The Laboratoire de Physique et Chimie de l'Environnement et de l'Espace |
| MAD | Mission Assumption Document |
| MAG | Magnetometer |
| MCP | MicroChannel Plate |
| MEX | Mars Express |
| MLA | Multi-Lateral Agreement |
| MLI | Multi-Layer Insulation |
| MMH | Monomethylhydrazine |
| MMS | Magnetospheric Multi-Scale |
| MOC | Mission Operations Centre |
| MON | Mixed Oxides of Nitrogen |
| MRD | Mission Requirements Documents |
| NASA | National Aeronautics and Space Administration |
| NAND | Not And |
| NRE | Non-Recurring Costs |
| OBDH | On Board Data Handling |
| PAD | Pitch Angle Distribution |
| PCA | Pressurant Control Assembly |
| PCDU | Power Control and Distribution Unit |
| PDD | Payload Definition Document |
| PF | Platform |
| PFM | Proto Flight Model |
| PI | Principle Investigator |
| PIA | Propellant Isolation Assembly |
| PL | Payload |
| PM | Processor Module |
| PRM | Propulsion Module |
| PL | Payload |
| PP | Planning Period |
| PSU | Power Supply Unit |
| QM | Qualification Model |
| RCS | Reaction Control System |
| RF | Radio Frequency |
| ROSCOSMOS | Russian Federal Space Agency |
| RTOS | Real Time Operating System |
| Rx | Recieve |
| SciRD | Science Requirements Document |
| SCOPE | cross Scale Coupling in Plasma universE |
| SD | Summary Data |
| SDC | Science Data Centre |
| SEL | Science Event List |
| SET | Science Event Team |
| SGS | Science Ground Segment |
| SLAMS | Short Large Amplitude Magnetic Structures |
| SMART-1 | Small Missions for Advanced Research in Technology |
| SMP | Science Management Plan |
| SMU | Satellite Management Unit |
| SOC | Science Operations Centre |
| SOAD | Science Operations Assumptions Document |
| SPACON | Spacecraft Controller |
| SPC | Science Programme Committee |
| Spw | SpaceWire |
| SSD | Solid State Detector |
| SSC | Swedish Space Corporation |
| SSR | System Requirements Review |
| SST | Science Study Team |
| STR | Star Tracker |
| SW | SoftWare |
| SWT | Science Working Team |
| TBC | To Be Confirmed |
| TDA | Technology Development Activity |
| TDMA | Time Division Multiple Access |
| THC | Temperature and Humidity Controls |
| TOF | Time Of Flight |
| THEMIS | Time History |
| TMTC | Telemetry and Telecommand module |
| RM | Reconfiguration Module |
| TRL | Technology Readiness Level |
| TT&C | Telemetry, Tracking and Control |
| TWT | Travelling Wave Tube |
| TWTA | Travelling Wave Tube Amplifiers |
| Tx | Transmit |
| VC | Virtual Channel |
| VEX | Venus EXpress |
| VHDL | (VHSIC Hardware Description Language) |
| VHSIC | Very High Speed Integrated Circuits |
| WAXS | Wave Analyzer for Cross-Scale |

# Index